\documentclass[aps,prd,preprintnumbers,showpacs,nofootinbib,amssymb]{
revtex4}
\usepackage{graphicx}
\usepackage{amsmath}
\usepackage{amssymb}
\usepackage{amsfonts}
\usepackage{bm}
\usepackage{color}

\def\be{\begin{equation}}
\def\ee{\end{equation}}
\def\bea{\begin{eqnarray}}
\def\eea{\end{eqnarray}}

\begin{document}

\title{Mass-radius relation of self-gravitating Bose-Einstein condensates \\
with
a central black hole}
\author{Pierre-Henri Chavanis}
\affiliation{Laboratoire de Physique Th\'eorique, Universit\'e de Toulouse,
CNRS, UPS, France}

\begin{abstract} 

We determine the mass-radius relation of self-gravitating Bose-Einstein
condensates with an attractive $-1/r$ external potential created by a central
mass. Following our previous work [P.H. Chavanis, Phys.
Rev. D {\bf 84}, 043531 (2011)], we use an analytical
approach based on a Gaussian
ansatz. We consider the case of noninteracting bosons as well as the case of
self-interacting bosons with a repulsive or an attractive self-interaction.
These results may find application in the context of dark matter halos made of
self-gravitating Bose-Einstein condensates. In that case, the central mass may
mimic a
supermassive
black hole. We apply our results to ultralight axions with an attractive
self-interaction. We determine how the central black hole  affects the
mass-radius relation and the
maximum mass of axionic halos found in
our previous papers. Our approximate analytical results based on the
Gaussian ansatz are compared with exact analytical results obtained in
particular limits. 
\end{abstract}


\maketitle

\section{Introduction}
\label{sec_intro}

The nature of dark matter remains one of the most important mysteries of
modern
cosmology. It has been proposed that dark matter could be made of bosons in
the form of Bose-Einstein condensates (BECs) and
that dark matter halos could correspond to giant self-gravitating
BECs \cite{baldeschi,khlopov,membrado,sin,jisin,leekoh,schunckpreprint,
matosguzman,
sahni,guzmanmatos,hu,peebles,goodman,mu,arbey1,silverman1,matosall,silverman,
lesgourgues,arbey,fm1,bohmer,fm2,bmn,fm3,sikivie,mvm,lee09,ch1,lee,prd1,prd2,
prd3,briscese,harkocosmo,harko,abrilMNRAS,aacosmo,velten,pires,park,rmbec,
rindler,chavharko,lora2,abrilJCAP,mhh,lensing,glgr1,ch2,ch3,shapiro,bettoni,
lora,mlbec,
madarassy,abrilph,playa,stiff,guth,souza,freitas,alexandre,schroven,pop,eby,
cembranos,braaten,davidson,schwabe,fan,calabrese,marsh,bectcoll,cotner,
chavmatos,helfer,hui,tkachevprl,abrilphas,shapironew,ggp,moczetal,eby2,phi6,
abriljeans,zhang,moczchavanis,desjacques,pdu,predictive} (see the
introduction of
\cite{prd1} for a short historic of this model). To
account for the mass and size of dark matter halos, the mass of the bosons must
be extraordinarily  small, between $10^{-3}-10^{-22}\, {\rm
eV/c^2}$ (see Appendix D of
\cite{abrilphas}). The quantum nature of the
bosonic particles may solve important problems that the standard cold dark
matter (CDM) model encounters at small (galactic) scales such as the cusp-core
problem \cite{cusp}, the missing
satellite problem \cite{satellites1,satellites2,satellites3,satellites4}, and
the too big to fail problem \cite{tbtf}. As a result, there is a
huge activity on the BEC dark matter (BECDM) model.
Apart
from its astrophysical
applications, this model is also interesting on a physical point of view since
it combines fundamental concepts of quantum mechanics (like Bose-Einstein
condensation or superfluidity) and gravity. It is fascinating to realize that
quantum
mechanics may manifest
itself at the scale of dark matter halos and that it may stabilize them in the
same manner that it stabilizes ordinary matter at atomic scales.

In Refs. \cite{prd1,prd2}, we have determined the mass-radius relation of
self-gravitating BECs  in Newtonian gravity described by the
Gross-Pitaevskii-Poisson (GPP) equations. We have considered the
possibility that the bosons are noninteracting or self-interacting with a
scattering length $a_s$. In Ref.
\cite{prd1} we
have used a Gaussian ansatz to obtain an approximate
analytical expression of the
mass-radius relation. In Ref. \cite{prd2} we have
compared our approximate analytical results with the
exact ones obtained by determining the ground state of the GPP equations
numerically. We found a reasonable agreement between the numerical and the
analyical results showing that the Gaussian ansatz
can provide a useful qualitative description of self-gravitating BECs at
equilibrium. Furthermore, it allows us to play easily with the parameters and to
incorporate new effects into the problem.

In the noninteracting case  ($a_s=0$), there exist equilibrium states for 
any
mass $M$ and they are stable. The mass-radius relation is given by
\cite{membrado,prd1,prd2}:
\begin{eqnarray}
\label{intro1}
R_{99}^{\rm exact}=9.946\frac{\hbar^2}{GMm^2},
\end{eqnarray}
where $R_{99}$  is the radius containing $99\%$
of the mass. The radius decreases as the mass increases.

When the self-interaction between bosons is repulsive ($a_s>0$), we found
\cite{prd1,prd2} 
that equilibrium states also exist for any mass $M$ and that they are stable.
The radius decreases with the mass but it remains always larger
than the gravitational Thomas-Fermi (TF)
radius \cite{leekoh,goodman,arbey,bohmer,prd1}:
\begin{eqnarray}
\label{intro2}
R_{\rm TF}^{\rm exact}=\pi\left (\frac{a_s\hbar^2}{Gm^3}\right )^{1/2}
\end{eqnarray}
obtained when $M\rightarrow +\infty$. Comparing
Eqs. (\ref{intro1}) and (\ref{intro2}), we obtain the mass scale \cite{prd1}:
\begin{eqnarray}
\label{intro3}
M_{s}\sim \frac{\hbar}{\sqrt{Gm a_s}}.
\end{eqnarray}
The noninteracting limit corresponds to $M\ll M_s$ and $R\gg R_{\rm TF}$.
The TF limit corresponds to $M\gg M_s$ and $R\sim R_{\rm TF}$. In that limit,
the equilibrium
states  have approximately the same radius $R_{\rm TF}$
 independently of their mass $M$.

When the self-interaction between bosons is attractive ($a_s<0$), we found
that equilibrium states exist only below a maximum
mass\footnote{This maximum mass can be expressed in various forms (depending
on the parameter used to measure the self-interaction of the bosons) as detailed
in Sec. IV of \cite{phi6}.}
\cite{prd1,prd2}:
\begin{eqnarray}
\label{intro4}
M_{\rm max}^{\rm exact}=1.012\frac{\hbar}{\sqrt{Gm|a_s|}}
\end{eqnarray}
corresponding to a radius
\begin{eqnarray}
\label{intro5}
(R_{99}^*)^{\rm exact}=5.5\left
(\frac{|a_s|\hbar^2}{Gm^3}\right )^{1/2}.
\end{eqnarray}
For $M<M_{\rm max}$ there are two branches of solutions on the mass-radius
relation $M(R)$. The equilibrium states
on the decreasing branch ($R>R_*$) are stable while the equilibrium states on
the increasing branch  ($R<R_*$) are unstable. Therefore, $R_*$ is the minimum
radius for stable equilibrium states.
The noninteracting limit corresponds to $M\ll M_{\rm max}$ and $R\gg R_*$. The
nongravitational limit corresponds to $M\ll M_{\rm max}$ and $R\ll R_*$. In that
case, the mass-radius relation is given by \cite{prd2}:
\begin{eqnarray}
\label{intro6}
R_{99}^{\rm exact}=3.64\frac{|a_s|}{m}M
\end{eqnarray}
but these equilibrium states are unstable.

One of the most serious dark matter particle candidates is the axion
\cite{marsh}. This is a
bosonic particle with an attractive self-interaction ($a_s<0$). As a result,
dilute axion stars (or more generally dilute axionic clusters) can exist only
below the maximum mass given by Eq. (\ref{intro4}). For QCD axions with
$m=10^{-4}\,
{\rm eV}/c^2$  and $a_s=-5.8\times 10^{-53}\, {\rm m}$,
we find $M_{\rm max}^{\rm
exact}=6.46\times 10^{-14}\,
M_{\odot}=1.29\times 10^{17}\, {\rm kg}=2.16\times 10^{-8}\,
M_{\oplus}$ and  $(R_{99}^*)^{\rm exact}=3.26\times 10^{-4}\,
R_{\odot}=227\, {\rm km}=3.56\times 10^{-2}\, R_{\oplus}$
which are of the order of the asteroids size. QCD axions can form mini ``axion
stars'' but they cannot form dark matter halos of relevant size. However, string
theory predicts
the existence of axions with a very small mass up to $10^{-34}\, {\rm eV/c^2}$
\cite{axiverse}. For ultralight axions (ULAs), the maximum mass given by
Eq. (\ref{intro4})
is
of the order of the galactic mass ($\sim 10^8\, M_{\odot}$ or
larger).\footnote{The precise characteristics of the dark matter particle are
not known.  For that reason, we prefer to remain general (and therefore
necessarily a bit vague) in order to cover all the possibilities. We refer to
Appendix D of \cite{abrilphas} and to Ref. \cite{phi6} for numerical
applications (see also 
\cite{baldeschi,khlopov,membrado,sin,jisin,leekoh,schunckpreprint,
matosguzman,
sahni,guzmanmatos,hu,peebles,goodman,mu,arbey1,silverman1,matosall,silverman,
lesgourgues,arbey,fm1,bohmer,fm2,bmn,fm3,sikivie,mvm,lee09,ch1,lee,prd1,prd2,
prd3,briscese,harkocosmo,harko,abrilMNRAS,aacosmo,velten,pires,park,rmbec,
rindler,chavharko,lora2,abrilJCAP,mhh,lensing,glgr1,ch2,ch3,shapiro,bettoni,
lora,mlbec,madarassy,abrilph,playa,stiff,guth,souza,freitas,alexandre,schroven,
pop,eby,
cembranos,braaten,davidson,schwabe,fan,calabrese,marsh,bectcoll,cotner,
chavmatos,helfer,hui,tkachevprl,abrilphas,shapironew,ggp,moczetal,eby2,phi6,
abriljeans,zhang,moczchavanis,desjacques,pdu,predictive}).} Therefore, ULAs can
form ``axionic clusters''
of the size of dark matter halos. For $M>M_{\rm max}$, the system undergoes a
gravitational collapse.\footnote{This may concern the solitonic core of large
dark matter halos as suggested in \cite{phi6}.} An estimate of the collapse time
has been obtained analytically in \cite{bectcoll} from the 
Gaussian ansatz. However, the Gaussian ansatz
is not able to describe the complex collapse dynamics of the system. A
detailed study of the collapse
process requires solving the GPP equations, or the Klein-Gordon-Einstein (KGE)
equations, numerically. It is then found that the system first
undergoes gravitational collapse (implosion) until collisions between axions
stop the collapse and lead to an explosion accompanied by  the 
emission of relativistic axions with a characteristic radiation
(bosenova) \cite{tkachevprl}. There is also the possibility to form dense axion
stars (or dense axionic clusters) \cite{braaten}.  Finally, the
collapse of very massive axion stars (or axionic clusters) can lead to the
formation of a
black hole \cite{helfer}. The phase transitions between dilute and dense
axion stars have been studied in Ref. \cite{phi6} with the Gaussian ansatz.
This analytical study is able to reproduce the numerical results of Braaten
{\it et al.} \cite{braaten} and to display a tricritical 
point between dilute axion stars, dense axion stars and black holes similar
to the one found by
Helfer {\it et al.} \cite{helfer}.

In this paper, we complete our former study \cite{prd1}. Using a
Gaussian ansatz, we study how the mass-radius relation of self-gravitating BECs
is modified when there is a massive object at the center of the system. In the
case of BECDM halos, the central object could represent a supermassive black
hole. Indeed, supermassive black holes are purported to exist at the centers
of the galaxies.

The paper is organized as follows. In Sec. \ref{sec_arsi} we present the exact
GPP equations describing self-gravitating BECs with a central black hole and the
approximate equations obtained from the Gaussian ansatz. In Sec. \ref{sec_part}
we consider particular cases of physical interest and identify characteristic
mass and length scales. In Sec. \ref{sec_pure} we treat the general case using
dimensionless variables. The Appendices provide additional results.
Dimensionless variables
are introduced in Appendix \ref{sec_cmmm}. In Appendix \ref{sec_g} we explain
how the general formalism
of self-gravitating BECs developed in Ref. \cite{ggp} can be generalized in the
presence of a central mass (black hole). In Appendix \ref{sec_theorem} we
derive a general expression of the
gravitational (potential) energy of a self-gravitating polytropic sphere in the
presence of an external potential, possibly created by a central black hole.
In Appendices \ref{sec_b}-\ref{sec_gft}, we derive the exact
mass-radius
relation of self-gravitating BECs with a central black hole in particular
limits of the theory. These exact results are compared with the approximate ones
obtained with the Gaussian ansatz usually giving a good
qualitative agreement.

\section{Self-gravitating BECs with a central black hole}
\label{sec_arsi}

\subsection{Gross-Pitaevskii-Poisson equations}
\label{sec_gpp}

We consider a self-gravitating BEC at $T=0$ whose complex
wavefunction $\psi({\bf r},t)$ is described by the GPP equations \cite{ggp}:
\begin{eqnarray}
\label{gpp1}
i\hbar \frac{\partial\psi}{\partial t}=-\frac{\hbar^2}{2m}\Delta\psi
+m(\Phi+\Phi_{\rm ext})\psi+m\frac{dV}{d|\psi|^2}\psi,
\end{eqnarray}
\begin{equation}
\label{gpp2}
\Delta\Phi=4\pi G |\psi|^2,
\end{equation}
where $\Phi({\bf r},t)$ is the gravitational potential produced by the system,
$\Phi_{\rm ext}({\bf r})$ is a fixed external potential, and $V(|\psi|^2)$ is
the self-interaction potential of the bosons. The mass density of the bosons is
$\rho=|\psi|^2$. The GPP equations conserve the total mass and the total energy
which can be written
as 
\begin{eqnarray}
\label{gpp3}
M=\int |\psi|^2\, d{\bf r},
\end{eqnarray}
\begin{eqnarray}
\label{gpp4}
E_{\rm tot}=\frac{\hbar^2}{2m^2}\int |\nabla\psi|^2\, d{\bf
r}+\frac{1}{2}\int |\psi|^2 \Phi\, d{\bf r}
+\int |\psi|^2 \Phi_{\rm
ext}\, d{\bf r}+\int V(|\psi|^2)\,
d{\bf r}.
\end{eqnarray}
The energy includes the kinetic energy $\Theta$, the gravitational energy $W$,
the potential
energy of the external potential $W_{\rm ext}$, and the internal energy $U$
\cite{ggp}.

In this paper, we consider a quartic self-interaction potential of the form
\begin{equation}
\label{gpp5}
V(|\psi|^2)=\frac{2\pi
a_s\hbar^2}{m^3}|\psi|^4.
\end{equation}
It corresponds to the effective potential of the axions expanded at second 
order in $|\psi|^2$ (see, e.g., Sec. III of \cite{phi6}). Since this term
dominates at low densities, it
describes
dilute axion stars (or dilute axionic clusters). It also corresponds to the
usual
$|\psi|^2\psi$ (cubic) nonlinearity present in the standard GP equation
\cite{revuebec}. It describes
short-range binary collisions between the bosons modeled by a pair contact
potential $u_{\rm SR}({\bf r}-{\bf r}')=(4\pi a_s\hbar^2/m^3)\delta({\bf r}-{\bf
r}')$ where $a_s$ is the scattering length (see, e.g., Sec. II.A. of
\cite{prd1}). When $a_s>0$ the self-interaction is repulsive and when
$a_s<0$ the self-interaction is attractive. When $a_s=0$ the bosons are
noninteracting. We shall consider these three
possibilities. We shall also assume that there is a mass at the center of the
system mimicking for example a supermassive black
hole or any other massive object.
Therefore, we consider an external potential of the form 
\begin{eqnarray}
\label{gpp6}
\Phi_{\rm BH}=-\frac{GM_{\rm BH}}{r}
\end{eqnarray}
that we shall call the BH potential (the corresponding force by unit of mass
created by the BH is
$-\nabla\Phi_{\rm BH}=-GM_{\rm BH}{\bf r}/r^3$). As a result, the GPP equations
considered
in the present paper can be written as
\begin{eqnarray}
\label{gpp7}
i\hbar \frac{\partial\psi}{\partial t}=-\frac{\hbar^2}{2m}\Delta\psi
+m\Phi\psi-\frac{GM_{\rm BH}m}{r}\psi+\frac{4\pi a_s\hbar^2}{m^2}|\psi|^2\psi,
\end{eqnarray}
\begin{equation}
\label{gpp8}
\Delta\Phi=4\pi G |\psi|^2.
\end{equation}

\subsection{Hydrodynamic representation}
\label{sec_mad}

Using the Madelung \cite{madelung} transformation
\begin{equation}
\label{mad1zz}
\psi({\bf r},t)=\sqrt{{\rho({\bf r},t)}} e^{iS({\bf r},t)/\hbar},\qquad
\rho=|\psi|^2,\qquad {\bf u}=\frac{\nabla S}{m},
\end{equation}
the GPP equations
(\ref{gpp1}) and (\ref{gpp2}) are equivalent to the hydrodynamic equations
\cite{ggp}:
\begin{equation}
\label{mad1}
\frac{\partial\rho}{\partial t}+\nabla\cdot (\rho {\bf u})=0,
\end{equation}
\begin{equation}
\label{mad2}
\frac{\partial {\bf u}}{\partial t}+({\bf u}\cdot \nabla){\bf
u}=-\frac{1}{\rho}\nabla P-\nabla\Phi-\nabla \Phi_{\rm ext}-\frac{1}{m}\nabla
Q,
\end{equation}
\begin{equation}
\label{mad3}
\Delta\Phi=4\pi G\rho,
\end{equation}
where
\begin{equation}
\label{mad4}
Q=-\frac{\hbar^2}{2m}\frac{\Delta
\sqrt{\rho}}{\sqrt{\rho}}=-\frac{\hbar^2}{4m}\left\lbrack
\frac{\Delta\rho}{\rho}-\frac{1}{2}\frac{(\nabla\rho)^2}{\rho^2}\right\rbrack
\end{equation}
is the quantum potential which takes into account the Heisenberg uncertainty
principle and $P$ is the pressure which is determined by the
self-interaction potential from the relation \cite{ggp}:
\begin{equation}
\label{mad5}
P(\rho)=\rho
V'(\rho)-V(\rho)=\rho^2\left\lbrack
\frac{V(\rho)}{\rho}\right\rbrack'.
\end{equation} 
Inversely, the self-interaction potential is related to the pressure by
\begin{equation}
\label{mad5b}
V(\rho)=\rho\int^{\rho}\frac{P(\rho')}{{\rho'}^2}\, d\rho'.
\end{equation} 
We have
\begin{equation}
\label{mad5bb}
V'(\rho)=\int^{\rho} \frac{P'(\rho')}{\rho'}\, d\rho',\qquad
V''(\rho)=\frac{P'(\rho)}{\rho}.
\end{equation} 
In the hydrodynamic representation, the mass (\ref{gpp3}) and the total energy
(\ref{gpp4}) can be written as 
\begin{eqnarray}
M=\int \rho\, d{\bf r},
\end{eqnarray}
\begin{eqnarray}
\label{mad6}
E_{\rm tot}=\int\rho \frac{{\bf u}^2}{2}\, d{\bf r}+\frac{1}{m}\int
\rho Q\, d{\bf r}
+\frac{1}{2}\int\rho\Phi\, d{\bf r}
+\int\rho\Phi_{\rm ext}\, d{\bf
r}+\int
V(\rho)\, d{\bf r}.
\end{eqnarray}
The total energy includes the classical kinetic energy $\Theta_c$, the quantum
kinetic
energy $\Theta_Q$, the
gravitational energy $W$, the potential energy of the external potential
$W_{\rm ext}$, and the
internal energy $U$ \cite{ggp}.

The self-interaction potential defined by Eq. (\ref{gpp5}) can be
rewritten as
\begin{equation}
V(\rho)=\frac{2\pi
a_s\hbar^2}{m^3}\rho^2.
\end{equation}
According to Eq. (\ref{mad5}) it generates a pressure
associated
with an
equation of state of the form
\begin{equation}
\label{mad7}
P(\rho)=\frac{2\pi
a_s\hbar^2}{m^3}\rho^2.
\end{equation}
We note that the pressure is positive when $a_s>0$ and negative  when
$a_s<0$. This equation of state can be written as
\begin{equation}
\label{mad8}
P(\rho)=K_2\rho^2\qquad {\rm with}\qquad K_2=\frac{2\pi a_s\hbar^2}{m^3}.
\end{equation}
This is a polytropic equation of state of the
form $P=K\rho^{\gamma}$ ($\gamma=1+1/n$) with index $\gamma=2$
($n=1$). We note that $P(\rho)=V(\rho)$.

\subsection{Equilibrium state}
\label{sec_eq}

In the hydrodynamic representation, an equilibrium state of the quantum
Euler equations (\ref{mad1}) and (\ref{mad2}), obtained by taking $\partial_t=0$
and ${\bf u}={\bf 0}$, satisfies 
\begin{equation}
\label{eq1}
\nabla P+\rho\nabla\Phi+\rho\nabla \Phi_{\rm ext}+\frac{\rho}{m}\nabla Q={\bf
0}.
\end{equation}
This equation can be interpreted as a condition of quantum hydrostatic
equilibrium. It
is equivalent to the stationary solution of the GPP equations (see \cite{ggp}
and Appendix \ref{sec_eigen}).  It describes the balance between the pressure
due to
short-range interactions (self-interaction), the gravitational force,
the
external force (black hole) and the quantum force arising from the Heisenberg
uncertainty principle.  Combining Eq. (\ref{eq1}) with the Poisson
equation (\ref{mad3}), we
obtain the fundamental differential equation of quantum hydrostatic equilibrium
\begin{equation}
\label{eq2}
-\nabla\cdot \left (\frac{\nabla P}{\rho}\right )+\frac{\hbar^2}{2m^2}\Delta
\left (\frac{\Delta\sqrt{\rho}}{\sqrt{\rho}}\right )=4\pi G\rho+\Delta\Phi_{\rm
ext}.
\end{equation}
For the BH potential (\ref{gpp6}), we have
\begin{eqnarray}
\label{eq3}
\Delta\Phi_{\rm BH}=4\pi GM_{\rm BH}\delta({\bf r})
\end{eqnarray}
and the foregoing equation can be rewritten as
\begin{equation}
\label{eq4}
-\nabla\cdot \left (\frac{\nabla P}{\rho}\right )+\frac{\hbar^2}{2m^2}\Delta
\left (\frac{\Delta\sqrt{\rho}}{\sqrt{\rho}}\right )=4\pi G\rho+4\pi GM_{\rm
BH}\delta({\bf
r}).
\end{equation}
For the quartic self-interaction potential (\ref{gpp5}), using Eq.
(\ref{mad7}), it takes the form
\begin{eqnarray}
\label{eq5}
-\frac{4\pi a_s\hbar^2}{m^3}\Delta\rho+\frac{\hbar^2}{2m^2}\Delta
\left (\frac{\Delta\sqrt{\rho}}{\sqrt{\rho}}\right )=4\pi G\rho+4\pi GM_{\rm
BH}\delta({\bf
r}).
\end{eqnarray}

\subsection{Exact equilibrium relations}
\label{sec_eer}

From now one, we restrict ourselves to the case where the external
potential is due to a central black hole [see Eq. (\ref{gpp6})] and to the case
of
a quartic self-interaction potential [see Eq. (\ref{gpp5})]. At equilibrium, the
total energy [see Eq. (\ref{gf4b})]  is given by
\begin{eqnarray}
\label{eer1}
E_{\rm tot}=\Theta_Q+W+W_{\rm BH}+U,
\end{eqnarray}
the eigenenergy [see Eq. (\ref{tigp4c})] is given by
\begin{equation}
\label{eer2}
NE=2W+2U+W_{\rm BH}+\Theta_Q,
\end{equation}
and the scalar virial theorem  [see Eq. (\ref{gv2})] is given by
\begin{equation}
\label{eer3}
2\Theta_Q+3U+W+W_{\rm BH}=0,
\end{equation}
where $U=\int P\, d{\bf r}$ with $P(\rho)$ given by Eq.
(\ref{mad7}). So far, the results are exact in the sense that they do not
rely on any approximation, at least with respect to the GPP equations
(\ref{gpp1}) and (\ref{gpp2})  that are our starting point. In the following
sections, we shall provide approximate analytical results of the GPP equations
based on a Gaussian ansatz.

\subsection{Gaussian ansatz}
\label{sec_gauss}

Making a Gaussian ansatz for the wavefunction \cite{ggp}, we
can write the total energy of the self-gravitating BEC as
\begin{eqnarray}
E_{\rm tot}=\frac{1}{2}\alpha
M\left (\frac{dR}{dt}\right
)^2+V(R),
\label{ag1}
\end{eqnarray}
where $R(t)$ is the typical radius of the BEC and $M$ is its
mass.\footnote{For a Gaussian density profile, the relation
between the radius
$R$ and the
radius $R_{99}$ containing $99\%$
of the mass is $R_{99}=2.38167 R$  \cite{prd1}. We must keep this relation
in mind when we compare the results from the Gaussian ansatz with the
exact results, i.e., we must compare the exact results with the approximate
ones expressed in terms of $R_{99}$, not in terms of $R$.} The first
term corresponds to the classical kinetic energy $\Theta_c$ and the second term
corresponds to the potential energy. The
conservation of
energy, $\dot E_{\rm tot}=0$, provides the following equation determining the
temporal
evolution of the radius of the BEC:
\begin{eqnarray}
\label{ag2}
\alpha M\frac{d^2R}{dt^2}=-\frac{d{V}}{dR}.
\end{eqnarray}
This is similar to the equation of motion of a fictive particle of mass $\alpha
M$ in a potential $V(R)$. The potential associated with
the equation of state (\ref{mad7}) and with the BH potential (\ref{gpp6}) is 
(see \cite{ggp} and Appendix \ref{sec_gg}):
\begin{eqnarray}
\label{ag3}
V(R)=\sigma\frac{\hbar^2M}{m^2R^2}-\nu\frac{GM^2}{R}+\zeta \frac{2\pi
a_s\hbar^2M^2}{m^3R^3}-\lambda\frac{GM_{\rm BH}M}{R}.
\end{eqnarray}
The first term is the quantum kinetic energy $\Theta_Q$, the second
term is the gravitational energy $W$, the third term is the internal energy
$U$ and the fourth term is the potential energy $W_{\rm BH}$ due to the BH. The
coefficients appearing in the foregoing equations are
\begin{eqnarray}
\label{ag4}
\alpha=\frac{3}{2}, \qquad
\sigma=\frac{3}{4},\qquad \zeta=\frac{1}{(2\pi)^{3/2}},\qquad
\nu=\frac{1}{\sqrt{2\pi}}, \qquad \lambda=\frac{2}{\sqrt{\pi}}.
\end{eqnarray}

At equilibrium, we have
\begin{eqnarray}
\label{ag3b}
E_{\rm tot}=\sigma\frac{\hbar^2M}{m^2R^2}-\nu\frac{GM^2}{R}+\zeta \frac{2\pi
a_s\hbar^2M^2}{m^3R^3}-\lambda\frac{GM_{\rm BH}M}{R}.
\end{eqnarray}
\begin{eqnarray}
\label{ag3c}
NE=\sigma\frac{\hbar^2M}{m^2R^2}-2\nu\frac{GM^2}{R}+2\zeta \frac{2\pi
a_s\hbar^2M^2}{m^3R^3}-\lambda\frac{GM_{\rm BH}M}{R}.
\end{eqnarray}

\subsection{The mass-radius relation}
\label{sec_mr}

An equilibrium state of the self-gravitating BEC is obtained by extremizing
$E_{\rm tot}(\dot
R,R)$ at fixed mass. From Eq. (\ref{ag1}), we first get the condition $\dot
R=0$
meaning that an
equilibrium state is static. We then obtain the condition $V'(R)=0$. Computing
the first derivative of
${V}(R)$ giving
\begin{eqnarray}
\label{mr1}
V'(R)=-2\sigma\frac{\hbar^2M}{m^2R^3}+\nu\frac{GM^2}{R^2}-3\zeta \frac{2\pi
a_s\hbar^2M^2}{m^3R^4}+\lambda\frac{GM_{\rm BH}M}{R^2},
\end{eqnarray}
and writing $V'(R)=0$, we obtain the mass-radius relation
\begin{eqnarray}
\label{mr2}
-2\sigma\frac{\hbar^2M}{m^2R^3}+\nu\frac{GM^2}{R^2}-6\pi\zeta \frac{
a_s\hbar^2M^2}{m^3R^4}+\lambda\frac{GM_{\rm BH}M}{R^2}=0
\end{eqnarray}
or, equivalently,
\begin{eqnarray}
\label{mr3}
M=\frac{2\sigma\frac{\hbar^2}{m^2R^3}-\lambda\frac{GM_{\rm BH}}{R^2}}{\nu\frac{
G } { R^2 }
-6\pi\zeta\frac{a_s\hbar^2}{m^3R^4}}.
\end{eqnarray}

\subsection{The pulsation}
\label{sec_puls}

An equilibrium state is stable if and only if it is a (local) minimum of $E_{\rm
tot}(\dot
R,R)$, or equivalently of $V(R)$, at fixed mass. Therefore, it must satisfy
the
condition ${V}''(R)>0$.
Computing the second derivative of ${V}(R)$, we get 
\begin{eqnarray}
V''(R)=6\sigma\frac{\hbar^2M}{m^2R^4}-2\nu\frac{GM^2}{R^3}
+24\pi\zeta\frac{a_s\hbar^2M^2}{m^3 R^{5}}
-2\lambda\frac{GM_{\rm BH}M}{R^{3}}.
\label{p1}
\end{eqnarray}
The square of the complex pulsation of the system about an equilibrium state is
given
by \cite{ggp}:
\begin{eqnarray}
\omega^2=\frac{1}{\alpha M}V''(R).
\label{p2}
\end{eqnarray}
Therefore
\begin{eqnarray}
\omega^2=\frac{6\sigma}{\alpha}\frac{\hbar^2}{m^2R^4}-\frac{2\nu}{\alpha}\frac{
GM} { R^3 }
+\frac{24\pi\zeta}{\alpha}\frac{a_s\hbar^2M}{m^3 R^{5}}
-\frac{2\lambda}{\alpha}\frac{GM_{\rm BH}}{R^{3}}.
\label{p3}
\end{eqnarray}
On the other hand, differentiating the mass-radius relation (\ref{mr2}) with
respect
to $R$
and using Eqs. (\ref{p1}) and (\ref{p2}), we obtain the identity (see Eq. (315)
of
\cite{ggp}):
\begin{eqnarray}
\omega^2=-\frac{1}{\alpha M}\left
(\frac{2\sigma\hbar^2}{m^2R^3}-\lambda\frac{GM_{\rm BH}}{R^{2}}\right
)\frac{dM}{dR}.
\label{p4}
\end{eqnarray}
This relation shows that the pulsation vanishes
($\omega=0$) at a turning point
of mass ($dM/dR=0$) in agreement with the Poincar\'e theory of linear series
of
equilibria \cite{poincare}. On the other hand, the term in parenthesis
vanishes at a turning point of radius ($dR/dM=0$).

\section{Particular cases}
\label{sec_part}

In this section, we consider particular cases of the mass-radius
relation (\ref{mr3}).

\subsection{Nongravitational $+$ noninteracting case}
\label{sec_ngni}

In the nongravitational $+$ noninteracting  case ($G=a_s=0$),
the equilibrium
states exist for
unique value of the radius
\begin{eqnarray}
\label{ngni1}
R_{\rm B}= \frac{2\sigma}{\lambda} \frac{\hbar^2}{GM_{\rm BH}m^2},
\end{eqnarray}
independent of their mass $M$.  The
prefactor is equal to $1.33$. This can be interpreted as a
gravitational Bohr radius. The pulsation is given by 
\begin{eqnarray}
\omega_B^2=\frac{\lambda}{\alpha}\frac{GM_{\rm
BH}}{R_B^{3}}=\frac{\lambda^4}{8\alpha\sigma^3}\frac{G^4M_{\rm
BH}^4m^6}{\hbar^{6}}.
\label{ngni2}
\end{eqnarray}
The prefactors are equal to $0.752$ and  $0.320$ consecutively. The equilibrium
states are
all stable ($\omega_B^2>0$).

{\it Remark:} The Schr\"odinger equation with an attractive (gravitational)
$1/r$ potential can be
solved analytically (see Appendix \ref{sec_b}). This corresponds to
the gravitational Bohr atom. The approximate results (\ref{ngni1}) and 
(\ref{ngni2}) can be compared to the
exact ones from Eqs. (\ref{b12}) and (\ref{ma7}).

\subsection{Nongravitational $+$ TF case}
\label{sec_ngtf}

In the nongravitational $+$ TF case ($G=\hbar=0$), the mass-radius
relation is given by
\begin{eqnarray}
\label{ngtf1}
M=\frac{\lambda}{6\pi\zeta} \frac{GM_{\rm BH}m^3R^2}{a_s\hbar^2}
\end{eqnarray}
provided that $a_s>0$ (there is no equilibrium state when $a_s<0$). The
prefactor is equal to $0.943$.
The radius increases as the mass increases. The pulsation is given by
\begin{eqnarray}
\omega^2=\frac{2\lambda}{\alpha}\frac{GM_{\rm
BH}}{R^{3}}.
\label{ngtf2}
\end{eqnarray}
The prefactor is equal to $1.50$. The equilibrium states are all
stable ($\omega^2>0$). 

{\it Remark:} When $G=\hbar=0$ the equation of hydrostatic
equilibrium (\ref{eq5}) can be solved analytically (see Appendix \ref{sec_ex}).
The approximate results (\ref{ngtf1}) and  (\ref{ngtf2}) can be compared to the
exact ones from Eqs. (\ref{ex11}) and (\ref{mer6}).

\subsection{Nongravitational $+$ no BH case}
\label{sec_ngngbh}

In the nongravitational $+$ no BH case ($G=M_{\rm BH}=0$), the
mass-radius relation is given by
\begin{eqnarray}
\label{ngngbh1}
M=\frac{\sigma}{3\pi\zeta} \frac{m}{|a_s|}R
\end{eqnarray}
provided that $a_s<0$ (there is no equilibrium state when $a_s>0$). The
prefactor is equal to $1.25$. The radius increases as the mass increases. The
pulsation is given by 
\begin{eqnarray}
\label{ngngbh2}
\omega^2= -\frac{2\sigma}{\alpha} \frac{\hbar^2}{m^2R^4}.
\end{eqnarray}
The
prefactor is equal to $1$.  The equilibrium states are all
unstable ($\omega^2<0$). 

{\it Remark:} When $G=M_{\rm BH}=0$ the wave function of the BEC is the solution
of the nongravitational GP equation with an attractive self-interaction
($a_s<0$). This equation has a stationary solution in the form of a soliton
which can
be obtained numerically \cite{prd2}. The exact
mass-radius relation is given by Eq.
(\ref{intro6}). These equilibrium states are unstable. Other exact results are
given in \cite{prd1,prd2} and in Appendix \ref{sec_engnbh}.

\subsection{TF $+$ noninteracting case}
\label{sec_tfni}

In the TF  $+$ noninteracting   case ($\hbar=a_s=0$), there is no
equilibrium state.

\subsection{Noninteracting $+$ no BH case}
\label{sec_ningbh}

In the noninteracting $+$ no BH case ($a_s=M_{\rm BH}=0$), the mass-radius
relation is given by
\begin{eqnarray}
\label{ningbh1}
M= \frac{2\sigma}{\nu} \frac{\hbar^2}{Gm^2R}.
\end{eqnarray}
The
prefactor is equal to $3.76$. The radius decreases as the mass increases. 
The pulsation is given by
\begin{eqnarray}
\omega^2= \frac{2\sigma}{\alpha}\frac{\hbar^2}{m^2R^4}.
\label{ningbh2}
\end{eqnarray}
The
prefactor is equal to $1$. The equilibrium states are all stable ($\omega^2>0$).

{\it Remark:} When $a_s=M_{\rm BH}=0$, the equation of hydrostatic
equilibrium (\ref{eq5}) can be solved numerically
\cite{membrado,prd1,prd2}
leading to the exact mass-radius relation from Eq.
(\ref{intro1}).  Other exact results are
given  in \cite{prd1,prd2} and in Appendix \ref{sec_eninbh}.

\subsection{TF $+$ no BH case}
\label{sec_tfngbh}

In the TF $+$ no BH case ($\hbar=M_{\rm BH}=0$), the equilibrium states exist
for
unique value of the radius
\begin{eqnarray}
\label{tfngbh1}
R_{\rm TF}=\left (\frac{6\pi\zeta}{\nu}\right )^{1/2}\left
(\frac{a_s\hbar^2}{Gm^3}\right )^{1/2}
\end{eqnarray}
independent of their mass $M$, provided that $a_s>0$ (there is no equilibrium
state when $a_s<0$). The
prefactor is equal to $1.73$.  Their pulsation is given by
\begin{eqnarray}
\omega^2=\frac{2\nu}{\alpha}\frac{GM}{R_{\rm TF}^3}.
\label{tfngbh2}
\end{eqnarray}
The
prefactor is equal to $0.532$.  The equilibrium states are all stable
($\omega^2>0$).

{\it Remark:} When $\hbar=M_{\rm BH}=0$, the BEC is equivalent to a classical
polytrope of index $n=1$ (see Appendix \ref{sec_tfnobh}). The equation of
hydrostatic
equilibrium (\ref{eq5}) can be solved
analytically \cite{leekoh,goodman,arbey,bohmer,prd1} following standard results
\cite{chandrabook}. The exact radius
is given by Eq. (\ref{intro2}). The pulsation can be obtained from the Ledoux
formula giving  $\omega_{\rm Ledoux}^2=0.123\,GM/R_{\rm TF}^3$ or by
numerically solving the Eddington equation of pulsation
giving $\omega_{\rm exact}^2=0.121\, GM/R_{\rm TF}^3$  \cite{prd1}.
These exact formulae can be compared to the  approximate
results from Eqs. (\ref{tfngbh1}) and (\ref{tfngbh2}).

\subsection{Nongravitational case}
\label{sec_ng}

In the nongravitational case ($G=0$), the mass-radius
relation is given by
\begin{eqnarray}
\label{ng1}
M=\frac{2\sigma\frac{\hbar^2}{m^2R^3}-\lambda\frac{GM_{\rm BH}}{R^2}}{
-6\pi\zeta\frac{a_s\hbar^2}{m^3R^4}}.
\end{eqnarray}
The pulsation can be written as
\begin{eqnarray}
\omega^2=\frac{6\sigma}{\alpha}\frac{\hbar^2}{m^2R^4}
+\frac{24\pi\zeta}{\alpha}\frac{a_s\hbar^2M}{m^3 R^{5}}
-\frac{2\lambda}{\alpha}\frac{GM_{\rm BH}}{R^{3}}.
\label{ng2}
\end{eqnarray}
Using the mass-radius relation (\ref{ng1}), the identity from Eq. (\ref{p4})
reduces to
\begin{eqnarray}
\omega^2=\frac{6\pi\zeta}{\alpha}\frac{a_s\hbar^2}{m^3R^4}\frac{dM}{dR}.
\label{ng3}
\end{eqnarray}

\subsubsection{Repulsive self-interaction}

\begin{figure}[h]
\scalebox{0.33}{\includegraphics{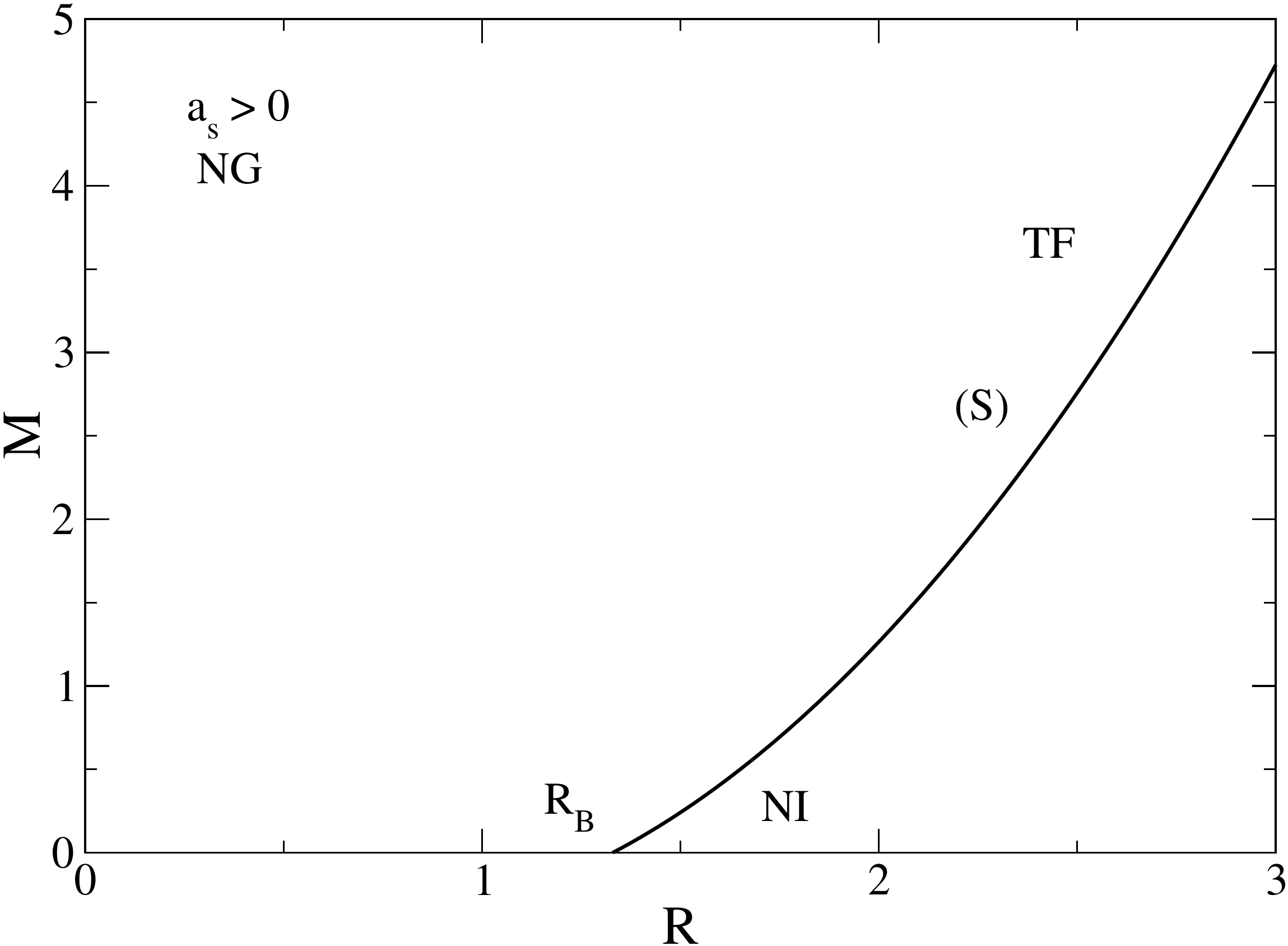}} 
\caption{Mass-radius relation of nongravitational ($G=0$) BECs with a
repulsive self-interaction ($a_s>0$) in the presence of a central
BH. We have
normalized the radius by $\hbar^2/GM_{\rm BH}m^2$ and
the mass by
$\hbar^2/GM_{\rm BH}m a_s$. This amounts to taking $\hbar=G=M_{\rm
BH}=m=a_s=1$ in the dimensional equations.}
\label{mrNGpos}
\end{figure}

When $a_s>0$, the mass-radius relation is represented in Fig. \ref{mrNGpos}. The
radius increases as the mass increases. There is a minimum radius $R_B$ given by
Eq. (\ref{ngni1}). According to Eq. (\ref{ng3}) the equilibrium states are all
stable (S) since $a_s>0$ and $M'(R)>0$ implying $\omega^2>0$.

For $R\rightarrow R_B^+$, the mass tends towards zero. This corresponds to
the nongravitational $+$ noninteracting limit (see Sec. \ref{sec_ngni}). In that
limit, the pulsation $\omega_B$ is given by Eq.
(\ref{ngni2}).

For $R\rightarrow
+\infty$, the mass-radius relation is given by Eq. (\ref{ngtf1}).
This corresponds to the nongravitational $+$ TF limit (see Sec. \ref{sec_ngtf}).
In that limit, the pulsation is
given by Eq. (\ref{ngtf2}).

Comparing Eqs. (\ref{ngni1}) and (\ref{ngtf1}), we obtain the mass scale
\begin{eqnarray}
\label{ngpos1}
M_{s}^{\rm NG}\sim \frac{\hbar^2}{GM_{\rm
BH}m a_s}.
\end{eqnarray}
The noninteracting limit is valid for $M\ll M_{s}^{\rm NG}$ and $R\sim
R_B$. The TF limit is valid for $M\gg M_{s}^{\rm NG}$ and $R\gg
R_B$. For a given mass $M$, the radius of the BEC is given by
\begin{eqnarray}
R=\frac{\sigma m\hbar^2+\sqrt{\sigma^2 m^2\hbar^4+6\pi\zeta\lambda
Gm^3a_s\hbar^2 M_{\rm BH}M}}{\lambda G M_{\rm BH} m^3}.
\end{eqnarray}
There is no equilibrium state without BH. For a given mass $M$, the radius
decreases as the BH mass increases. We have
\begin{eqnarray}
R\sim \frac{2\sigma \hbar^2}{\lambda
G m^2 M_{\rm BH}} \qquad (M_{\rm BH}\rightarrow 0), 
\end{eqnarray}
\begin{eqnarray}
R\sim \left (\frac{6\pi\zeta}{\lambda}\right )^{1/2}\left
(\frac{a_s\hbar^2M}{G m^3 M_{\rm BH}}\right )^{1/2} \qquad (M_{\rm
BH}\rightarrow +\infty). 
\end{eqnarray}

\subsubsection{Attractive self-interaction}
\label{sec_asi}

\begin{figure}[h]
\scalebox{0.33}{\includegraphics{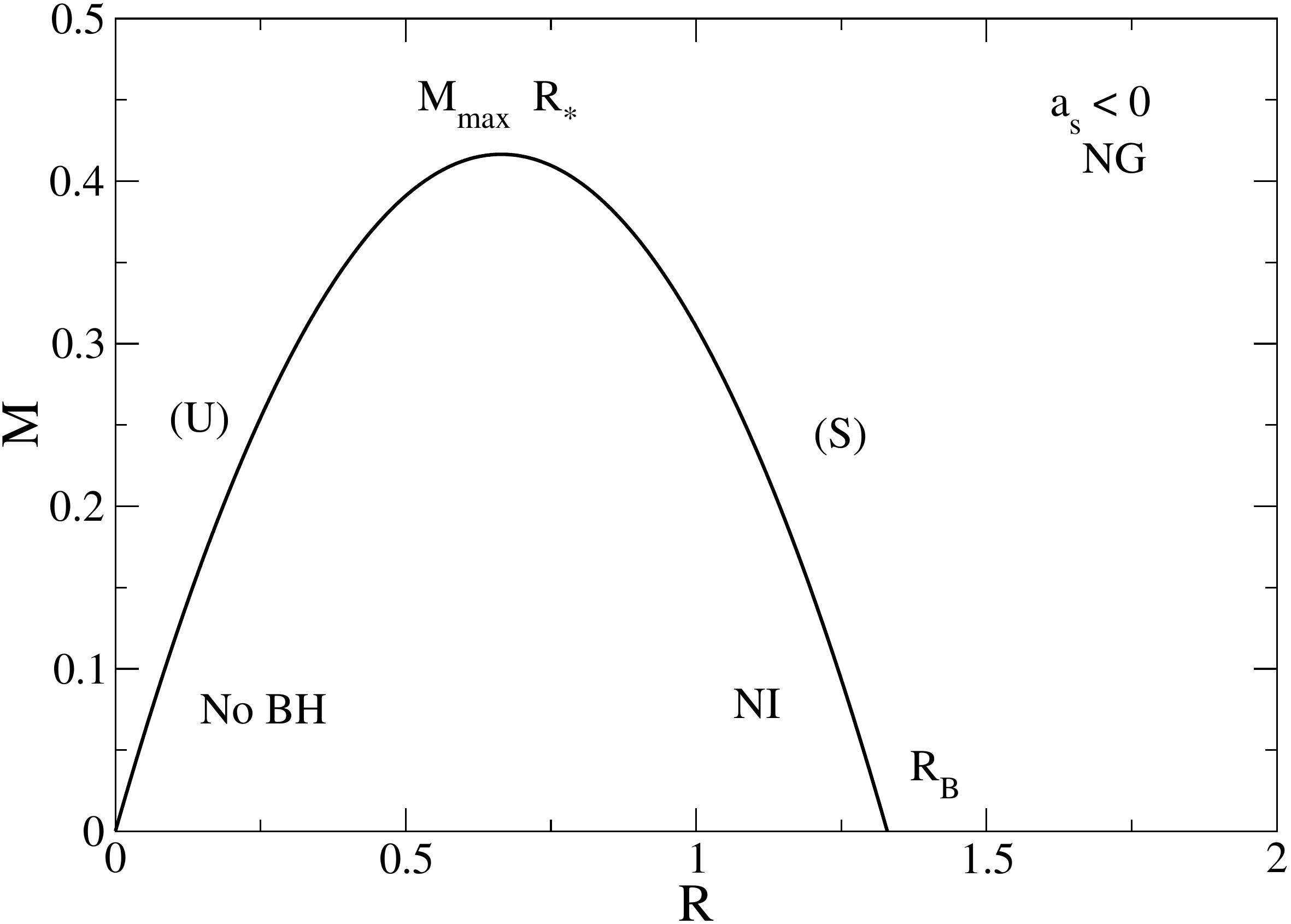}} 
\caption{Mass-radius relation of nongravitational ($G=0$) BECs with an
attractive self-interaction ($a_s<0$) in the presence of a central
BH. We have normalized the radius by
$\hbar^2/GM_{\rm BH}m^2$ and the mass by
$\hbar^2/GM_{\rm BH}m|a_s|$. This amounts to taking $\hbar=G=M_{\rm
BH}=m=|a_s|=1$ in the dimensional equations.}
\label{mrNGneg}
\end{figure}

When $a_s<0$, the mass-radius relation is represented in Fig. \ref{mrNGneg}.
There is a
maximum mass $M_{\rm max}^{\rm NG}$ at $R_*^{\rm NG}$ and a maximum radius 
$R_B$ given by Eq. (\ref{ngni1}). According to Eq.
(\ref{ng3}) the branch where $M(R)$ is decreasing corresponds to stable (S)
equilibrium states since  $a_s<0$ and $M'(R)<0$ implying $\omega^2>0$ while the
branch where $M(R)$
is increasing corresponds to
unstable (U) equilibrium
states since  $a_s<0$ and $M'(R)>0$ implying $\omega^2<0$.

For $R\rightarrow 0$, the mass-radius relation is given by Eq. (\ref{ngngbh1}).
This corresponds to the nongravitational $+$ no BH limit (see Sec.
\ref{sec_ngngbh}). In that limit, the
pulsation is given by Eq. (\ref{ngngbh2}).

For $R\rightarrow R_B^-$, the mass tends towards zero. This corresponds to
the nongravitational $+$ noninteracting limit (see Sec. \ref{sec_ngni}). In that
limit, the pulsation $\omega_B$ is given by Eq.
(\ref{ngni2}).

The maximum mass $M^{\rm NG}_{\rm max}$ and the corresponding radius $R_{*}^{\rm
NG}$ are given by
\begin{eqnarray}
\label{ngneg2}
M_{\rm max}^{\rm NG}= \frac{\sigma^2}{6\pi\zeta\lambda} \frac{\hbar^2}{GM_{\rm
BH}m |a_s|},
\end{eqnarray}
\begin{eqnarray}
\label{ngneg1}
R_{*}^{\rm NG}= \frac{\sigma}{\lambda} \frac{\hbar^2}{GM_{\rm
BH}m^2}=\frac{R_B}{2}.
\end{eqnarray}
We note the identity
\begin{eqnarray}
\label{ngneg3}
M_{\rm max}^{\rm NG}= \frac{\sigma}{6\pi\zeta} \frac{m}{|a_s|}R_*^{\rm NG}.
\end{eqnarray}
According to Eq.
(\ref{ng3}) the pulsation vanishes ($\omega^2=0$) at the maximum mass
($M'(R)=0$). The turning point of mass separates stable from unstable
equilibrium states in agreement with the Poincar\'e
criterion. On the other hand, we find that there is a maximum pulsation
\begin{eqnarray}
\omega_{\rm max}^2=0.540\frac{G^4M_{\rm BH}^4m^6}{\hbar^6}
\qquad {\rm at}\qquad R_{\omega}=0.886 \frac{\hbar^2}{GM_{\rm
BH}m^2}.
\label{mex2}
\end{eqnarray}

The no BH limit is valid for $M\ll M_{\rm max}^{\rm NG}$ and $R\ll R_B$.
The noninteracting limit is valid for $M\ll M_{\rm max}^{\rm
NG}$ and $R\sim R_B$. For a given mass $M$, the radius of the BEC with or
without central black hole is given by
\begin{eqnarray}
R=\frac{\sigma m\hbar^2\pm\sqrt{\sigma^2 m^2\hbar^4-6\pi\zeta\lambda
Gm^3|a_s|\hbar^2 M_{\rm BH}M}}{\lambda G M_{\rm BH} m^3},\qquad
R_0=\frac{3\pi\zeta|a_s|M}{\sigma m}.
\end{eqnarray}
The relative deviation is
\begin{eqnarray}
\frac{\Delta
R}{R_0}=\frac{R-R_0}{R_0}=\frac{\sigma^2\hbar^2}{3\pi\zeta\lambda|a_s|GM_{\rm
BH}Mm}\left\lbrack1\pm\sqrt{1-\frac{6\pi\zeta\lambda Gm|a_s|M_{\rm
BH}M}{\sigma^2\hbar^2}}\right\rbrack -1.
\end{eqnarray}
For a given mass $M$, there is an
equilibrium state only for
\begin{eqnarray}
M_{\rm BH}\le (M_{\rm BH})_{\rm max}(M)=\frac{\sigma^2\hbar^2}{6\pi\zeta\lambda
Gm|a_s|M}.
\end{eqnarray}
There is no stable solution without BH. On the stable branch,  the radius
decreases as the BH mass increases. On the unstable branch, the radius increases
as the BH mass increases. We have
\begin{eqnarray}
R\sim \frac{2\sigma\hbar^2}{\lambda Gm^2M_{\rm BH}}\qquad (M_{\rm
BH}\rightarrow 0, \,\,{\rm stable \,\, branch}),
\end{eqnarray}
\begin{eqnarray}
R\rightarrow  \frac{3\pi\zeta |a_s|M}{\sigma m}\qquad (M_{\rm
BH}\rightarrow 0, \,\,{\rm unstable \,\, branch}),
\end{eqnarray}
\begin{eqnarray}
R\rightarrow  \frac{6\pi\zeta|a_s|M}{\sigma m}\qquad (M_{\rm
BH}\rightarrow (M_{\rm BH})_{\rm max}(M)).
\end{eqnarray}

\subsection{Noninteracting case}
\label{sec_ni}

In the noninteracting case ($a_s=0$), the mass-radius relation is given by
\begin{eqnarray}
\label{ni1}
M=\frac{2\sigma\frac{\hbar^2}{m^2R^3}-\lambda\frac{GM_{\rm BH}}{R^2}}{\nu\frac{
G } { R^2 }}
\end{eqnarray}
and the pulsation by
\begin{eqnarray}
\omega^2=\frac{6\sigma}{\alpha}\frac{\hbar^2}{m^2R^4}-\frac{2\nu}{\alpha}\frac{
GM} { R^3 }
-\frac{2\lambda}{\alpha}\frac{GM_{\rm BH}}{R^{3}}.
\label{ni2}
\end{eqnarray}
Using the mass-radius relation (\ref{ni1}), the identity
from Eq. (\ref{p4})
reduces to
\begin{eqnarray}
\omega^2=-\frac{\nu}{\alpha}\frac{G}{R^2}\frac{dM}{dR}.
\label{ni3}
\end{eqnarray}

\begin{figure}[h]
\scalebox{0.33}{\includegraphics{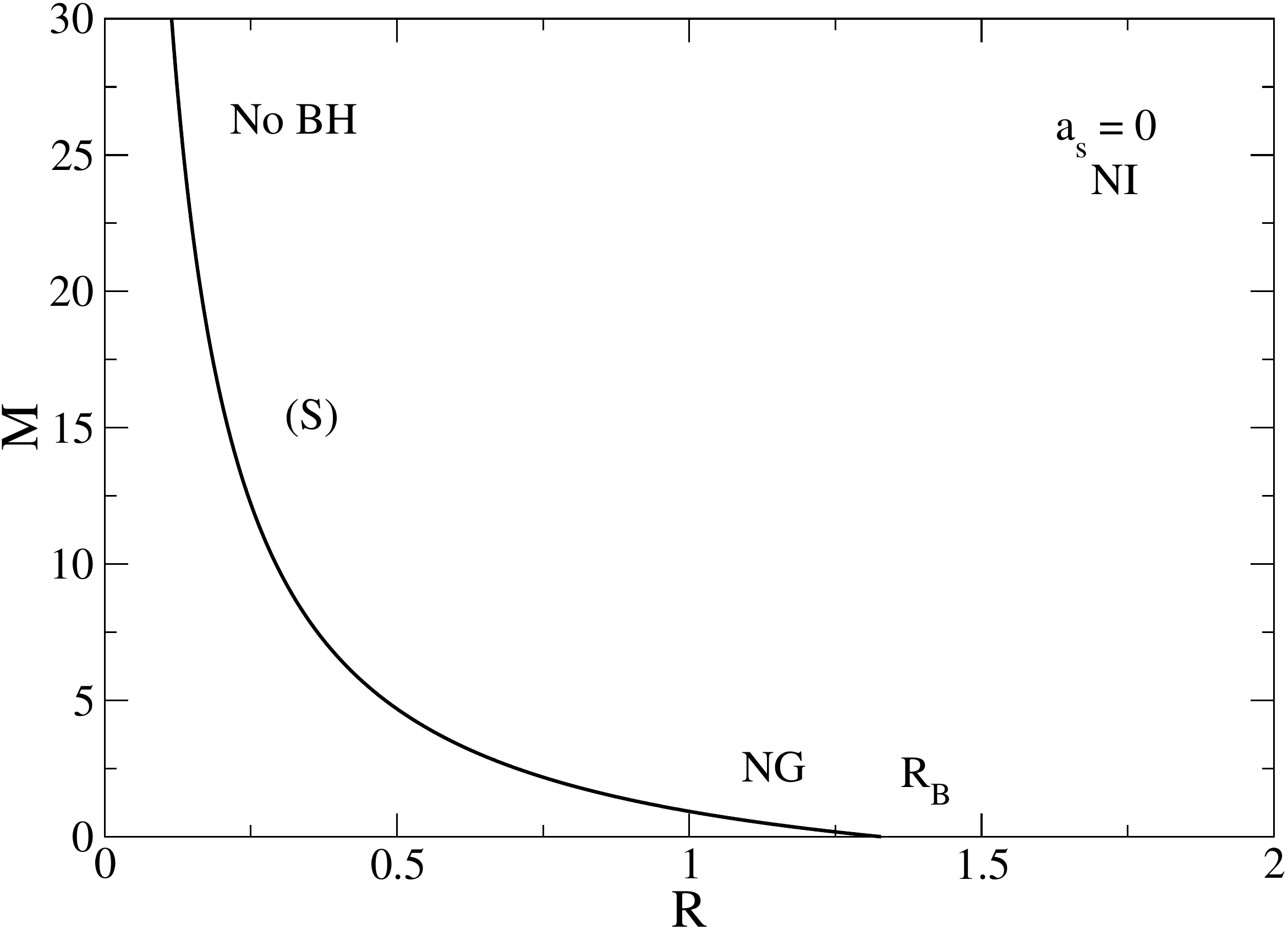}} 
\caption{Mass-radius relation of noninteracting ($a_s=0$)
self-gravitating BECs in the presene of  a central
BH. We have
normalized the
radius by $\hbar^2/GM_{\rm BH}m^2$ and the mass by
$M_{\rm BH}$. This amounts to taking $\hbar=G=M_{\rm
BH}=m=1$ in the dimensional equations.}
\label{mrNI}
\end{figure}

The mass-radius relation is represented in Fig. \ref{mrNI}. The
radius decreases as the mass increases. There is a maximum radius  $R_B$ given
by Eq. (\ref{ngni1}).
According to Eq. (\ref{ni3}) the equilibrium states are all stable (S) since
$M'(R)<0$ implying $\omega^2>0$. 

For $R\rightarrow 0$ the mass-radius is given by Eq.
(\ref{ningbh1}) and the mass tends towards $+\infty$. 
This corresponds to the noninteracting $+$ no BH limit (see Sec.
\ref{sec_ningbh}). In that limit, the
pulsation is given by  Eq.
(\ref{ningbh2}). 

For $R\rightarrow R_B^-$, the mass tends towards zero. This
corresponds to the noninteracting $+$ nongravitational limit (see Sec.
\ref{sec_ngni}). In that limit, the pulsation $\omega_B$ is given
by Eq. (\ref{ngni2}).

Comparing Eqs. (\ref{ngni1}) and (\ref{ningbh1}), we obtain the
BH mass scale $M_{\rm BH}$. The no BH limit is valid for  $M\gg M_{\rm BH}$
and $R\ll
R_B$.
The nongravitational limit is valid for $M\ll M_{\rm BH}$ and $R\sim
R_B$. For a given mass $M$, the
radius of the BEC with or without  central BH is given by
\begin{eqnarray}
R=\frac{2\sigma}{\nu}\frac{\hbar^2}{Gm^2\left (M+\frac{\lambda}{\nu}M_{\rm
BH}\right )},\qquad
R_0=\frac{2\sigma}{\nu}\frac{\hbar^2}{Gm^2M}.
\end{eqnarray}
The relative deviation is
\begin{eqnarray}
\frac{\Delta
R}{R_0}=\frac{R-R_0}{R_0}=-\frac{\lambda}{\nu}\frac{M_{\rm
BH}}{M+\frac{\lambda}{\nu}M_{\rm
BH}}.
\end{eqnarray}
For a given mass $M$, the radius decreases as the BH mass increases. We
have
\begin{eqnarray}
R\rightarrow  \frac{2\sigma}{\nu}\frac{\hbar^2}{Gm^2M} \qquad (M_{\rm
BH}\rightarrow 0), 
\end{eqnarray}
\begin{eqnarray}
R\sim    \frac{2\sigma}{\lambda}\frac{\hbar^2}{Gm^2M_{\rm BH}} \qquad (M_{\rm
BH}\rightarrow +\infty). 
\end{eqnarray}

\subsection{TF case}
\label{sec_tf}

\begin{figure}[h]
\scalebox{0.33}{\includegraphics{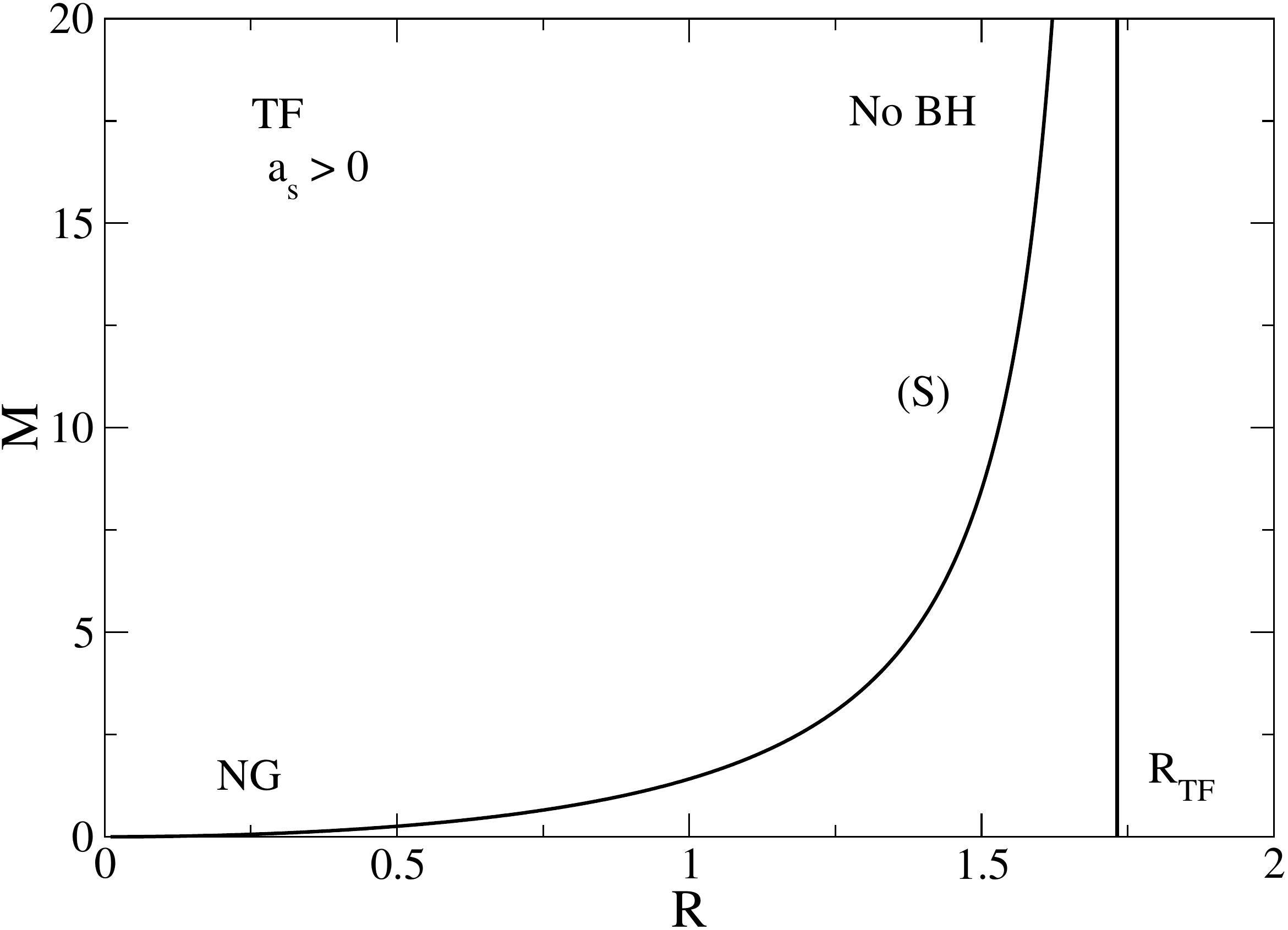}} 
\caption{Mass-radius relation of self-gravitating BECs with a repulsive
self-interaction ($a_s>0$) in the TF
limit ($\hbar=0$) in the presence of a central black hole. We have
normalized
the
radius by $(a_s\hbar^2/Gm^3)^{1/2}$ and the mass by
$M_{\rm BH}$. This amounts to taking $\hbar=G=M_{\rm
BH}=m=a_s=1$ in the dimensional equations.}
\label{mrTF}
\end{figure}

In the TF case ($\hbar=0$), the mass-radius relation is given by
\begin{eqnarray}
\label{tf1}
M=\frac{-\lambda\frac{GM_{\rm BH}}{R^2}}{\nu\frac{
G } { R^2 }
-6\pi\zeta\frac{a_s\hbar^2}{m^3R^4}}
\end{eqnarray}
provided that $a_s>0$ (there is no equilibrium state when $a_s<0$). 
The pulsation is given by
\begin{eqnarray}
\omega^2=-\frac{2\nu}{\alpha}\frac{
GM} { R^3 }
+\frac{24\pi\zeta}{\alpha}\frac{a_s\hbar^2M}{m^3 R^{5}}
-\frac{2\lambda}{\alpha}\frac{GM_{\rm BH}}{R^{3}}.
\label{tf2}
\end{eqnarray}
The identity
from Eq. (\ref{p4})
reduces to
\begin{eqnarray}
\omega^2=\frac{\lambda}{\alpha
M}\frac{GM_{\rm BH}}{R^2}\frac{dM}{dR}.
\label{tf3}
\end{eqnarray}
The
mass-radius relation is
represented in Fig. \ref{mrTF}.  The
radius increases as the mass increases. There is a maximum radius  $R_{\rm TF}$
given by Eq. (\ref{tfngbh1}).
According to Eq. (\ref{tf3}) the equilibrium states are all stable since
$M'(R)>0$ implying $\omega^2>0$. 

For $R\rightarrow 0$, the mass-radius relation is given by Eq. (\ref{ngtf1}).
This corresponds to the TF $+$ nongravitational limit (see Sec.
\ref{sec_ngtf}). In that limit, the pulsation is given by Eq.
(\ref{ngtf2}).

For $R\rightarrow R_{\rm TF}^-$, the mass
tends towards $+\infty$ as
\begin{eqnarray}
\label{tf4}
M\sim \frac{\lambda}{2\nu}\frac{M_{\rm BH}R_{\rm TF}}{R_{\rm TF}-R}.
\end{eqnarray}
This corresponds to the TF $+$ no BH limit (see Sec. \ref{sec_tfngbh}). The
pulsation is given by Eq.
(\ref{tfngbh2}). It behaves as
\begin{eqnarray}
\omega^2\sim \frac{\lambda}{\alpha}\frac{GM_{\rm BH}}{R_{\rm
TF}^2}\frac{1}{R_{\rm TF}-R}.
\label{tf5}
\end{eqnarray}

Comparing Eqs. (\ref{ngtf1}) and (\ref{tfngbh1}), we obtain the
BH mass scale $M_{\rm BH}$. The nongravitational limit is valid for  $M\ll
M_{\rm BH}$ and $R\ll
R_{\rm TF}$.
The no BH limit is valid for $M\gg M_{\rm BH}$ and $R\sim
R_{\rm TF}$. For a given mass $M$, the
radius of the BEC with or without  central BH is given by
\begin{eqnarray}
R=\left (\frac{6\pi\zeta}{\nu}\right
)^{1/2} \left
(\frac{a_s\hbar^2}{Gm^3}\right
)^{1/2} \frac{1}{\sqrt{1+\frac{\lambda}{\nu}\frac{M_{\rm BH}}{M}}},\qquad
R_0=\left (\frac{6\pi\zeta}{\nu}\right
)^{1/2} \left
(\frac{a_s\hbar^2}{Gm^3}\right
)^{1/2}=R_{\rm TF}.
\end{eqnarray}
The relative deviation is
\begin{eqnarray}
\frac{\Delta
R}{R_0}=\frac{R-R_0}{R_0}=\frac{1}{\sqrt{1+\frac{\lambda}{\nu}\frac{M_{\rm
BH}}{M}}}-1.
\end{eqnarray}
For a given mass $M$, the radius decreases as the BH mass increases. We
have
\begin{eqnarray}
R\rightarrow  \left (\frac{6\pi\zeta}{\nu}\right
)^{1/2} \left
(\frac{a_s\hbar^2}{Gm^3}\right
)^{1/2}=R_{\rm TF} \qquad (M_{\rm
BH}\rightarrow 0), 
\end{eqnarray}
\begin{eqnarray}
R\sim   \left (\frac{6\pi\zeta}{\lambda}\right
)^{1/2} \left
(\frac{a_s\hbar^2M}{Gm^3M_{\rm BH}}\right
)^{1/2}  \qquad (M_{\rm
BH}\rightarrow +\infty). 
\end{eqnarray}

{\it Remark:} When $\hbar=0$, the BEC is equivalent to a
classical
polytrope of index $n=1$ in the presence of a central BH. The equation of
hydrostatic
equilibrium (\ref{eq5}) can be solved analytically (see Appendix \ref{sec_etf}).
The exact mass-radius relation is given by Eq. (\ref{etf15}) and the pulsation
by Eq. (\ref{al8}).   There is a minimum pulsation
$\omega_{\rm min}$ (see Fig. \ref{omega2TF} in Appendix \ref{sec_etf}). These
exact formulae can be compared to the approximate
results from Eqs. (\ref{tf1}) and (\ref{tf2}). From the Gaussian
ansatz there is a minimum pulsation
\begin{eqnarray}
\label{mex1}
\omega_{\rm min}^2=1.56 \left (\frac{Gm^3}{a_s\hbar^2}\right )^{3/2}GM_{\rm
BH}\qquad {\rm at}\qquad R'_{\omega}= 1.34 \left
(\frac{a_s\hbar^2}{Gm^3}\right
)^{1/2}.
\end{eqnarray}

\subsection{No BH case}

This case has been treated in detail in Refs. \cite{prd1,prd2}.

\section{Dimensionless study in the general case}
\label{sec_pure}

In this section, we consider the general case. We use the dimensionless
variables introduced in our previous papers \cite{bectcoll,phi6}. For
convenience, they are recalled
in Appendix \ref{sec_cmmm}.

\subsection{The effective potential}
\label{sec_pep}

In terms of the dimensionless variables, the total energy of the
self-gravitating BEC, and the equation determining the
temporal evolution of its
typical radius, are given by
\begin{equation}
E_{\rm tot}=\frac{1}{2}M\left (\frac{dR}{dt}\right )^2+V(R)
\label{pep1}
\end{equation}
and
\begin{equation}
M\frac{d^2R}{dt^2}=-V'(R).
\label{pep2}
\end{equation}
The effective potential is given by
\begin{eqnarray}
\label{pep3}
V(R)=\frac{M}{R^2}-\frac{M^2}{R}\pm \frac{M^2}{3R^3}-\frac{\mu M}{R},
\end{eqnarray}
where 
\begin{eqnarray}
\label{pep4}
\mu=\frac{\lambda}{\nu}M_{\rm BH}.
\end{eqnarray}
We stress that the BH mass has also been normalized by the mass scale from Eq.
(\ref{cmmm1}). Here and in the following, the upper sign corresponds to a
repulsive
self-interaction ($a_s>0$) and the lower sign corresponds to an attractive
self-interaction ($a_s<0$).

\subsection{The mass-radius relation}
\label{sec_pmr}

Cancelling the first derivative of the effective potential given by
\begin{eqnarray}
\label{pmr1}
V'(R)=-\frac{2M}{R^3}+\frac{M^2}{R^2}\mp \frac{M^2}{R^4}+\frac{\mu M}{R^2},
\end{eqnarray}
we obtain the mass-radius relation
\begin{equation}
M=\frac{2R-\mu R^2}{R^2\mp 1}.
\label{pmr2}
\end{equation}

\subsection{The pulsation}
\label{sec_pp}

The pulsation is given by 
\begin{equation}
\omega^2=\frac{V''(R)}{M}.
\label{pp1}
\end{equation}
Since
\begin{eqnarray}
\label{pp2}
V''(R)=\frac{6M}{R^4}-\frac{2M^2}{R^3}\pm \frac{4M^2}{R^5}-\frac{2\mu M}{R^3},
\end{eqnarray}
we obtain
\begin{eqnarray}
\label{pp3}
\omega^2=\frac{6}{R^4}-\frac{2M}{R^3}\pm \frac{4M}{R^5}-\frac{2\mu}{R^3}.
\end{eqnarray}
We also
have the identity:
\begin{equation}
\omega^2=-\frac{\mu}{MR^3}\left (\frac{2}{\mu}-R\right
)\frac{dM}{dR}.
\label{pp5}
\end{equation}

\subsection{Repulsive self-interaction}
\label{sec_pos}

For a repulsive self-interaction ($a_s>0$), the mass-radius relation and the
pulsation are given by
\begin{eqnarray}
\label{pos1}
M=\frac{2R-\mu R^2}{R^2-1},\qquad \omega^2=\frac{6}{R^4}-\frac{2M}{R^3}+
\frac{4M}{R^5}-\frac{2\mu}{R^3}.
\end{eqnarray}
The mass vanishes at the gravitational Bohr radius 
\begin{eqnarray}
\label{pos2}
R_{\rm B}=\frac{2}{\mu}
\end{eqnarray}
and is infinite at the TF radius 
\begin{eqnarray}
\label{pos3}
R_{\rm TF}=1.
\end{eqnarray}
There is a critical BH mass corresponding to
\begin{eqnarray}
\label{pos4}
\mu_*=2
\end{eqnarray}
at which $R_{\rm B}=R_{\rm TF}=1$. In this very special case, the radius of the
BEC exists at a unique value $R=1$ whatever its  mass $M$.

\subsubsection{$\mu<\mu_*$}
\label{sec_posa}

\begin{figure}[h]
\scalebox{0.33}{\includegraphics{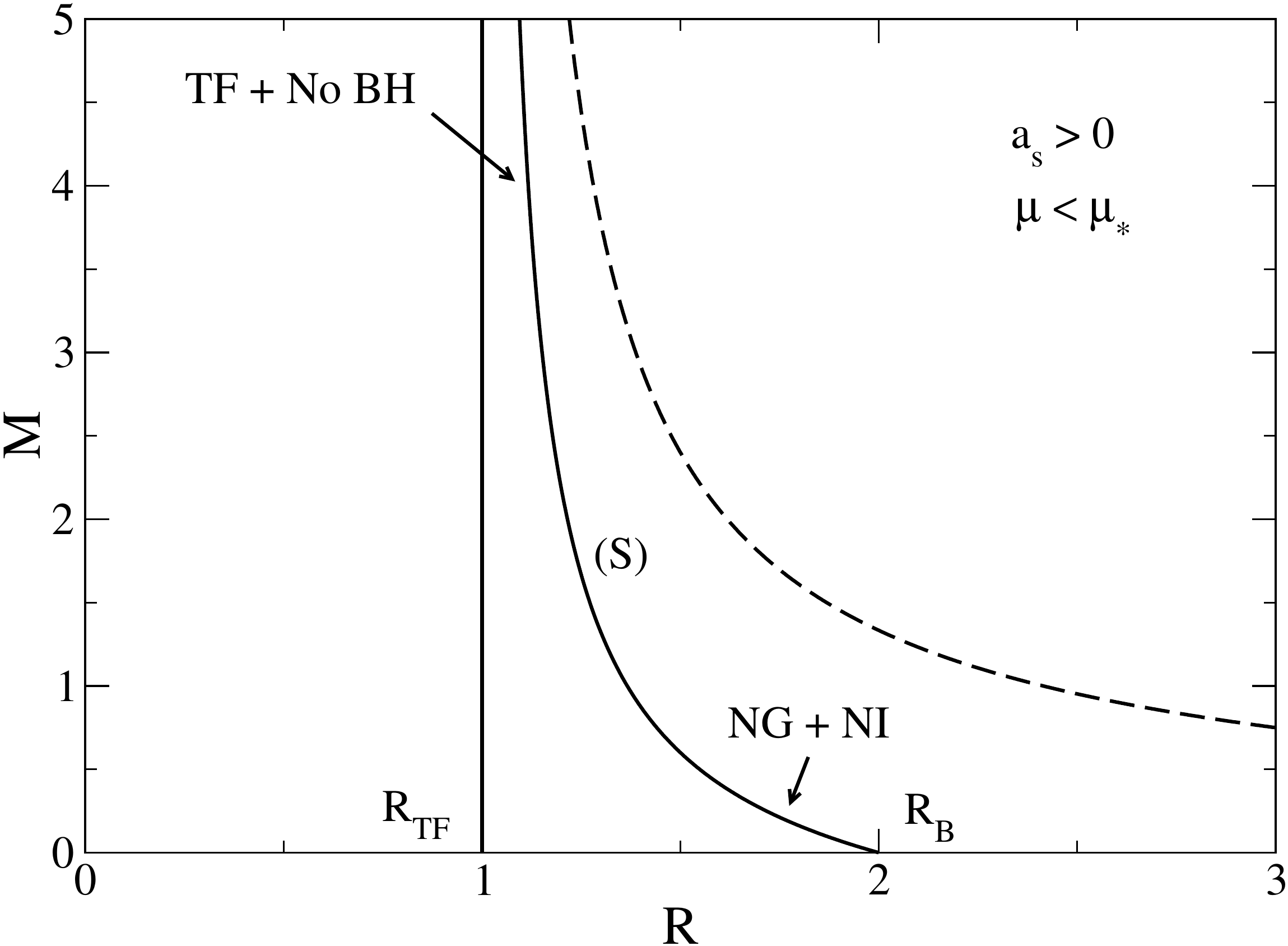}} 
\caption{Mass-radius relation of self-gravitating BECs with a repulsive
self-interaction ($a_s>0$) in the presence of a BH with a mass $\mu<\mu_*$
(specifically $\mu=1$). The dashed line corresponds to $\mu=0$ (no black hole)
\cite{prd1}.}
\label{mrpos1}
\end{figure}

When $\mu<\mu_*$ we are in the situation where $R_B>R_{\rm TF}$. The mass-radius
relation is plotted
in Fig. \ref{mrpos1}. The radius decreases as the mass increases.\footnote{When
$\mu<\mu_*$ the mass-radius relation has no extremum. The condition 
$M'(R)=0$ yields
the second degree equation $R^2-\mu R+1=0$ whose discriminant
$\Delta=\mu^2-4$ is negative ($\Delta<0$).} 
There is a maximum radius $R_B$ and a minimum radius $R_{\rm TF}$.
According to Eq.
(\ref{pp5}), the equilibrium states are all
stable (S) since $R<R_B$ and $M'(R)<0$, implying $\omega^2>0$.

For $R\rightarrow R_{\rm TF}$, the mass tends towards infinity
as
\begin{eqnarray}
\label{pos4b}
M\sim \frac{2-\mu}{2(R-R_{\rm TF})}.
\end{eqnarray}
This corresponds to the TF $+$ no BH limit. In that limit the pulsation
tends  towards infinity as
\begin{eqnarray}
\label{pos4bb}
\omega^2\sim \frac{2-\mu}{R-R_{\rm TF}}.
\end{eqnarray}

For $R\rightarrow R_{B}$, the mass tends towards zero as
\begin{eqnarray}
\label{pos4c}
M\sim \frac{2\mu^2}{4-\mu^2}(R_{\rm B}-R).
\end{eqnarray}
 This
corresponds to the nongravitational $+$ noninteracting limit. In that limit the
pulsation tends towards
\begin{eqnarray}
\label{pos4cb2}
\omega_{\rm B}=\frac{\mu^2}{2\sqrt{2}}.
\end{eqnarray}

Substituting $R_{\rm B}$ given by Eq. (\ref{pos2}) into Eq. (\ref{pos4b})
or substituting $R_{\rm TF}$ given by Eq. (\ref{pos3}) into Eq. (\ref{pos4c}),
we find that
the transition
between these two regimes occurs for $M_t\sim M_{\rm BH}$. The TF $+$ no BH
limit is valid for $M\gg M_{\rm BH}$ and $R\sim R_{\rm TF}$. The 
nongravitational $+$ noninteracting limit is valid for $M\ll M_{\rm BH}$ and
$R\sim R_{\rm B}$.

\subsubsection{$\mu>\mu_*$}
\label{sec_posb}

\begin{figure}[h]
\scalebox{0.33}{\includegraphics{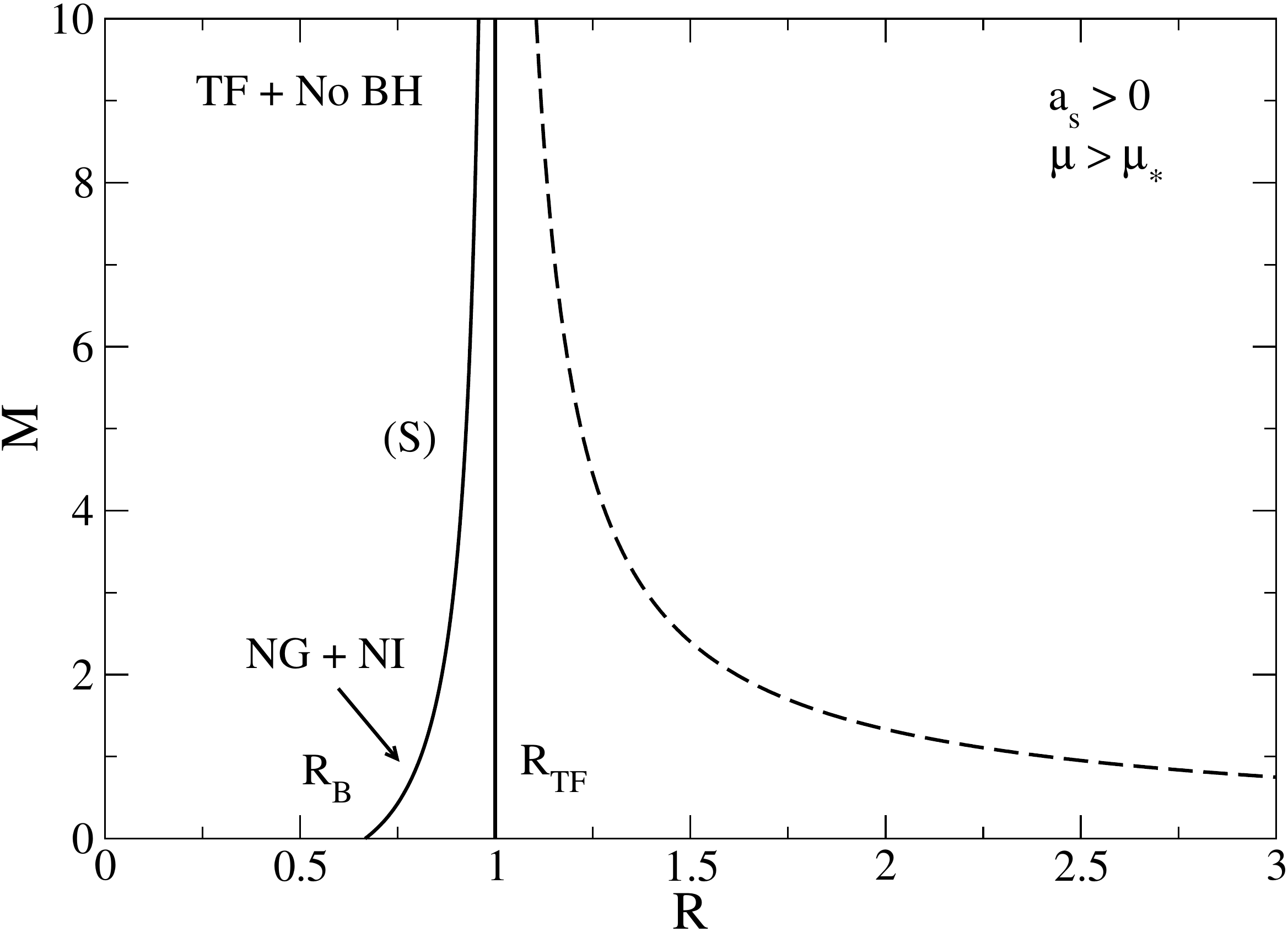}} 
\caption{Mass-radius relation of self-gravitating BECs with a repulsive
self-interaction in the presence of a BH with a mass $\mu>\mu_*$
(specifically $\mu=3$). The dashed line corresponds to $\mu=0$ (no black hole)
\cite{prd1}.}
\label{mrpos2}
\end{figure}

When $\mu>\mu_*$ we are in the situation where $R_B<R_{\rm TF}$. The
mass-radius
relation is plotted
in Fig. \ref{mrpos2}. The radius increases as the mass increases.\footnote{When
$\mu>\mu_*$, the mass-radius relation has no extremum in the physical range
$[R_B,R_{\rm TF}]$ where the mass
is positive. The condition $M'(R_e)=0$ yields
the second degree equation $R_e^2-\mu R_e+1=0$, with a positive  discriminant
$\Delta=\mu^2-4>0$, which determines the extrema of mass $M_e$. Combining
the equation $R^2-\mu R+1=0$ with Eq. (\ref{pos1}) we get $M_e=-R_e<0$.
Therefore, the extrema of mass correspond to an unphysical negative mass.}
There is a
minimum radius $R_{B}$ and a maximum radius $R_{\rm TF}$. According to Eq.
(\ref{pp5}), the equilibrium states are all
stable (S) since $R>R_B$ and $M'(R)>0$ implying $\omega^2>0$.

For $R\rightarrow R_{B}$, the mass tends towards zero as
\begin{eqnarray}
\label{wpos4c}
M\sim \frac{2\mu^2}{\mu^2-4}(R-R_{\rm B}).
\end{eqnarray}
 This
corresponds to the nongravitational $+$ noninteracting limit. In that limit the
pulsation tends towards
\begin{eqnarray}
\label{wpos4cb}
\omega_{\rm B}=\frac{\mu^2}{2\sqrt{2}}.
\end{eqnarray}

For $R\rightarrow R_{\rm TF}$, the mass tends towards infinity
as
\begin{eqnarray}
\label{wpos4b}
M\sim \frac{\mu-2}{2(R_{\rm TF}-R)}.
\end{eqnarray}
This corresponds to the TF $+$ no BH limit. In that limit the pulsation tends
towards infinity
as
\begin{eqnarray}
\label{wpos4bb}
\omega^2\sim \frac{\mu-2}{R_{\rm TF}-R}.
\end{eqnarray}

The nongravitational $+$ noninteracting limit is valid for $M\ll M_{\rm BH}$ and
$R\sim R_{\rm B}$. The TF $+$ no BH limit is valid for $M\gg M_{\rm BH}$ and
$R\sim R_{\rm TF}$. 

\subsubsection{General results}

\begin{figure}[h]
\scalebox{0.33}{\includegraphics{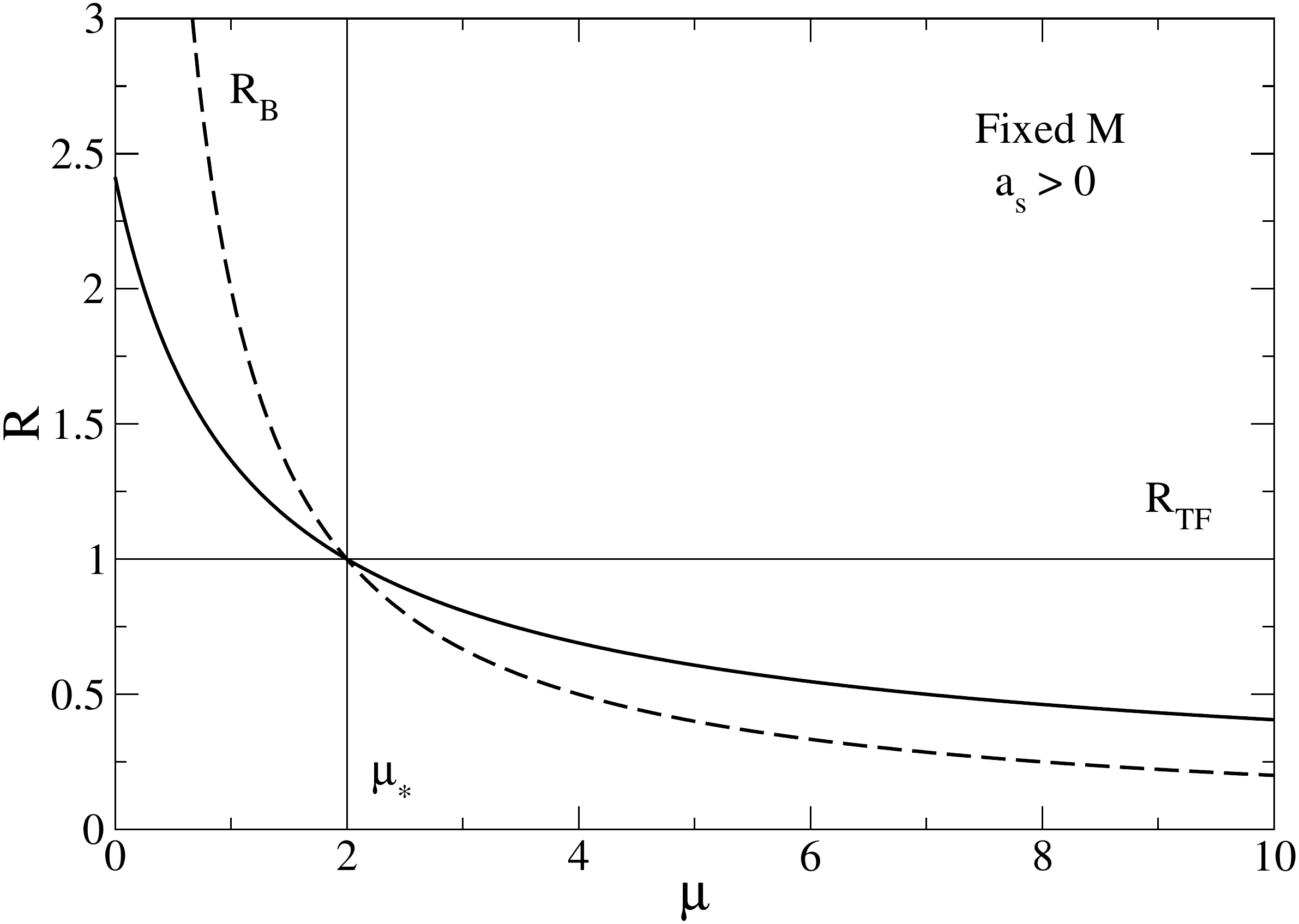}} 
\caption{Radius $R$ of the BEC with a repulsive self-interaction as a function
of the BH mass $\mu$ for
a fixed value of the mass $M$ (specifically $M=1$). For
$\mu<\mu_*$ the radius of the BEC is larger than $R_{\rm TF}$ and smaller than
$R_B$. For
$\mu>\mu_*$ the radius of the BEC is larger than $R_{B}$ and smaller than
$R_{\rm TF}$. }
\label{muRfixedMrep}
\end{figure}

For a given mass $M$, the radius of the BEC with or without central BH is
given by
\begin{equation}
R=\frac{1+\sqrt{1+(M+\mu)M}}{M+\mu},\qquad R_0=\frac{1+\sqrt{1+ M^2}}{M}.
\end{equation}
The relative deviation is 
\begin{eqnarray}
\frac{\Delta
R}{R_0}=\frac{R-R_0}{R_0}=\frac{1+\sqrt{1+
(M+\mu)M}}{1+
\sqrt{1+ M^2}}\frac{M}{M+\mu}-1.
\end{eqnarray}
For a given mass $M$, the radius decreases as the BH mass increases (see Fig.
\ref{muRfixedMrep}). We have
\begin{eqnarray}
R\rightarrow \frac{1+\sqrt{1+ M^2}}{M}\qquad (\mu\rightarrow 0), 
\end{eqnarray}
\begin{eqnarray}
R\sim \left (\frac{M}{\mu}\right )^{1/2} \qquad (\mu\rightarrow +\infty). 
\end{eqnarray}

\begin{figure}[h]
\scalebox{0.33}{\includegraphics{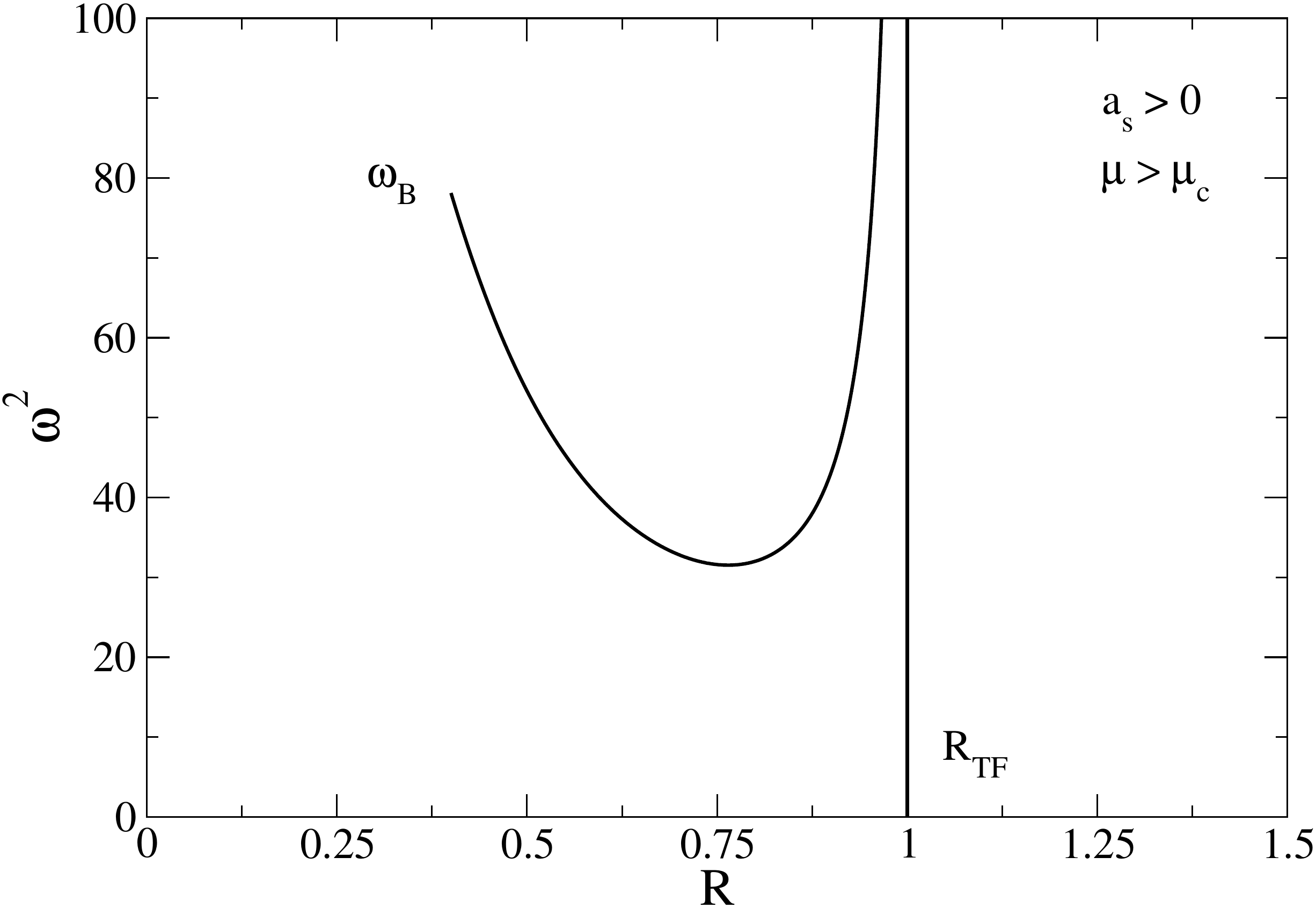}} 
\caption{Square complex pulsation $\omega^2$ as a function of
the radius $R$ for self-gravitating BECs with a repulsive
self-interaction ($a_s>0$) in the presence of a BH with a mass
$\mu>\mu_c\simeq 2.83$
(specifically $\mu=5$).}
\label{rw2rep}
\end{figure}

\begin{figure}[h]
\scalebox{0.33}{\includegraphics{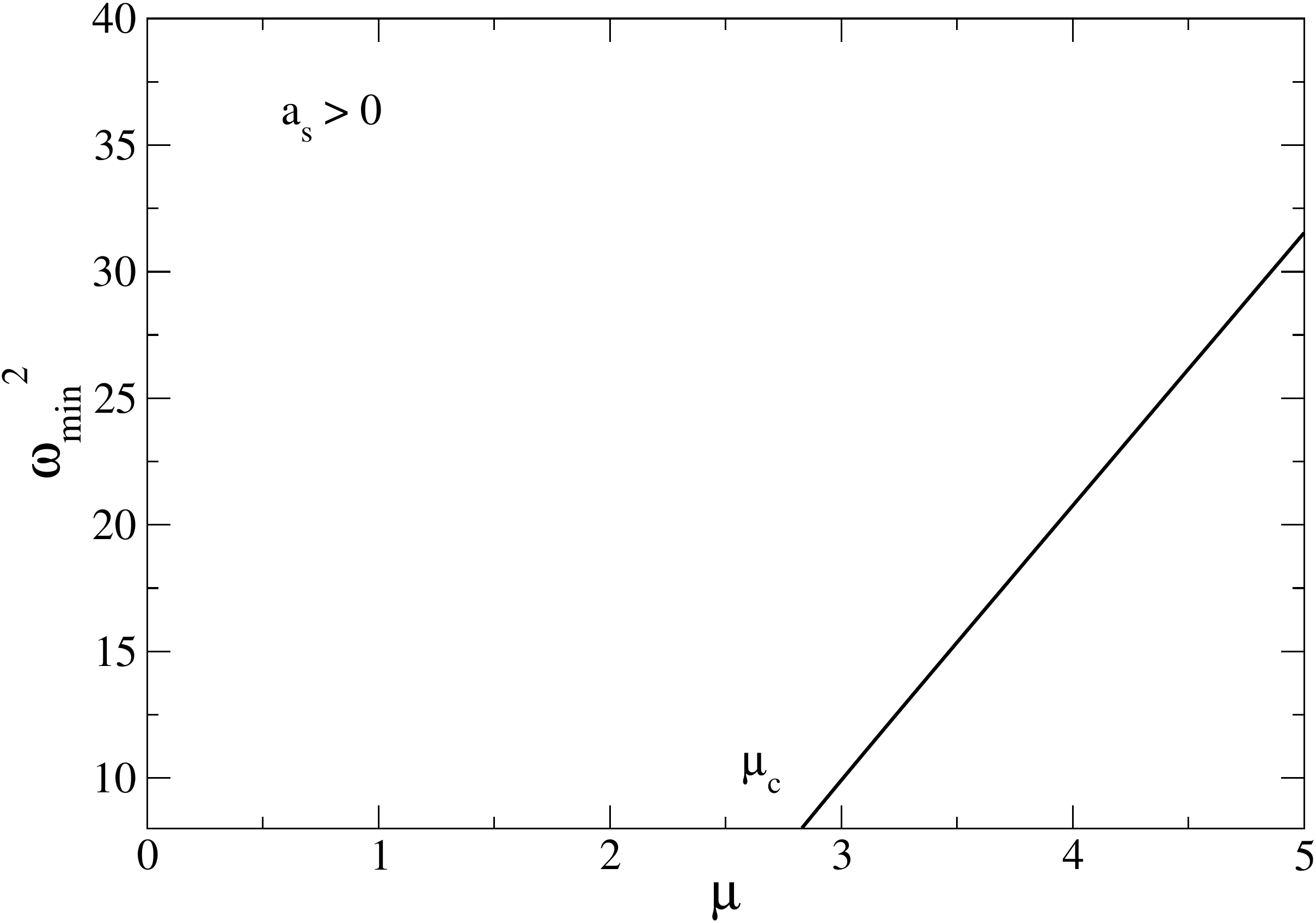}} 
\caption{Minimum square pulsation $\omega_{\rm min}$ as a function of the BH
mass $\mu$ (a linear fit gives $\omega_{\rm min}^2=10.8\mu-22.5$).}
\label{muMinw2}
\end{figure}

The square complex pulsation $\omega^2$ is plotted as a function of the
radius $R$ in Fig. \ref{rw2rep} for $\mu>\mu_c\simeq 2.83$. It starts from
$\omega_B^2$ at $R=R_B$, decreases, reaches a minimum $\omega^2_{\rm
min}(\mu)$, and increases towards infinity as
 $R\rightarrow R_{\rm TF}$. The existence of a minimum
pulsation $\omega_{\rm min}(\mu)$ for sufficiently large values of $\mu$ is
consistent with the exact results obtained in the TF approximation (see
Appendix \ref{sec_etf}). We find that the minimum square pulsation increases
linearly with the BH mass (see Fig. \ref{muMinw2}). Its
asymptotic behavior for $\mu\rightarrow +\infty$ can be obtained from Eq.
(\ref{mex1}) yielding $\omega_{\rm min}^2\sim 10.8\mu$. For $\mu_*<\mu<\mu_c$,
the
square pulsation increases monotonically from $\omega_B^2$ at $R=R_B$ to
infinity as $R\rightarrow R_{\rm TF}$. For $\mu<\mu_*$, the
square pulsation decreases monotonically from infinity as $R\rightarrow R_{\rm
TF}$ to $\omega_B^2$ at $R=R_B$.

\subsection{Attractive self-interaction}
\label{sec_neg}

\begin{figure}[h]
\scalebox{0.33}{\includegraphics{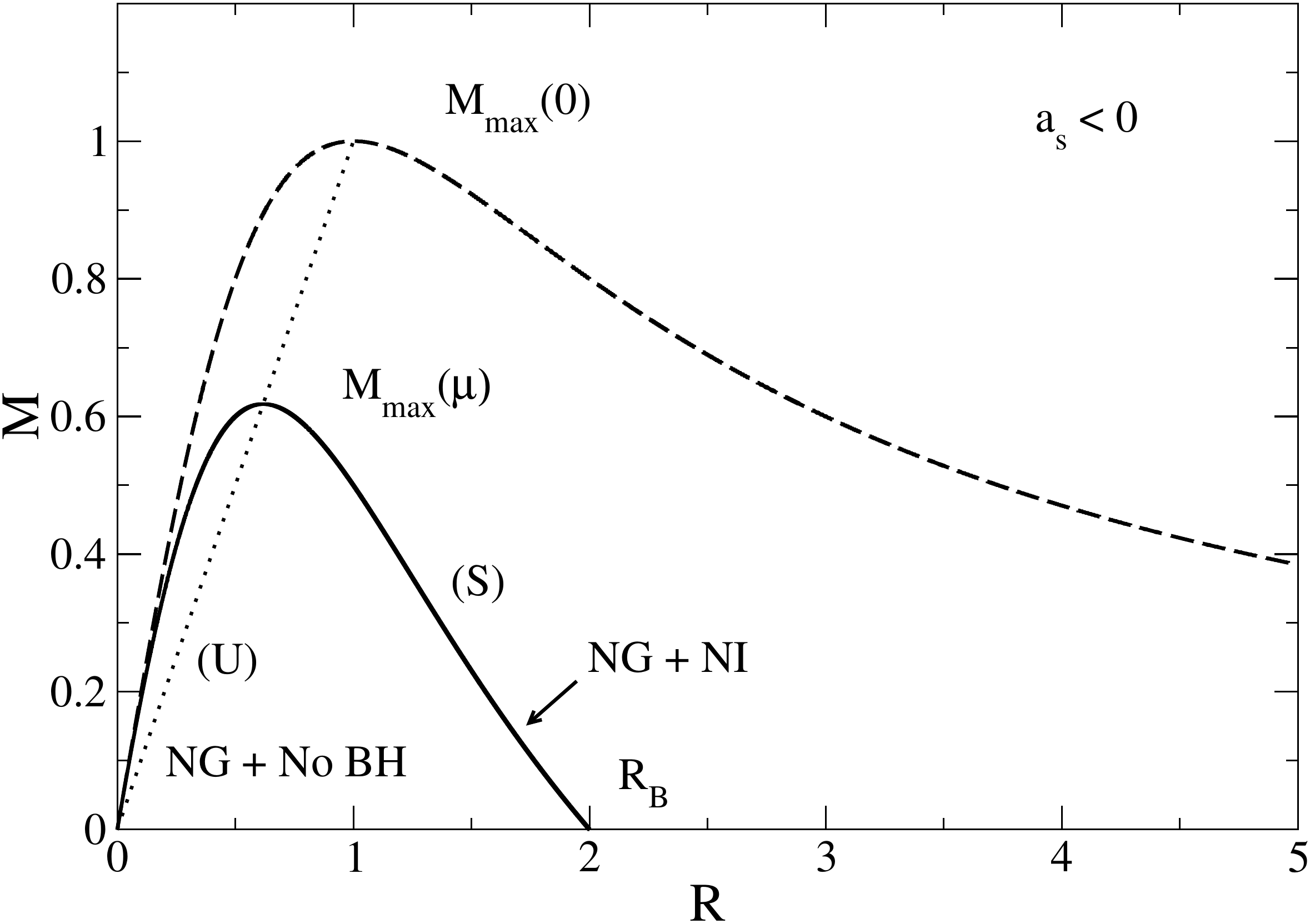}} 
\caption{Mass-radius relation of self-gravitating BECs with an attractive
self-interaction ($a_s<0$) in the presence of a BH with a mass $\mu$
(specifically $\mu=1$). The dashed line corresponds to $\mu=0$ (no black hole)
\cite{prd1}. The maximum ($R_*(\mu), M_{\rm max}(\mu)$) of the
curve $M(R)$ lies on the straight line $M=R$ (dotted line) parametrized by $\mu$
going from zero (top) to infinity (bottom). }
\label{mrneg}
\end{figure}

For an attractive self-interaction ($a_s<0$) the mass-radius relation and the
pulsation are given
by 
\begin{eqnarray}
\label{neg1}
M=\frac{2R-\mu
R^2}{R^2+1},\qquad \omega^2=\frac{6}{R^4}-\frac{2M}{R^3}-
\frac{4M}{R^5}-\frac{2\mu}{R^3}.
\end{eqnarray}
The mass vanishes at $R=0$ and at the gravitational Bohr radius $R_B$ given by
Eq.
(\ref{pos2}). The mass-radius relation is plotted
in Fig. \ref{mrneg}. There is a maximum mass
$M_{\rm max}(\mu)$ and a maximum
radius
$R_B$. For a given mass $M<M_{\rm max}$ there are two branches of
solution. According to Eq.
(\ref{pp5}), the equilibrium states with $R>R_*$
are stable (S) since $R<R_B$ and $M'(R)<0$, implying $\omega^2>0$, while the 
equilibrium states with
$R<R_*$
are unstable (U) since $R<R_B$ and $M'(R)>0$, implying $\omega^2<0$. 

\begin{figure}[h]
\scalebox{0.33}{\includegraphics{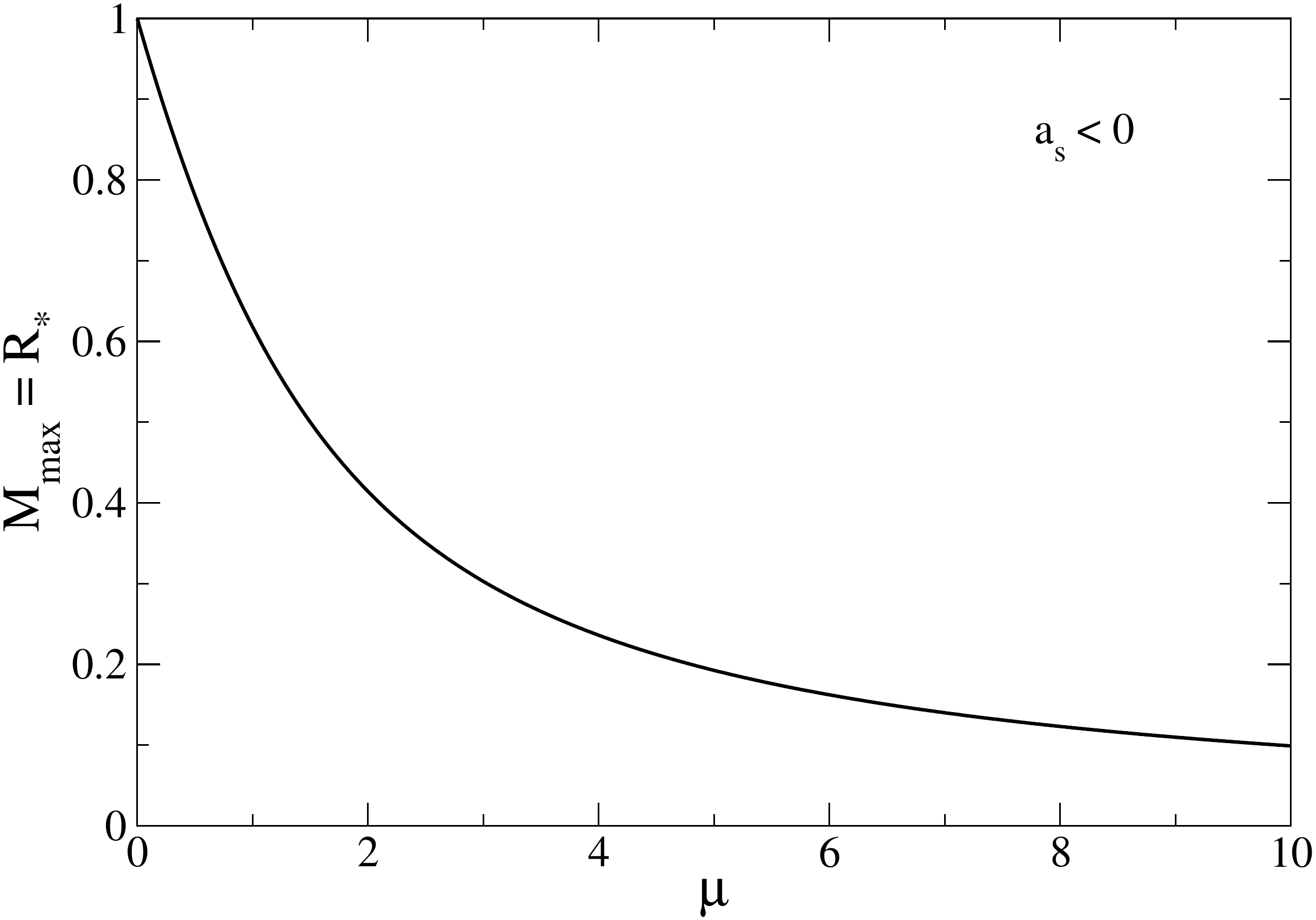}} 
\caption{Maximum mass and minimum radius of self-gravitating BECs with an
attractive
self-interaction ($a_s<0$) as a function of the BH mass $\mu$.}
\label{muMmaxneg}
\end{figure}

For $R\rightarrow 0$, the mass tends towards zero as
\begin{eqnarray}
\label{neg2}
M\sim 2R.
\end{eqnarray}
This corresponds to the nongravitational $+$ no BH limit. In that limit the
square pulsation tends towards $-\infty$ as 
\begin{eqnarray}
\label{neg2c}
\omega^2\sim -\frac{2}{R^4}.
\end{eqnarray}
As $R$ increases, the mass $M(R)$ reaches a
maximum $M_{\rm max}(\mu)$ at $R_*(\mu)$ then
decreases and vanishes at 
$R=R_{\rm B}$. 
For $R\rightarrow R_{B}$, the mass tends towards zero as
\begin{eqnarray}
\label{neg2b}
M\sim \frac{2\mu^2}{4+\mu^2}(R_{\rm B}-R).
\end{eqnarray}
 This
corresponds to the nongravitational $+$ noninteracting limit. In that limit the
pulsation tends towards
\begin{eqnarray}
\label{pos4cb}
\omega_{\rm B}=\frac{\mu^2}{2\sqrt{2}}.
\end{eqnarray}
The condition
$M'(R_*)=0$ yields the second degree equation
\begin{eqnarray}
\label{neg3}
R_*^2+\mu R_*-1=0.
\end{eqnarray}
By eliminating $\mu$
between Eqs. (\ref{neg1}) and (\ref{neg3}) we find that
\begin{eqnarray}
\label{neg4}
M_{\rm max}(\mu)=R_*(\mu).
\end{eqnarray}
By solving Eq. (\ref{neg3}) we get
\begin{eqnarray}
\label{neg5}
M_{\rm max}(\mu)=R_*(\mu)=-\frac{\mu}{2}+\frac{1}{2}\sqrt{\mu^2+4}.
\end{eqnarray}
The maximum mass (or minimum radius) is plotted as a function of $\mu$ in Fig.
\ref{muMmaxneg}.
For
$\mu\rightarrow 0$:
\begin{eqnarray}
\label{neg6}
M_{\rm max}(\mu)=R_*(\mu) \simeq
1-\frac{\mu}{2}+\frac{\mu^2}{8}-\frac{\mu^4}{128}+...
\end{eqnarray}
For
$\mu\rightarrow +\infty$:
\begin{eqnarray}
\label{neg7}
M_{\rm max}(\mu)=R_*(\mu) \sim \frac{1}{\mu}.
\end{eqnarray}
We note that $R_*(\mu) \sim R_{\rm B}(\mu)/2$. This returns the results of
Sec. \ref{sec_asi} valid in the nongravitational limit.
The maximum mass and the
minimum radius are smaller than the  values they would have in the absence of a
central BH. The nongravitational $+$ no BH
limit is valid for $M\ll M_{\rm
max}$  and $R\ll R_*$. The nongravitational $+$ noninteracting limit
is valid for $M\ll M_{\rm
max}$  and $R\sim R_{\rm B}$.

\begin{figure}[h]
\scalebox{0.33}{\includegraphics{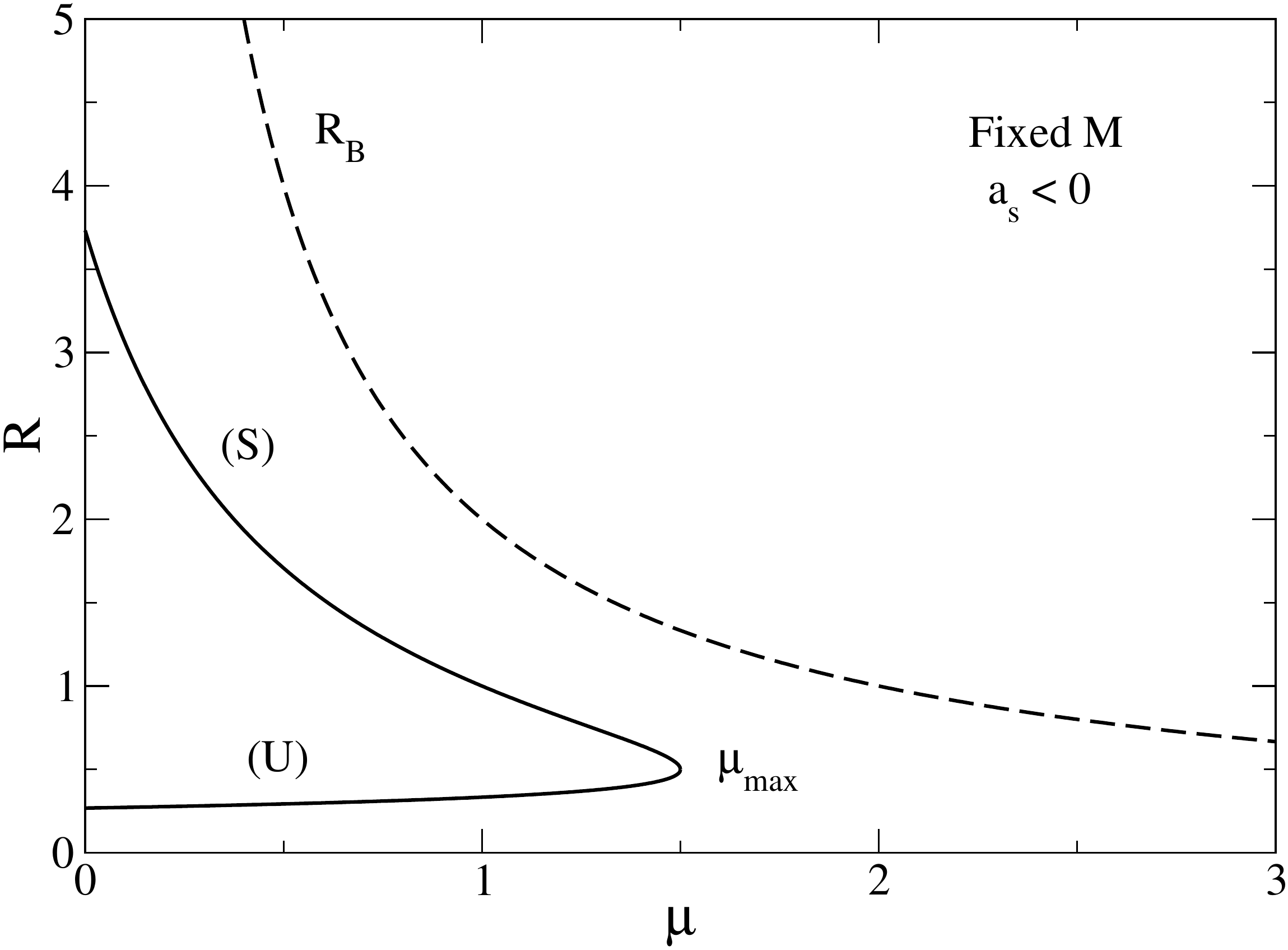}} 
\caption{Radius $R$ of the BEC with an attractive
self-interaction as a function
of the BH mass $\mu$ for
a fixed value of the mass $M\le 1$ (specifically $M=0.5$).}
\label{muRfixedMatt}
\end{figure}

For a given mass $M$, the radius of the BEC with or without  central BH is
given by
\begin{equation}
R=\frac{1\pm \sqrt{1- (M+\mu)M}}{M+\mu},\qquad R_0=\frac{1\pm
\sqrt{1-M^2}}{M}.
\label{pmr2b}
\end{equation}
The relative deviation is
\begin{eqnarray}
\frac{\Delta
R}{R_0}=\frac{R-R_0}{R_0}=\frac{1\pm \sqrt{1-
(M+\mu)M}}{1\pm
\sqrt{1- M^2}}\frac{M}{M+\mu}-1.
\end{eqnarray}
For a given mass $M\le 1$, there is an equilibrium state only for
\begin{eqnarray}
\mu\le \mu_{\rm max}(M)=\frac{1-M^2}{M}.
\end{eqnarray}
On the stable branch,  the radius
decreases as the BH mass increases. On the unstable branch, the radius increases
as the BH mass increases (see Fig. \ref{muRfixedMatt}). We
have
\begin{eqnarray}
R\rightarrow \frac{1+\sqrt{1-M^2}}{M}\qquad  (\mu\rightarrow 0, \,\,{\rm stable
\,\, branch}),
\end{eqnarray}
\begin{eqnarray}
R\rightarrow  \frac{1-
\sqrt{1-M^2}}{M}\qquad (\mu\rightarrow 0, \,\,{\rm unstable \,\, branch}),
\end{eqnarray}
\begin{eqnarray}
R\rightarrow M\qquad (\mu\rightarrow \mu_{\rm max}(M)).
\end{eqnarray}

\begin{figure}[h]
\scalebox{0.33}{\includegraphics{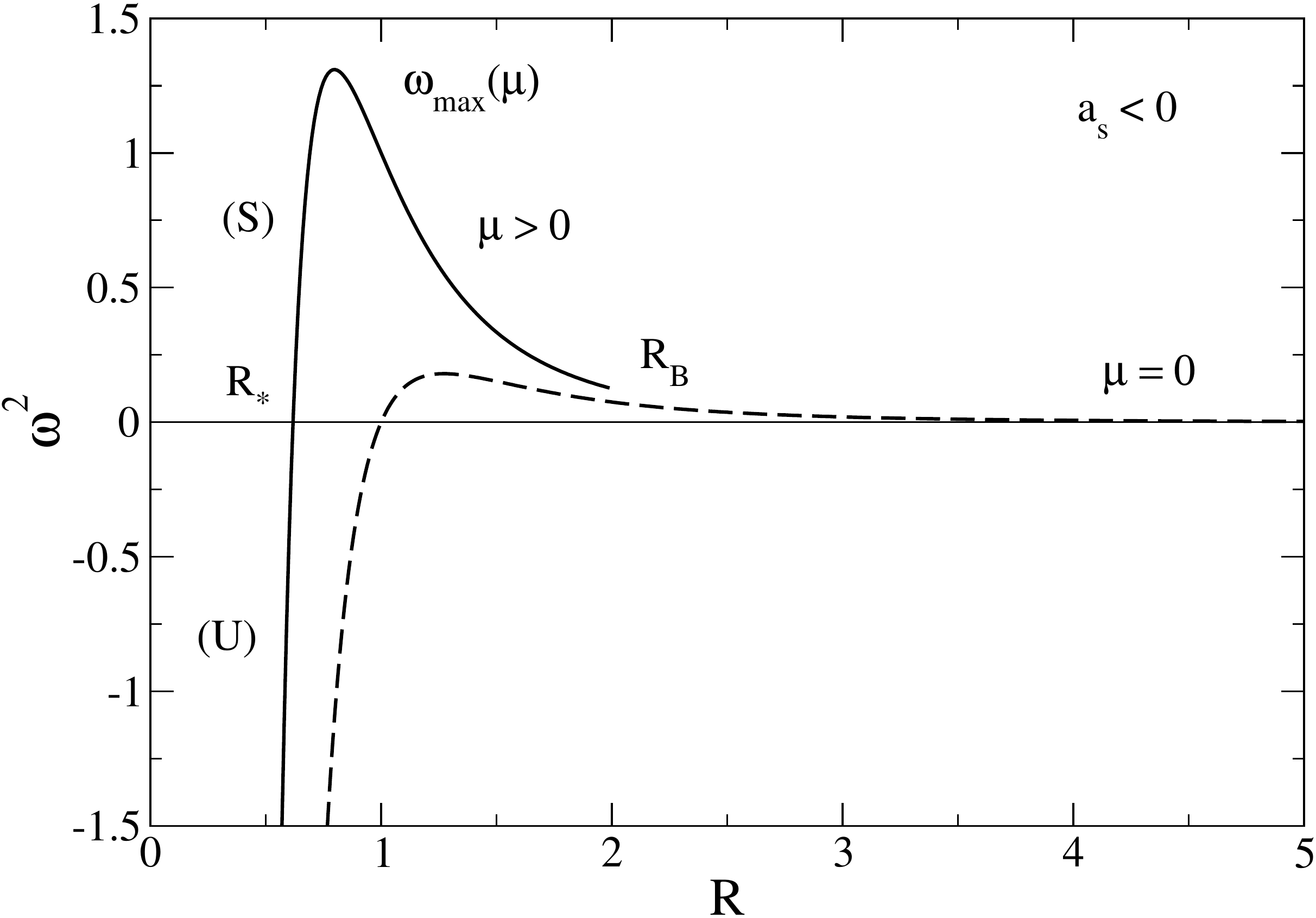}} 
\caption{Square complex pulsation $\omega^2$ as a function of
the radius $R$ for self-gravitating BECs with an attractive
self-interaction ($a_s<0$) in the presence of a BH with a mass $\mu$
(specifically $\mu=1$). The dashed line corresponds to $\mu=0$ (no black hole)
\cite{prd1}. In that case $\omega_{\rm max}(0)=0.4246$ corresponding to
$R=1.272$ and $M=0.9717$ (see Appendix I of \cite{phi6}).}
\label{rw2}
\end{figure}

The square complex pulsation $\omega^2$ is plotted as a function of the radius
$R$ in Fig. \ref{rw2}. It starts from $-\infty$ as $R\rightarrow 0$, vanishes at
$R_*$ (corresponding to the maximum mass), reaches a maximum  $\omega^2_{\rm
max}(\mu)$, and decreases towards $\omega_B^2$ as
 $R\rightarrow R_B$. The existence of a maximum
pulsation $\omega_{\rm max}$ was  previously noted  in the absence of
a BH \cite{prd1,prd2,bectcoll,phi6}. We find that the maximum
pulsation increases with the BH mass (see Fig. \ref{muMaxw2}). For
$\mu=0$, we have  $\omega_{\rm max}(0)=0.4246$ (see Appendix I
of \cite{phi6}). For $\mu\rightarrow +\infty$, using Eq. (\ref{mex2}) valid in
the nongravitational limit, we get $\omega_{\rm max}^2\sim 0.211\mu^4$.

\begin{figure}[h]
\scalebox{0.33}{\includegraphics{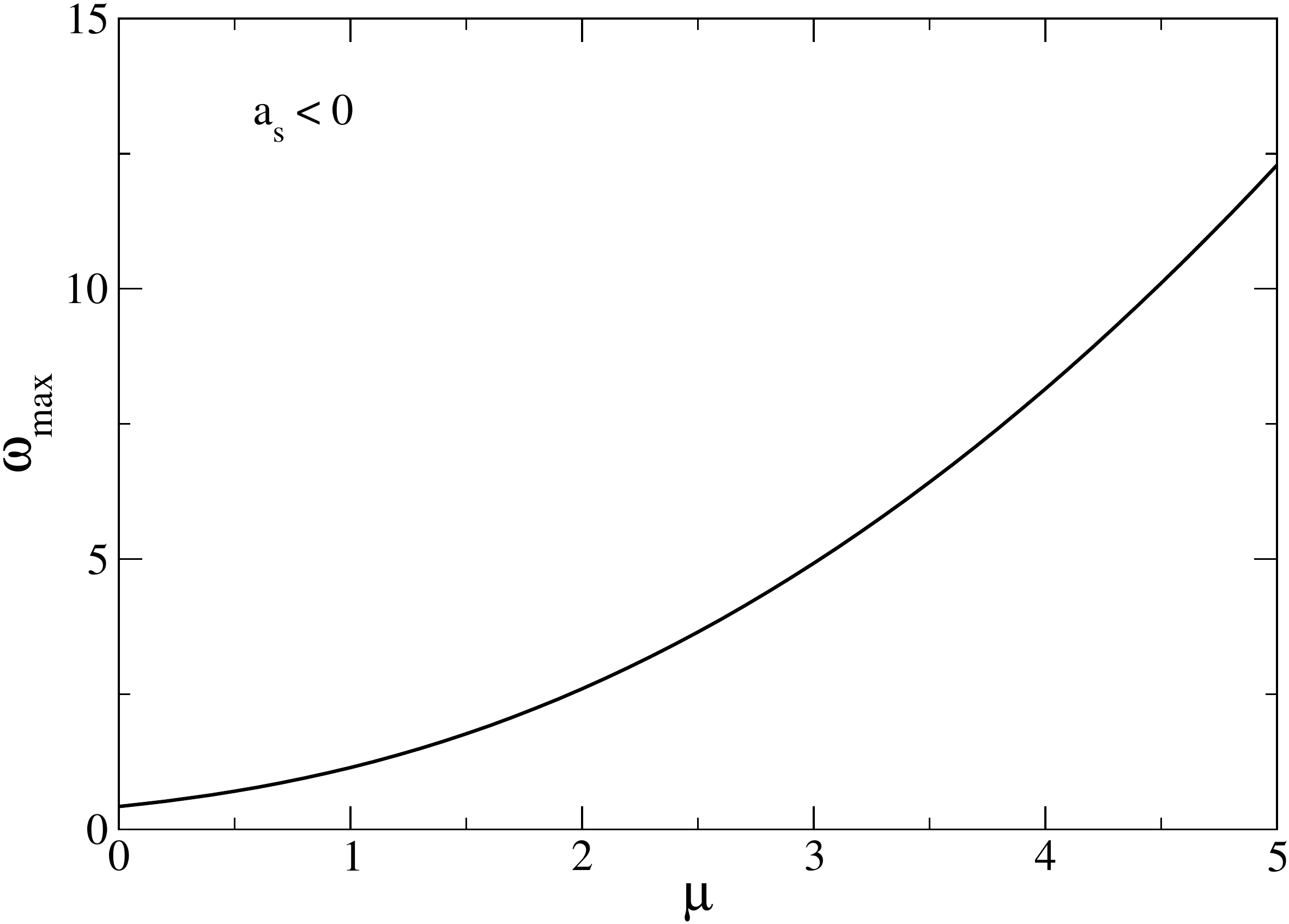}} 
\caption{Maximum pulsation $\omega_{\rm max}$ as a function of
the BH mass $\mu$.}
\label{muMaxw2}
\end{figure}

\section{Conclusion}
\label{sec_conclusion}

In this paper, we have studied the influence of a black hole that would be
present at the center of a
self-gravitating BEC representing a dark matter halo. In particular, we have
studied how the black hole
modifies the mass-radius relation of self-gravitating BECs
obtained in our previous papers \cite{prd1,prd2}. These results may find
applications in the context of dark matter halos made of bosons like
ultralight axions. Using a Gaussian ansatz, we have obtained general analytical
results valid for noninteracting and self-interacting bosons. 

In the noninteracting case ($a_s=0$), there exists a stable equilibrium state
for any mass $M$. The radius of the BEC decreases as the mass increases and
remains always smaller than the gravitational Bohr radius $R_B$.

For a repulsive self-interaction ($a_s>0$), there exists a stable equilibrium
state for any mass $M$. When $\mu<2$, the radius
of the BEC decreases as the mass increases and remains between $R_{\rm TF}$ and
$R_B$. When $\mu>2$ the radius of the BEC increases
as the mass increases and remains between $R_B$ and $R_{\rm TF}$.

For an attractive self-interaction ($a_s<0$), there exists a stable equilibrium
state only below a maximum mass $M_{\rm max}(M_{\rm BH})$ generalizing the
maximum mass found in \cite{prd1,prd2}. The radius of the BEC
decreases as the mass increases and remains between $R_*(M_{\rm BH})$ and $R_B$.
On the other hand, the maximum mass $M_{\rm max}(M_{\rm BH})$  decreases as the
black hole mass
increases.

We have resisted the temptation to make numerical applications because they
would be too speculative at this stage (see footnote 2). Indeed, we do not know
the characteristics
($m$, $a_s$) of the dark matter particle, nor the mass of the solitonic core
$M_c$ as a function of the halo mass $M_h$ for self-interacting
bosons in the presence of a central black hole (work is in progress in
this direction). Therefore, we have provided
general analytical formulae that can be used easily to make numerical
applications once these quantities will be determined. This will be the object
of future research.

\appendix

\section{Dimensionless variables}
\label{sec_cmmm}

In this Appendix, we recall the expression of the dimensionless variables
introduced in our
previous papers \cite{bectcoll,phi6} and used in Sec. \ref{sec_pure}. Let us
consider a self-gravitating BEC in
the absence of a central BH. When the self-interaction of the bosons is
attractive ($a_s<0$), the maximum mass and the corresponding radius of the
system are given, within the Gaussian ansatz, by \cite{prd1}:
\begin{eqnarray}
\label{cmmm1}
M_{\rm max}=\left
(\frac{\sigma^2}{6\pi\zeta\nu}\right
)^{1/2}\frac{\hbar}{\sqrt{Gm|a_s|}},
\end{eqnarray}
\begin{eqnarray}
\label{cmmm2}
R_{*}=\left (\frac{6\pi \zeta}{\nu}\right
)^{1/2}\left
(\frac{|a_s|\hbar^2}{Gm^3}\right )^{1/2}.
\end{eqnarray}
When the self-interaction of the bosons is repulsive ($a_s<0$), these scales
typically determine the transition between the noninteracting regime and the
TF regime (see \cite{prd1} and the Introduction). 

We define a density scale, a
pressure scale, an energy scale
and a dynamical time scale by
\begin{eqnarray}
\label{cmmm3}
\rho_{0}=\frac{\sigma
\nu}{(6\pi\zeta)^2}\frac{Gm^4}{a_s^2\hbar^2},\qquad
P_{0}=\frac{2\pi\sigma^2\nu^2 } { (6\pi\zeta)^4}\frac{G^2m^5}{|a_s|^3\hbar^2},
\end{eqnarray}
\begin{eqnarray}
V_0=\frac{\sigma^2\nu^{1/2}}{(6\pi\zeta)^{3/2}}\frac{\hbar
m^{1/2}G^{1/2}}{|a_s|^{3/2}},\qquad t_D=\frac{6\pi\zeta}{\nu}\left
(\frac{\alpha}{\sigma}\right
)^{1/2}\frac{|a_s|\hbar}{Gm^2}.
\label{cmmm5}
\end{eqnarray}
We note the
identities
\begin{eqnarray}
M_{\rm max}=\frac{\sigma}{\nu}\frac{\hbar^2}{Gm^2 R_*},\qquad
\rho_{0}=\frac{M_{\rm max}}{R_{*}^3},
\label{cmmm7}
\end{eqnarray}
\begin{eqnarray}
V_0=\nu \frac{GM_{\rm max}^2}{R_*},\qquad t_D=\left (\frac{\alpha}{\nu}\right
)^{1/2}\frac{1}{\sqrt{G\rho_0}},
\label{cmmm8}
\end{eqnarray}
\begin{eqnarray}
\frac{M_{\rm max}}{R_*}=\frac{\sigma}{6\pi\zeta}\frac{m}{|a_s|}.
\end{eqnarray}

Using the scales from Eqs. (\ref{cmmm1})-(\ref{cmmm5}), we introduce the
dimensionless variables
\begin{eqnarray}
\hat M=\frac{M}{M_{\rm max}},\qquad \hat R=\frac{R}{R_*},\qquad  \hat
\rho=\frac{\rho}{\rho_{0}},
\label{cmmm9}
\end{eqnarray}
\begin{eqnarray}
 \hat P
=\frac{P}{P_{0}},\qquad \hat
V=\frac{V}{V_0}, \qquad \hat t=\frac{t}{t_D},\qquad \hat\omega=\omega t_D.
\label{cmmm10}
\end{eqnarray}
In Sec. \ref{sec_pure} we work with these
dimensionless variables but, in order to
simplify the notations, we do not write the ``hats''.

\section{Generalized GPP equations with an algebraic and a logarithmic
external potential}
\label{sec_g}

In Ref. \cite{ggp} we have developed a general formalism applying to
dissipative self-gravitating
BECs in $d$ dimensions  described by the generalized GPP equations
\begin{eqnarray}
\label{g1}
i\hbar \frac{\partial\psi}{\partial t}=-\frac{\hbar^2}{2m}\Delta\psi
+m\left\lbrack\Phi
+\frac{dV}{d|\psi|^2}+\Phi_{\rm ext}\right\rbrack\psi
-i\frac{\hbar}{2}\xi\left\lbrack \ln\left (\frac{\psi}{\psi^*}\right
)-\left\langle \ln\left (\frac{\psi}{\psi^*}\right
)\right\rangle\right\rbrack\psi,
\end{eqnarray}
\begin{equation}
\label{g2}
\Delta\Phi=S_d G |\psi|^2,
\end{equation}
where $\xi$ is the friction coefficient and $\Phi_{\rm ext}$ an arbitrary
external potential. In this Appendix, we make some of our results more
explicit in
the case where the external potential $\Phi_{\rm ext}$ is algebraic
or logarithmic. We give the results without derivation and  refer to our paper
\cite{ggp} for technical details.

\subsection{Algebraic and logarithmic external potentials}
\label{sec_ga}

We consider an algebraic external potential of the form
\begin{eqnarray}
\label{ga1}
\Phi_{A}=-\frac{A}{r^{s}}
\end{eqnarray}
and a logarithmic  external potential of the form
\begin{eqnarray}
\label{ga2}
\Phi_{L}=B\ln r.
\end{eqnarray}
The harmonic potential 
\begin{eqnarray}
\label{ga3}
\Phi_{H}=\frac{1}{2}\omega_0^2 r^2
\end{eqnarray}
is a  particular case of Eq. (\ref{ga1}) corresponding to $s=-2$ and
$A=-\omega_0^2/2$. The BH (central point mass) potential is given by
\begin{eqnarray}
\label{ga4n1}
\Phi_{\rm BH}=-\frac{1}{d-2}\frac{GM_{\rm BH}}{r^{d-2}}\qquad (d\neq 2),
\end{eqnarray}
\begin{eqnarray}
\label{ga4n2}
\Phi_{\rm BH}=GM_{\rm BH}\ln r \qquad (d=2).
\end{eqnarray}
The BH potential in $d\neq 2$ dimensions  is a  particular
case of Eq. (\ref{ga1}) corresponding to $s=d-2$ and
$A=GM_{\rm BH}/(d-2)$ (for $d=3$, we obtain $s=1$ and
$A=GM_{\rm BH}$). The BH potential in $d=2$ dimensions  is a  particular
case of Eq. (\ref{ga2}) corresponding to $B=GM_{\rm BH}$. In the
following, to make the formulae as  general and useful
as possible, we assume that the system is submitted to  an arbitrary algebraic
potential
(\ref{ga1}), a logarithmic potential (\ref{ga2}), and a harmonic potential
(\ref{ga3}).

\subsection{Free energy}
\label{sec_gf}

The free energy associated with the generalized GPP equations
(\ref{g1}) and (\ref{g2}) is
\begin{eqnarray}
\label{gf1}
F=\int\rho \frac{{\bf u}^2}{2}\, d{\bf r}+\frac{1}{m}\int
\rho Q\, d{\bf r}
+\frac{1}{2}\int\rho\Phi\, d{\bf r}
+\int\rho\Phi_{\rm ext}\, d{\bf
r}+\int
V(\rho)\, d{\bf r}.
\end{eqnarray}
This is the sum of the classical kinetic energy $\Theta_c$, the quantum kinetic
energy $\Theta_Q$, the gravitational potential energy $W$, the external
potential energy $W_{\rm ext}$, and the internal energy $U$. The potential
energies
associated with the algebraic
potential (\ref{ga1}) and with the logarithmic
potential (\ref{ga2}) are
\begin{eqnarray}
\label{gf1b}
W_A=\int \rho \Phi_A\, d{\bf r}\qquad{\rm and}\qquad  W_L=\int \rho \Phi_L\,
d{\bf r}.
\end{eqnarray}
The potential energy
associated with the harmonic
potential (\ref{ga3}) can be written as
\begin{eqnarray}
\label{gf2}
W_H=\frac{1}{2}\omega_0^2 I, \qquad {\rm where}\qquad I=\int \rho r^2\, d{\bf r}
\end{eqnarray}
is the moment of intertia. The potential energy
associated with the BH
potential  (\ref{ga4n1}) or (\ref{ga4n2}) can be written as
\begin{eqnarray}
\label{gf2n1}
W_{\rm BH}=\int \rho \Phi_{\rm BH}\,
d{\bf r}.
\end{eqnarray}
The free energy is therefore given by
\begin{eqnarray}
\label{gf4}
F=\Theta_c+\Theta_Q
+W+W_H+W_A+W_L+U.
\end{eqnarray}
At equilibrium, it reduces to
\begin{eqnarray}
\label{gf4b}
F=\Theta_Q+W+W_H+W_A+W_L+U.
\end{eqnarray}

{\it Remark:} In the case of a power-law potential
$V(\rho)=K\rho^{\gamma}/(\gamma-1)$, leading to a polytropic equation of state
$P=K\rho^{\gamma}$, the internal energy is given by
\begin{equation}
\label{inten}
U=\frac{1}{\gamma-1}\int P\, d{\bf r}.
\end{equation}

\subsection{$H$-theorem and equilibrium states}
\label{sec_htes}

The generalized GPP equations (\ref{g1}) and
(\ref{g2}) satisfy an $H$-theorem for the free
energy (\ref{gf1}) (see \cite{ggp} for details). An equilibrium state
extremizes $F$ at fixed mass $M$. Writing the variational principle as $\delta
F-(\mu/m)\delta M=0$, where $\mu$ (chemical potential) is a Lagrange multiplier
taking into the conservation of mass,
we obtain
\begin{equation}
\label{htes1}
m\Phi+mV'(\rho)+m\Phi_{\rm ext}+Q=\mu.
\end{equation}
Using Eq. (\ref{mad5bb}), the foregoing equation can be
rewritten as
\begin{equation}
\label{tigp4b}
\Phi+\int^{\rho} \frac{P'(\rho')}{\rho'}\, d\rho'+\Phi_{\rm
ext}+\frac{Q}{m}=\frac{\mu}{m}.
\end{equation}
Taking the gradient of this relation, one recovers the condition of quantum
hydrostatic equilibrium from Eq. (\ref{eq1}). An equilibrium state is
(linearly) stable if, and only if, it is a (local) minimum of $F$ at fixed mass
$M$.

\subsection{Virial theorem}
\label{sec_gv}

The time-dependent scalar virial theorem  can be written as
\begin{equation}
\label{gv1b}
\frac{1}{2}\ddot I+\frac{1}{2}\xi\dot I=2(\Theta_c+\Theta_Q)+d\int
P\, d{\bf
r}+W_{ii}+W_{ii}^{\rm ext}.
\end{equation}
At equilibrium ($\ddot I=\dot I=\Theta_c=0$), the scalar virial theorem becomes
\begin{equation}
\label{gv2b}
2\Theta_Q+d\int P\, d{\bf r}+W_{ii}+W_{ii}^{\rm ext}=0.
\end{equation}
The virial of the external force is defined by
\begin{equation}
\label{gv3}
W_{ii}^{\rm ext}=-\int \rho {\bf r}\cdot\nabla\Phi_{\rm ext}\, d{\bf r}.
\end{equation}
For the algebraic potential (\ref{ga1}) and for the logarithmic potential
(\ref{ga2}), we
obtain
\begin{eqnarray}
\label{gv4}
W_{ii}^{A}=s W_{A}\qquad {\rm and}\qquad W_{ii}^{L}=-BM.
\end{eqnarray}
For the harmonic potential (\ref{ga3}), we
get
\begin{eqnarray}
\label{gv6}
W_{ii}^{H}=-2 W_{H}=-\omega_0^2 I.
\end{eqnarray}
For the  BH potential  (\ref{ga4n1}) and (\ref{ga4n2}), we
get 
\begin{eqnarray}
\label{gv7n1}
W_{ii}^{\rm BH}=(d-2)W_{\rm BH}\qquad (d\neq 2),
\end{eqnarray}
\begin{eqnarray}
\label{gv7n2}
W_{ii}^{\rm BH}=-GM_{\rm BH}M\qquad (d=2).
\end{eqnarray}
In particular, in $d=3$ dimensions, we have $W_{ii}^{\rm BH}=W_{\rm
BH}$. Using the foregoing relations, the time-dependent scalar virial theorem
and the scalar virial theorem
can be rewritten as
\begin{equation}
\label{gv1}
\frac{1}{2}\ddot I+\frac{1}{2}\xi\dot I+\omega_0^2 I=2(\Theta_c+\Theta_Q)+d\int
P\, d{\bf
r}+W_{ii}+s W_A-BM,
\end{equation}
\begin{equation}
\label{gv2}
2\Theta_Q+d\int P\, d{\bf r}+W_{ii}-\omega_0^2 I+s W_A-BM=0.
\end{equation}
In the case of a polytropic equation of state, the integral $\int P\, d{\bf r}$
appearing in the virial theorem can be related to the internal energy $U$ by 
using Eq. (\ref{inten}).

{\it Remark:} If we consider the nongravitational limit
($G=0$) and the dissipationless case
($\xi=0$) where the free energy
$F$ is conserved, we can combine Eqs. (\ref{gf4}) and
(\ref{gv1}) to obtain
\begin{equation}
\frac{1}{2}\ddot I+2\omega_0^2 I=2F-2U+d\int
P\, d{\bf r}+(s-2)W_A-2W_L-BM.
\end{equation}
For a polytropic equation of state $P=K\rho^{\gamma}$, using Eq.
(\ref{inten}),  we get
\begin{equation}
\frac{1}{2}\ddot I+2\omega_0^2 I=2F+\lbrack
d(\gamma-1)-2\rbrack U+(s-2)W_A-2W_L-BM.
\end{equation}
For the critical polytropic index $\gamma_c=1+2/d$ (i.e. $n_c=d/2$)
\cite{sulem}, for an
algebraic potential of index $s=2$ (this potential turns out to
have special properties), and in the absence of logarithmic
potential ($B=0$),
we get the
closed equation
\begin{equation}
\label{closed}
\frac{1}{2}\ddot I+2\omega_0^2 I=2F.
\end{equation}
It has the solution $I(t)=A\cos(2\omega_0 t+\phi)+F/\omega_0^2$. This
result is valid for a repulsive ($K>0$) or an attractive ($K<0$)
self-interaction. In $d=2$ dimensions, the critical
index is $\gamma_c=2$, corresponding to the standard BEC.

{\it Remark:} If we consider the strong friction limit ($\xi\rightarrow
+\infty$), the TF approximation ($\hbar=0$), a two-dimensional system ($d=2$)
and an isothermal equation of state $P=\rho k_B T/m$, the scalar virial theorem
(\ref{gv1}) reduces,  in the absence of algebraic potential ($A=0$), to
\begin{equation}
\frac{1}{2}\xi\dot I+\omega_0^2 I=2Nk_B T-\frac{GM^2}{2}-BM.
\end{equation}
This is a closed equation. At equilibrium, we get the identity
\begin{equation}
\label{closed2}
\omega_0^2 I=2Nk_B T-\frac{GM^2}{2}-BM.
\end{equation}
With respect to the study performed in Sec. 5.1.5 of Ref. \cite{ggp}, the
external logarithmic potential
simply shifts the critical temperature to the value
\begin{equation}
k_B T_c=\frac{GMm}{4}+\frac{Bm}{2}.
\end{equation}

\subsection{Eigenenergy}
\label{sec_eigen}

If we consider a wave function of the form
\begin{equation}
\label{tigp1}
\psi({\bf r},t)=\phi({\bf r})e^{-i E t/\hbar},
\end{equation}
where $\phi({\bf r})=\sqrt{\rho({\bf r})}$ is real, and substitute Eq.
(\ref{tigp1}) into Eqs. (\ref{g1}) and (\ref{g2}), we obtain the 
time-independent generalized GPP equations
\begin{eqnarray}
\label{tigp2}
-\frac{\hbar^2}{2m}\Delta\phi+m(\Phi+V'(\rho)+\Phi_{\rm ext})\phi=E\phi,
\end{eqnarray}
\begin{equation}
\label{tigp2b}
\Delta\Phi=S_d G \phi^2.
\end{equation}
Equations (\ref{tigp2}) and (\ref{tigp2b}) define a nonlinear
eigenvalue problem for the 
wave function $\phi({\bf r})$ where the eigenvalue $E$ is the energy
(eigenenergy). 
Dividing Eq. (\ref{tigp2}) by $\phi({\bf r})$ and using $\rho=\phi^2$,  we get
\begin{equation}
\label{tigp4}
m\Phi+mV'(\rho)+m\Phi_{\rm ext}+Q=E.
\end{equation}
This equation coincides with Eq. (\ref{htes1}) provided that we make the
identification $E=\mu$.
Taking the gradient of this relation, one recovers the condition of quantum
hydrostatic equilibrium from Eq. (\ref{eq1}). Finally, multiplying Eq.
(\ref{tigp4}) by $\rho$ and integrating over the whole configuration, we
obtain the identity
\begin{equation}
\label{tigp4c}
NE=2W+\int \rho V'(\rho)\, d{\bf r}+W_{\rm ext}+\Theta_Q.
\end{equation}

{\it Remark:} In the case of a power-law potential
$V(\rho)=K\rho^{\gamma}/(\gamma-1)$, leading to a polytropic equation of state
$P=K\rho^{\gamma}$, Eqs. (\ref{tigp4}) and (\ref{tigp4c}) take the form
\begin{equation}
\label{tigp4bb}
\Phi+\frac{K\gamma}{\gamma-1}\rho^{\gamma-1}+\Phi_{\rm
ext}+\frac{Q}{m}=\frac{E}{m},
\end{equation}
\begin{equation}
\label{tigp4cc}
NE=2W+\gamma U+W_{\rm ext}+\Theta_Q,
\end{equation}
where $U$ is given by Eq. (\ref{inten}).

\subsection{The Gaussian ansatz}
\label{sec_gg}

Making a Gaussian ansatz for the wave function, we find
that the potential energy associated with the algebraic potential (\ref{ga1}) is
given by
\begin{eqnarray}
\label{gg1}
W_{A}=-\lambda\frac{MA}{R^{s}}
\end{eqnarray}
with
\begin{equation}
\label{gg2}
\lambda=\frac{1}{\Gamma(d/2)}\int_0^{+\infty}e^{-t}t^{(d-2-s)/2}
\, dt=\frac{\Gamma\lbrack(d-s)/2\rbrack}{\Gamma(d/2)}.
\end{equation}
For the logarithmic potential (\ref{ga2}), we find that
\begin{eqnarray}
\label{gg5}
W_{L}=MB\ln R+D
\end{eqnarray}
with
\begin{equation}
\label{gg6}
D=\frac{MB}{2\Gamma(d/2)}\int_0^{+\infty}e^{-t}\ln(t)t^{(d-2)/2}\, dt=\frac{MB}{
2}\psi\left (\frac{d}{2}\right ),
\end{equation}
where $\psi(z)$ is the digamma function. We recall that
$\psi(1)=-\gamma_E=-0.577216...$ where $\gamma_E$ is the Euler constant.
For the harmonic potential (\ref{ga3}), denoting $\lambda_H$ by $\alpha$ in
order to be consistent with the notations from Ref. \cite{ggp}, we get
\begin{eqnarray}
W_{H}=\frac{1}{2}\alpha \omega_0^2 MR^2\qquad {\rm
with}\qquad \alpha=\frac{d}{2}.
\end{eqnarray}
For the BH potential (\ref{ga4n1}) and (\ref{ga4n2}), we get 
\begin{eqnarray}
\label{gg3}
W_{\rm BH}=-\frac{\lambda_{\rm BH}}{d-2}\frac{GM_{\rm BH}M}{R^{d-2}}\qquad {\rm
with}\qquad
\lambda_{\rm BH}=\frac{1}{\Gamma(d/2)} \qquad (d\neq 2),
\end{eqnarray}
\begin{eqnarray}
\label{gg3b}
W_{\rm BH}=GM_{\rm BH}M\ln R+D\qquad {\rm
with}\qquad
D=\frac{GM_{\rm BH}M}{2}\psi(1) \qquad (d=2).
\end{eqnarray}
In particular, in $d=3$, we get $W_{\rm BH}=-\lambda_{\rm BH} GM_{\rm
BH}M/R$ with $\lambda_{\rm BH}={2}/{\pi^{1/2}}$.

We now consider the generalized model of BECDM halos of Ref. \cite{ggp}
corresponding to an equation of state which is the sum of an isothermal
equation of state and a
polytropic equation of state: $P=\rho k_B T/m+K\rho^{\gamma}$.
The free energy
functional (\ref{gf4}) can be written
as a function of $R$ and $\dot R$  (for a fixed mass $M$) as
\begin{eqnarray}
\label{a10}
F=\frac{1}{2}\alpha M\left (\frac{dR}{dt}\right )^2+V(R),
\end{eqnarray}
with
\begin{equation}
\label{a11}
V(R)=\sigma \frac{\hbar^2M}{m^2R^2}-\frac{\nu}{d-2}
\frac{GM^2}{R^{d-2}}+\frac{1}{2}\omega_0^2\alpha MR^2
+\frac{\zeta}{\gamma-1}\frac{KM^{\gamma}}{R^{d(\gamma-1)}}-d\frac{M k_B T}{m}\ln
R+C-\lambda \frac{MA}{R^{s}}+MB\ln R+D\qquad (d\neq 2),
\end{equation}
\begin{equation}
\label{a12}
V(R)=\sigma \frac{\hbar^2M}{m^2R^2}+\frac{1}{2}
GM^2\ln R+\frac{1}{2}\omega_0^2\alpha MR^2
+\frac{\zeta}{\gamma-1}\frac{KM^{\gamma}}{R^{2(\gamma-1)}}-2\frac{M k_B T}{m}\ln
R+W_0-\lambda \frac{MA}{R^{s}}+MB\ln R+D\qquad (d=2).
\end{equation}
Equation (\ref{a10}) can be interpreted as the total energy of a
fictive   particle with effective mass $\alpha M$ and position $R$  moving in a
potential
$V(R)$. The first term is the classical kinetic energy $\Theta_c$ and the second
term is
the potential energy $V$ including the quantum kinetic energy $\Theta_Q$, the
gravitational potential energy $W$, the potential energy $W_{H}$ associated
with the harmonic external potential, the internal energy $U$ associated with
the polytropic equation of state, the internal energy  $U_B$ associated with the
isothermal equation of state, the potential energy $W_{A}$ associated
with the algebraic external potential, and the potential energy $W_{L}$
associated
with the logarithmic external potential. An equilibrium state is an extremum of
$V(R)$. This leads to the general mass-radius relation
\begin{eqnarray}
\label{a13b}
-2\sigma \frac{\hbar^2M}{m^2R^3}+\nu \frac{GM^2}{R^{d-1}}+\omega_0^2\alpha
MR
-d \zeta \frac{KM^{\gamma}}{R^{d(\gamma-1)+1}}-d \frac{M k_B T}{m
R}+\lambda s \frac{MA}{R^{s+1}}+\frac{MB}{R}=0.
\end{eqnarray}
This relation can also be obtained from the virial theorem \cite{ggp}.
The complex pulsation $\omega^2=(1/\alpha M)V''(R)$ describing the
evolution of a small perturbation about equilibrium is
given by
\begin{eqnarray}
\label{a29}
\omega^2=\omega_0^2+\frac{6\sigma}{\alpha}\frac{\hbar^2}{m^2 R^4}+\lbrack
d(\gamma-1)+1\rbrack \frac{d\zeta}{\alpha} \frac{K
M^{\gamma-1}}{R^{d(\gamma-1)+2}}
-\frac{(d-1)\nu}{\alpha}\frac{GM}{R^d
} +\frac { d } { \alpha}\frac{k_B T}{m
R^2}-\frac{\lambda}{\alpha}s(s+1)\frac{A}{R^{s+2}}-\frac{B}{\alpha R^2}.
\end{eqnarray}
It can be expressed under the form 
\begin{eqnarray}
\label{a30}
\omega^2=\frac{6\Theta_Q+\lbrack d(\gamma-1)+1\rbrack d(\gamma-1) U
+(d-1) W_{ii}+\omega_0^2 I+dNk_B T+s(s+1)W_A-MB}{I}.
\end{eqnarray}
Alternative expressions of the pulsation can be obtained by combining Eq.
(\ref{a30}) with the
equilibrium free energy (\ref{gf4}) in the case where the free energy is
conserved ($\xi=0$), or with the equilibrium virial theorem (\ref{gv2}). We
also note
the identity
\begin{eqnarray}
\label{turning2}
\omega^2(R)=-\frac{1}{\alpha M}\left
(\frac{2\sigma\hbar^2}{m^2R^3}-\omega_0^2\alpha R+\frac{d K\zeta(2-\gamma)
M^{\gamma-1}}{R^{d(\gamma-1)+1}}+\frac{dk_B T}{m
R}-\lambda s\frac{A}{R^{s+1}}-\frac{B}{R}\right )\frac{d M}{d R},
\end{eqnarray}
which is related to the Poincar\'e turning point criterion. Let us
consider particular
cases of Eq. (\ref{a30}).

For classical polytropes ($\Theta_Q=T=0$), the virial theorem
reduces to
$d(\gamma-1)U+W_{ii}-\omega_0^2I+sW_A-BM=0$ and the complex pulsation can be
written
as
\begin{eqnarray}
\label{a34}
\omega^2=\frac{(2d-2-d\gamma)W_{ii}+(d\gamma-d+2)\omega_0^2
I+s(s+d-d\gamma)W_A+d(\gamma-1)MB}{I}.
\end{eqnarray}
For
$d=3$ and $\omega_0=A=B=0$, using $W_{ii}=W$, we recover the
Ledoux
formula $\omega^2=(4-3\gamma){W}/{I}$ \cite{ledoux} (see Refs.
\cite{prd,csledoux} for generalizations).

For classical isothermal spheres ($\Theta_Q=U=0$),  the virial
theorem reduces
to $W_{ii}-\omega_0^2 I+dNk_B T+s W_A-BM=0$ and the complex pulsation can be
written
as 
\begin{eqnarray}
\label{a35}
\omega^2=\frac{(d-2)W_{ii}+2\omega_0^2 I+s^2 W_A}{I}\qquad {\rm or}\qquad
\omega^2=\frac{(2-d)dNk_BT+d\omega_0^2 I+(s+2-d)s W_A+(d-2)MB}{I}.
\end{eqnarray}
For $d=2$ and $A=0$, we obtain
$\omega^2=2\omega_0^2$ and
the virial theorem leads to identity (\ref{closed2}).

In the noninteracting case ($U=0$), the virial theorem reduces
to $2\Theta_Q+W_{ii}-\omega_0^2 I+dNk_B T+sW_A-BM=0$ and the complex pulsation
can be
written
as
\begin{eqnarray}
\label{a36}
\omega^2=\frac{(d-4)W_{ii}+4\omega_0^2I-2dNk_B T+s(s-2)W_A+2BM}{I}.
\end{eqnarray}

For nongravitational ($G=0$) polytropes ($T=0$), the virial
theorem reduces
to $2\Theta_Q+d(\gamma-1)U-\omega_0^2 I+sW_A-BM=0$ and the complex
pulsation can be
written
as
\begin{eqnarray}
\label{a37}
\omega^2=\frac{d(\gamma-1)\lbrack
d(\gamma-1)-2\rbrack U+4\omega_0^2I+s(s-2)W_A+2BM}{I}.
\end{eqnarray}
For the critical index $\gamma_c=1+2/d$ \cite{sulem}, for  $s=2$
 and for $B=0$, we obtain
$\omega^2=4\omega_0^2$ in
agreement with Eq. (\ref{closed}). In the TF approximation ($\Theta_Q=0$), the
virial theorem reduces to $d(\gamma-1)U-\omega_0^2 I+sW_A-BM=0$ and the
complex pulsation becomes
\begin{eqnarray}
\label{a37b}
\omega^2=\frac{\lbrack d(\gamma-1)+2\rbrack\omega_0^2 I+s\lbrack
s-d(\gamma-1)\rbrack W_A+d(\gamma-1)BM}{I}.
\end{eqnarray}

\section{Gravitational energy of a polytropic sphere with an external
potential}
\label{sec_theorem}

A simple analytical formula due to Betti and Ritter \cite{chandrabook} can be
obtained for the gravitational energy of a
polytropic sphere: 
\begin{eqnarray}
W=-\frac{3}{5-n}\frac{GM^2}{R}\qquad (d=3).
\end{eqnarray}
In this
Appendix, we determine the proper generalization of this formula in the case
where the polytrope is submitted to an arbitrary external potential.
Simplifications are given for the algebraic potential (including the harmonic
potential and the BH potential) and for the logarithmic potential. 

\subsection{General expression}

For classical self-gravitating systems, or for
self-gravitating BECs in the TF approximation, the condition of hydrostatic
equilibrium can be written as
\begin{eqnarray}
\label{t1}
\nabla P+\rho\nabla\Phi+\rho\nabla\Phi_{\rm ext}={\bf 0}.
\end{eqnarray}
For a polytropic equation of
state of the form
\begin{eqnarray}
\label{t3}
P=K\rho^{\gamma}\qquad {\rm with}\qquad \gamma=1+\frac{1}{n},
\end{eqnarray}
we have
\begin{eqnarray}
\label{t4}
\frac{\nabla P}{\rho}=(n+1)\nabla\left (\frac{P}{\rho}\right
).
\end{eqnarray}
As a result, the condition of hydrostatic equilibrium (\ref{t1}) can be
integrated into
\begin{eqnarray}
\label{t5}
(n+1)\frac{P}{\rho}+\Phi+\Phi_{\rm ext}=\frac{E}{m},
\end{eqnarray}
where $E$ is a constant of integration representing the eigenenergy of the BEC
(see Sec. \ref{sec_eigen}). Multiplying Eq. (\ref{t5}) by
$\rho$ and
integrating over the
whole
configuration, we obtain the identity
\begin{eqnarray}
\label{t6}
NE=(n+1)\int P\, d{\bf r}+2W+W_{\rm ext}.
\end{eqnarray}
Assuming $\gamma>1$ (i.e.
$0\le n<+\infty$) so that $P/\rho=0$ on the boundary of the system $r=R$ where
the density vanishes, we find from Eq. (\ref{t5}) that
\begin{eqnarray}
\label{t7}
\frac{E}{m}=\Phi(R)+\Phi_{\rm ext}(R).
\end{eqnarray}
This equation determines the
eigenenergy $E$ if we recall that \cite{ggp} 
\begin{eqnarray}
\label{t8}
\Phi(R)=-\frac{1}{d-2}\frac{GM}{R^{d-2}}\qquad (d\neq 2),
\end{eqnarray}
\begin{eqnarray}
\label{t9}
\Phi(R)=GM\ln R\qquad (d= 2).
\end{eqnarray}
As a result, Eq. (\ref{t6}) can be rewritten as 
\begin{eqnarray}
\label{t10}
(n+1)\int P\, d{\bf r}=M\Phi(R)+M\Phi_{\rm ext}(R)-2W-W_{\rm ext}.
\end{eqnarray}
Combining this relation with the equilibrium scalar virial theorem (see
Sec. \ref{sec_gv})
\begin{eqnarray}
\label{t11}
d\int P\, d{\bf r}+W_{ii}+W_{ii}^{\rm ext}=0,
\end{eqnarray}
we obtain the general identity
\begin{eqnarray}
\label{t12}
(n+1)W_{ii}-2dW=-dM\Phi(R)-dM\Phi_{\rm ext}(R)
+d  W_{\rm ext}-(n+1)W_{ii}^{\rm ext}
\end{eqnarray}
determining the gravitational energy $W$ of a classical  polytropic sphere
submitted to an external potential. More explicit expressions are given below.

\subsection{$d\neq 2$}

When $d\neq 2$, using $W_{ii}=(d-2)W$ \cite{ggp} and Eq. (\ref{t8}),
we find from Eq. (\ref{t12}) that the gravitational energy is given by
\begin{eqnarray}
\label{t13}
W=\frac{1}{(d-2)n-(d+2)}\left\lbrack \frac{d}{d-2}\frac{GM^2}{R^{d-2}}-dM\Phi_{
\rm ext}(R)
+d  W_{\rm ext}-(n+1)W_{ii}^{\rm ext}\right\rbrack.
\end{eqnarray}
In the absence of external potential, we recover the Betti-Ritter formula in
$d$ dimensions \cite{lang}:
\begin{eqnarray}
\label{t13b}
W=\frac{d}{(d-2)n-(d+2)}\frac{GM^2}{(d-2)R^{d-2}}.
\end{eqnarray}
For the algebraic potential
(\ref{ga1}), using Eq. (\ref{gv4}), Eq.
(\ref{t13}) takes the form
\begin{eqnarray}
\label{t14}
W=\frac{1}{(d-2)n-(d+2)}\left\lbrack
\frac{d}{d-2}\frac{GM^2}{R^{d-2}}+\frac{dMA} {R^{s}}
+\left\lbrack
d-s(n+1)\right\rbrack W_{A}\right\rbrack.
\end{eqnarray}
In particular, for the harmonic potential (\ref{ga3}), using Eq. (\ref{gf2}),
we get
\begin{eqnarray}
\label{t15}
W=\frac{1}{(d-2)n-(d+2)}\left\lbrack \frac{d}{d-2}\frac{GM^2}{R^{d-2}}-\frac{
dM\omega_0^2 } { 2 } R^{2}
+\frac{1}{2}(d+2+2n)\omega_0^2 I\right\rbrack.
\end{eqnarray}
For the BH potential (\ref{ga4n1}) we get 
\begin{eqnarray}
\label{t16n1}
W=\frac{1}{(d-2)n-(d+2)}\left\lbrack
\frac{d}{d-2}\frac{GM^2}{R^{d-2}}+\frac{d}{d-2}\frac{GM_{\rm BH}M}{R^{d-2}}
+\left\lbrack
d-(d-2)(n+1)\right\rbrack W_{\rm BH}\right\rbrack.
\end{eqnarray}
In particular, in $d=3$, we obtain
\begin{eqnarray}
\label{t16}
W=\frac{1}{n-5}\left\lbrack \frac{3GM^2}{R}+\frac{3GM_{\rm BH}M}{R}+(2-n)W_{\rm
BH}\right\rbrack.
\end{eqnarray}
A closed expression is obtained for $n=2$. Finally, for the logarithmic
potential
(\ref{ga2}), using Eq. (\ref{gv4}),
Eq.
(\ref{t13}) takes the form
\begin{eqnarray}
\label{t17}
W=\frac{1}{(d-2)n-(d+2)}\left\lbrack \frac{d}{d-2}\frac{GM^2}{R^{d-2}}-dMB\ln R
+d  W_{L}+(n+1)BM\right\rbrack.
\end{eqnarray}

\subsection{$d=2$}

When $d=2$, using $W_{ii}=-GM^2/2$ \cite{ggp} and Eq. (\ref{t9}),
we find from Eq. (\ref{t12}) that the gravitational energy is given by
\begin{eqnarray}
\label{t19}
W=-\frac{n+1}{8}GM^2+\frac{1}{2}GM^2\ln R
+\frac{1}{2}M\Phi_{\rm ext}(R)-\frac{1}{2}W_{\rm
ext}+\frac{1}{4}(n+1)W_{ii}^{\rm ext}.
\end{eqnarray}
For the algebraic potential (\ref{ga1}), using Eq. (\ref{gv4}), it takes
the form
\begin{eqnarray}
\label{t20}
W=-\frac{n+1}{8}GM^2+\frac{1}{2}GM^2\ln R
-\frac{MA}{2R^{s}}-\frac{1}{4}\left\lbrack
2-s(n+1)\right\rbrack W_{A}.
\end{eqnarray}
In particular, for  the harmonic potential  (\ref{ga3}), using Eq.
(\ref{gf2}), we get
\begin{eqnarray}
\label{t21}
W=-\frac{n+1}{8}GM^2+\frac{1}{2}GM^2\ln R
+\frac{M\omega_0^2}{4}R^{2}-\frac{1}{4}
(2+n)\omega_0^2 I.
\end{eqnarray}
On the other hand, for the logarithmic potential (\ref{ga2}), using Eq.
(\ref{gv4}), we obtain
\begin{eqnarray}
\label{t19n1}
W=-\frac{n+1}{8}GM^2+\frac{1}{2}GM^2\ln R
+\frac{1}{2}MB\ln R-\frac{1}{2}W_{\rm
L}-\frac{1}{4}(n+1)BM.
\end{eqnarray}
In particular, for  the BH potential  (\ref{ga4n2}), we get
\begin{eqnarray}
\label{t19n2}
W=-\frac{n+1}{8}GM^2+\frac{1}{2}GM^2\ln R
+\frac{1}{2}GM_{\rm BH}M\ln R-\frac{1}{2}W_{\rm
BH}-\frac{1}{4}(n+1)GM_{\rm BH}M.
\end{eqnarray}

\section{Exact nongravitational $+$ noninteracting case with the BH
potential}
\label{sec_b}

In the nongravitational $+$ noninteracting case ($G=a_s=0$), the wave function
of the BEC is determined by the
Schr\"odinger equation with the BH potential:
\begin{eqnarray}
\label{b1}
i\hbar\frac{\partial\psi}{\partial
t}=-\frac{\hbar^2}{2m}\Delta\psi-\frac{GM_{\rm BH}m}{r}\psi.
\end{eqnarray}
This equation is similar to the Schr\"odinger equation with a
Coulombian
potential
describing the Bohr atom. Therefore, a BEC in the nongravitational $+$
noninteracting case is equivalent to a gravitational Bohr atom. Considering  a
stationary solution of the form
\begin{eqnarray}
\label{b2}
\psi({\bf r},t)=\phi({\bf r})\, e^{-iEt/\hbar},
\end{eqnarray}
where  $E$ is real we obtain the eigenvalue equation
\begin{eqnarray}
\label{b3}
-\frac{\hbar^2}{2m}\Delta\phi-\frac{GM_{\rm
BH}m}{r}\phi=E\phi
\end{eqnarray}
or, equivalently,
\begin{eqnarray}
\label{b4}
\Delta\phi+\frac{2m}{\hbar^2}\left (E+\frac{GM_{\rm
BH}m}{r}\right )\phi=0.
\end{eqnarray}
Looking for a solution of the form 
\begin{eqnarray}
\label{b5}
\phi=Ae^{-\gamma r},
\end{eqnarray}
we find that 
\begin{eqnarray}
\label{b7}
\gamma=\frac{GM_{\rm BH}m^2}{\hbar^2}\qquad {\rm and}\qquad
E=-\frac{\hbar^2\gamma^2}{2m}=-\frac{G^2M_{\rm
BH}^2m^3}{2\hbar^2}.
\end{eqnarray}
Since the wave function (\ref{b5}) has no node, it corresponds to the ground
state of the
gravitational Bohr atom. The total
mass of the BEC is given by
\begin{eqnarray}
\label{b8}
M=\int_0^{+\infty} \phi^2 4\pi r^2\, dr,
\end{eqnarray}
implying
\begin{eqnarray}
\label{b9}
A=\left (\frac{M\gamma^3}{\pi}\right )^{1/2}\qquad {\rm and}\qquad \phi=\left
(\frac{M}{\pi}\right )^{1/2}\gamma^{3/2}e^{-\gamma r}.
\end{eqnarray}
The density of the BEC is
\begin{eqnarray}
\label{b10}
\rho=\frac{M}{\pi}\gamma^{3}e^{-2\gamma r}.
\end{eqnarray}
The central density is
\begin{eqnarray}
\label{b10b}
\rho_0=\frac{M}{\pi}\gamma^{3}=\frac{G^3M_{\rm BH}^3m^6M}{\pi\hbar^6}.
\end{eqnarray}
Let $R_{99}$ denote the radius containing $99\%$ of the
mass. It is given by $R_{99}=\xi_{99}/(2\gamma)$ where $\xi_{99}$ is determined
by the equation
\begin{eqnarray}
\label{b11}
\frac{\int_0^{\xi_{99}}e^{-\xi}\xi^2\,
d\xi}{\int_0^{+\infty}e^{-\xi}\xi^2\, d\xi}=0.99,
\end{eqnarray}
giving $\xi_{99}=8.406...$. Therefore, the exact
radius of
the BEC is given by
\begin{eqnarray}
\label{b12}
R_{99}=4.203\,\frac{\hbar^2}{GM_{\rm
BH}m^2}.
\end{eqnarray}
This is the gravitational
Bohr radius. We note that it is independent of the mass $M$ of the BEC. It
can be compared to the expression  (\ref{ngni1}) obtained from the Gaussian
ansatz. We note that the exact density profile $\rho$ is exponential instead of
being Gaussian.

In the nongravitational $+$ noninteracting case, the
exact equilibrium relations from Sec. \ref{sec_eer} reduce to
\begin{eqnarray}
\label{ma1}
E_{\rm tot}=\Theta_Q+W_{\rm BH},\qquad NE=W_{\rm BH}+\Theta_Q,\qquad
2\Theta_Q+W_{\rm BH}=0.
\end{eqnarray}
We note that $E_{\rm tot}=NE=-\Theta_Q=W_{\rm BH}/2$. A direct calculation
using Eq. (\ref{b10}) gives
\begin{equation}
\label{ma2}
W_{\rm
BH}=-GM_{\rm BH}M\gamma=-\frac{G^2M_{\rm BH}^2Mm^2}{\hbar^2},
\end{equation}
\begin{equation}
\label{ma3}
\Theta_Q=\frac{\hbar^2M\gamma^2}{2m^2}=\frac{G^2M_{\rm BH}^2Mm^2}{2\hbar^2},
\end{equation}
\begin{equation}
\label{ma4}
I=\frac{3M}{\gamma^2}=\frac{3M\hbar^4}{G^2M_{\rm BH}^2m^4}.
\end{equation}
We can check that the 
relations of Eq. (\ref{ma1}) are satisfied. On the other
hand, Eq. (\ref{a30}) reduces to 
\begin{equation}
\label{ma5}
\omega^2=\frac{6\Theta_Q+2W_{\rm BH}}{I}.
\end{equation}
Using the virial theorem from Eq. (\ref{ma1}), we obtain
\begin{equation}
\label{ma6}
\omega^2=-\frac{W_{\rm BH}}{I}=\frac{2\Theta_Q}{I}.
\end{equation}
Finally, using Eq. (\ref{ma4}), we get
\begin{equation}
\label{ma7}
\omega^2=\frac{GM_{\rm BH}\gamma^3}{3}=\frac{G^4M_{\rm BH}^4m^6}{3\hbar^6}.
\end{equation}

\section{Exact nongravitational $+$ TF case with the BH potential}
\label{sec_ex}

In the nongravitational $+$ TF case ($G=\hbar=0$), the condition of hydrostatic
equilibrium can be written as 
\begin{eqnarray}
\label{ex1}
\nabla P+\rho\nabla\Phi_{\rm ext}={\bf 0}.
\end{eqnarray}
For a polytropic equation of state of the form
\begin{eqnarray}
\label{ex2}
P=K\rho^{\gamma}\qquad {\rm with}\qquad \gamma=1+\frac{1}{n}
\end{eqnarray}
we have
\begin{eqnarray}
\label{ex2b}
\frac{\nabla P}{\rho}=(n+1)\nabla\left (\frac{P}{\rho}\right
).
\end{eqnarray}
As a result, the condition of hydrostatic
equilibrium (\ref{ex1}) can be
integrated into
\begin{eqnarray}
\label{ex3}
(n+1)K\rho^{1/n}+\Phi_{\rm ext}=\frac{E}{m},
\end{eqnarray}
where $E$ is a constant of integration representing the eigenenergy of the
BEC (see Sec. \ref{sec_eigen}).  Assuming $\gamma>1$ (i.e.
$0\le n<+\infty$) so that $P/\rho=0$ on the boundary of the system $r=R$ where
the density vanishes, we find from Eq. (\ref{ex3}) that
\begin{eqnarray}
\label{ex3b}
\frac{E}{m}=\Phi_{\rm ext}(R).
\end{eqnarray}
As a result, Eq. (\ref{ex3}) can be rewritten as
\begin{eqnarray}
\label{ex3c}
\rho(r)=\left\lbrack \frac{\Phi_{\rm ext}(R)-\Phi_{\rm
ext}(r)}{(n+1)K}\right\rbrack^n.
\end{eqnarray}
This equation determines the density profile $\rho({r})$ of the BEC in the TF
approximation for an
arbitrary external potential. For the  BH potential given by Eq.
(\ref{gpp6}), assuming $K>0$, we obtain
\begin{eqnarray}
\label{ex7}
\rho(r)=\left\lbrack \frac{GM_{\rm BH}}{K(n+1)}\right \rbrack^n\left
(\frac{1}{r}-\frac{1}{R}\right )^n.
\end{eqnarray}
For $r\rightarrow 0$, the density behaves as
\begin{eqnarray}
\label{ex8}
\rho(r)\sim\left\lbrack \frac{GM_{\rm BH}}{K(n+1)}\right
\rbrack^n\frac{1}{r^n}.
\end{eqnarray}
It is normalisable provided that $n<3$. Multiplying Eq. (\ref{ex7}) by $4\pi
r^2$ and
integrating over the sphere of radius $R$ we obtain the exact mass-radius
relation
\begin{eqnarray}
\label{ex9}
M=\frac{2}{3}(n-2)(n-1)
\frac{n\pi^2}{\sin(n\pi)}\left\lbrack \frac{GM_{\rm BH}}{K(n+1)}\right
\rbrack^n R^{3-n},
\end{eqnarray}
where we have used the identity
\begin{eqnarray}
\label{ex10}
\int_0^1 \left
(\frac{1}{x}-1\right )^n x^2\, dx=\frac{1}{6}(n-2)(n-1)
\frac{n\pi}{\sin(n\pi)}\qquad (-1<n<3).
\end{eqnarray}

For the usual BEC corresponding to
$n=1$ and $K=2\pi a_s\hbar^2/m^3$, the density profile  is
\begin{equation}
\label{mer1}
\rho(r)=\frac{GM_{\rm BH}m^3}{4\pi a_s\hbar^2}\left
(\frac{1}{r}-\frac{1}{R}\right ).
\end{equation} 
On the other hand, the integral in Eq. (\ref{ex10}) is equal
to $1/6$ so the exact
mass-radius relation (\ref{ex9}) reduces to
\begin{eqnarray}
\label{ex11}
M=\frac{GM_{\rm BH}m^3}{6a_s\hbar^2}R^2.
\end{eqnarray}
This can be compared to the relation (\ref{ngtf1}) obtained from the
Gaussian ansatz. We
note that the density profile $\rho(r)$ is very different from a Gaussian in the
present case. In the nongravitational $+$ TF case, the exact
equilibrium relations from Sec. \ref{sec_eer} reduce to
\begin{eqnarray}
\label{mer2}
E_{\rm tot}=W_{\rm BH}+U,\qquad NE=2U+W_{\rm BH},\qquad 3U+W_{\rm BH}=0.
\end{eqnarray}
We note that $E_{\rm
tot}=2NE=-2U=(2/3)W_{\rm BH}$. A direct calculation using Eq.
(\ref{mer1}) gives
\begin{equation}
\label{mer3}
U=\frac{G^2M_{\rm BH}^2m^3R}{6a_s\hbar^2},\qquad W_{\rm BH}=-\frac{G^2M_{\rm
BH}^2m^3R}{2a_s\hbar^2},\qquad NE=-\frac{GM_{\rm BH}M}{R},\qquad
I=\frac{GM_{\rm
BH}m^3R^4}{20 a_s\hbar^2}.
\end{equation}
We can check that the relations of
Eq. (\ref{mer2}) are satisfied. On
the other hand, Eq. (\ref{a30}) reduces to
\begin{equation}
\label{mer4}
\omega^2=\frac{12U+2W_{\rm BH}}{I}.
\end{equation}
Using the virial theorem from Eq. (\ref{mer2}), we obtain
\begin{equation}
\label{mer5}
\omega^2=-\frac{2W_{\rm BH}}{I}=\frac{6U}{I}.
\end{equation}
Finally, using Eq. (\ref{mer3}), we get 
\begin{equation}
\label{mer6}
\omega^2=\frac{20GM_{\rm BH}}{R^3}.
\end{equation}

\section{Exact nongravitational case without BH}
\label{sec_engnbh}

In the nongravitational case without
BH, the exact
equilibrium relations from Sec. \ref{sec_eer} reduce to
\begin{eqnarray}
\label{zmer2}
E_{\rm tot}=\Theta_Q+U,\qquad NE=2U+\Theta_Q,\qquad 2\Theta_Q+3U=0.
\end{eqnarray}
We note that $E_{\rm
tot}=-NE=-U/2=\Theta_Q/3$. On
the other hand, Eq. (\ref{a30}) reduces to 
\begin{equation}
\label{zmer4}
\omega^2=\frac{6\Theta_Q+12U}{I}.
\end{equation}
Using the virial theorem from Eq. (\ref{zmer2}), we obtain
\begin{equation}
\label{zmer5}
\omega^2=-\frac{2\Theta_Q}{I}=\frac{3U}{I}.
\end{equation}
We also
recall the exact results \cite{prd2}:
\begin{equation}
R_{99}=3.64\frac{|a_s|}{m}M,\qquad
NE=-0.435\frac{\hbar^2}{Ma_s^2}.
\end{equation}

\section{Exact noninteracting case without BH}
\label{sec_eninbh}

In the noninteracting case without
BH, the exact
equilibrium relations from Sec. \ref{sec_eer} reduce to
\begin{eqnarray}
\label{xzmer2}
E_{\rm tot}=\Theta_Q+W,\qquad NE=2W+\Theta_Q,\qquad 2\Theta_Q+W=0.
\end{eqnarray}
We note that $E_{\rm
tot}=NE/3=W/2=-\Theta_Q$. On
the other hand, Eq. (\ref{a30}) reduces to
\begin{equation}
\label{xzmer4}
\omega^2=\frac{6\Theta_Q+2W}{I}.
\end{equation}
Using the virial theorem from Eq. (\ref{xzmer2}), we obtain
\begin{equation}
\label{xzmer5}
\omega^2=\frac{2\Theta_Q}{I}=-\frac{W}{I}.
\end{equation}
We also recall the exact results \cite{membrado,prd1,prd2}:
\begin{equation}
\label{xzmer6}
R_{99}=9.946\frac{\hbar^2}{GMm^2},\qquad
NE=-0.1628\frac{G^2M^3m^2}{\hbar^2}.
\end{equation}

\section{Exact TF limit without BH}
\label{sec_tfnobh}

In the TF approximation ($\hbar=0$), the differential equation determining the
density profile of the BEC without central BH is (see Sec. \ref{sec_eq})
\begin{eqnarray}
\label{netf2}
\Delta\rho+\frac{Gm^3}{a_s\hbar^2}\rho=0.
\end{eqnarray}
The solution of this equation is
\begin{eqnarray}
\label{netf3}
\rho(r)=\rho_0\frac{\sin (\pi r/R)}{\pi r/R},
\end{eqnarray}
where 
\begin{eqnarray}
\label{netf4}
R=\pi \left (\frac{a_s\hbar^2}{Gm^3}\right )^{1/2}
\end{eqnarray}
is the radius at which the density vanishes and $\rho_0$ is the central density.
It is determined by the mass according to the relation
\begin{eqnarray}
\label{netf15}
\rho_0=\frac{\pi M}{4R^3}=\frac{M}{4\pi^2}\left (\frac{Gm^3}{a_s\hbar^2}\right
)^{3/2}.
\end{eqnarray}
We note that the radius has a constant value independent of the
mass. In the TF limit without BH, the exact
equilibrium relations from Sec. \ref{sec_eer} reduce to
\begin{eqnarray}
\label{er1}
E_{\rm tot}=W+U,\qquad NE=2W+2U,\qquad 3U+W=0.
\end{eqnarray}
We note that $E_{\rm tot}=(1/2)NE=-2U=(2/3)W$. We can determine the
eigenenergy $E$ by applying the
relation [see Eq. (\ref{tigp4bb}) with $Q=\Phi_{\rm ext}=0$]
\begin{equation}
\label{er2}
\Phi+\frac{4\pi a_s\hbar^2}{m^3}\rho=\frac{E}{m}
\end{equation}
at $r=R$, giving
\begin{equation}
\label{er3}
NE=-\frac{GM^2}{R}.
\end{equation}
On the other hand, a direct calculation using Eq. (\ref{netf3}) gives
\begin{equation}
\label{er4}
U=\frac{GM^2}{4R},\qquad I=\frac{(\pi^2-6)MR^2}{\pi^2}.
\end{equation}
Finally, according to the results of  Appendix \ref{sec_theorem}, we have
\begin{equation}
\label{er5}
W=-\frac{3GM^2}{4R}. 
\end{equation}
We can check that the relations from Eq.
(\ref{er1}) are satisfied. On
the other hand, Eq. (\ref{a30}) reduces to 
\begin{equation}
\label{er6}
\omega^2=\frac{12U+2W}{I}.
\end{equation}
Using the virial theorem from Eq. (\ref{er1}) we obtain
\begin{equation}
\label{er7}
\omega^2=\frac{6U}{I}=-\frac{2W}{I}.
\end{equation}
Finally, using Eq. (\ref{er4}), we get
\begin{equation}
\label{er8}
\omega^2=\frac{3\pi^2GM}{2(\pi^2-6)R^3}.
\end{equation}
This returns the results from \cite{prd1}.

\section{Exact TF limit with the BH potential}
\label{sec_etf}

\begin{figure}[h]
\scalebox{0.33}{\includegraphics{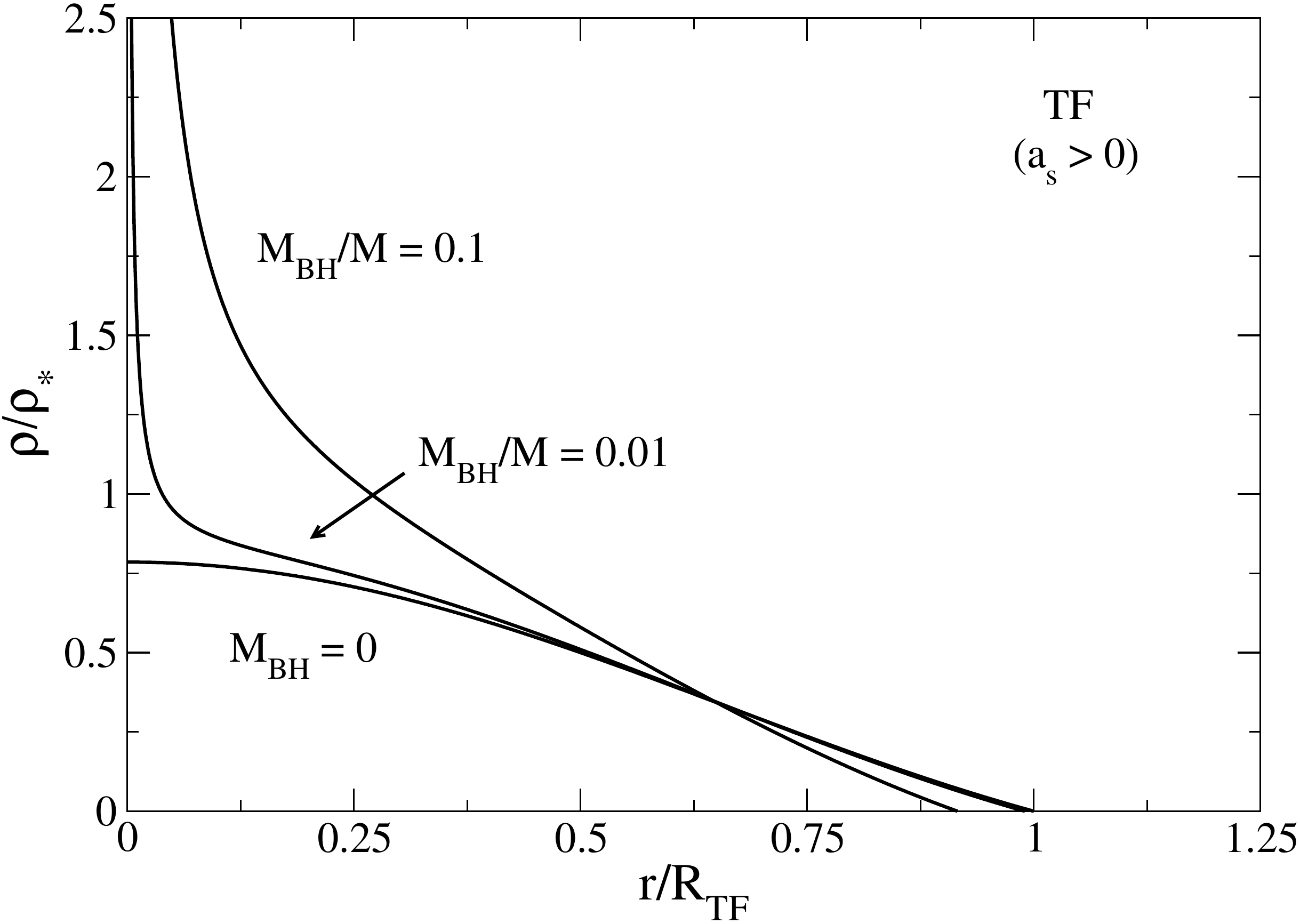}} 
\caption{Density profile of self-gravitating BECs with a repulsive
self-interaction ($a_s>0$) in the TF
limit ($\hbar=0$) in the presence of a central black hole. We have
normalized the
radius by $R_{\rm TF}$ given by Eq. (\ref{intro2}) and
the density by $\rho_*=M/R_{\rm TF}^3$. We have plotted the
profile  without black hole [see Eq. (\ref{netf3})] for comparison.}
\label{rhoTF}
\end{figure}

\begin{figure}[h]
\scalebox{0.33}{\includegraphics{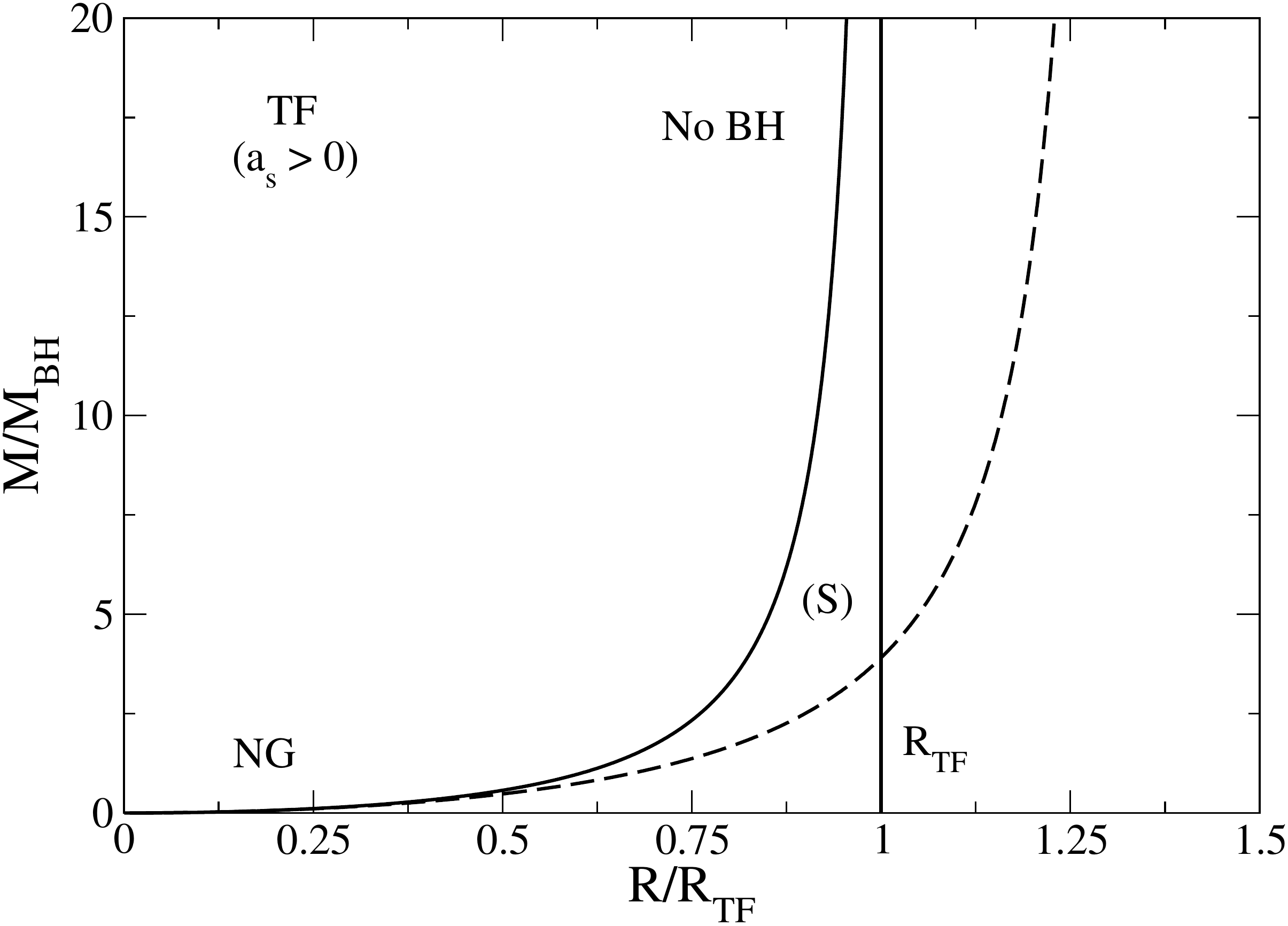}} 
\caption{Mass-radius relation of self-gravitating BECs with a repulsive
self-interaction ($a_s>0$) in the TF
limit ($\hbar=0$) in the presence of a central black hole. We have
normalized the radius by $R_{\rm TF}$  given by Eq. (\ref{intro2}) and
the mass by $M_{\rm BH}$. The solid line is
the exact mass-radius relation from
Eq. (\ref{etf15}) and the dashed line is the approximate mass-radius relation
(\ref{tf1}) obtained from the Gaussian ansatz (in that case,
$R$ on the figure represents the radius $R_{99}$ containing $99\%$ of the
mass).}
\label{mrTFexact}
\end{figure}

In the TF approximation ($\hbar=0$), the differential equation determining the
density
profile of the BEC in the presence of a central BH is (see Sec. \ref{sec_eq})
\begin{eqnarray}
\label{etf1}
-\frac{4\pi a_s\hbar^2}{m^3}\Delta\rho=4\pi G\rho+4\pi GM_{\rm BH}\delta({\bf
r}).
\end{eqnarray}
For $r\neq 0$, it reduces to
\begin{eqnarray}
\label{etf2}
\Delta\rho+\frac{Gm^3}{a_s\hbar^2}\rho=0.
\end{eqnarray}
The general solution of this equation is
\begin{eqnarray}
\label{etf3}
\rho=A\frac{\sin(kr)}{r}+B\frac{\cos(kr)}{r},
\end{eqnarray}
where we have defined
\begin{eqnarray}
\label{etf4}
k=\left (\frac{Gm^3}{a_s\hbar^2}\right )^{1/2}.
\end{eqnarray}
Integrating Eq. (\ref{etf1}) over a sphere of radius $r$, using
the Gauss-Ostrogradsky theorem  to
convert a volume integral into a surface integral, and letting
$r\rightarrow 0$, we get
\begin{eqnarray}
\label{etf5}
-\frac{4\pi a_s\hbar^2}{m^3}\oint_{S_r}\nabla\rho\cdot d{\bf S}=4\pi GM_{\rm
BH},
\end{eqnarray}
implying
\begin{eqnarray}
\label{etf6}
\frac{d\rho}{dr}\sim -\frac{Gm^3M_{\rm BH}}{4\pi a_s\hbar^2 r^2}\qquad
(r\rightarrow 0).
\end{eqnarray}
Therefore, when $r\rightarrow 0$, the density behaves
as\footnote{We can also
obtain this result by taking the limit $r\rightarrow 0$ in Eq.
(\ref{al2}), using  $\Phi\rightarrow 0$ when $r\rightarrow 0$.}
\begin{eqnarray}
\label{etf7}
\rho\sim \frac{Gm^3M_{\rm BH}}{4\pi a_s\hbar^2 r}.
\end{eqnarray}
This diverging behavior determines the constant $B$ in Eq. (\ref{etf3}). We get 
\begin{eqnarray}
\label{etf9}
B=\frac{Gm^3 M_{\rm BH}}{4\pi a_s\hbar^2}=\frac{k^2 M_{\rm BH}}{4\pi}.
\end{eqnarray}
On the other hand, if we call $R$ the value of the radial distance at which the
density vanishes, we find that the   constant $A$ in Eq. (\ref{etf3}) is given
by 
\begin{eqnarray}
\label{etf10}
A=-\frac{B}{\tan(kR)}.
\end{eqnarray}
As a result, the density profile can be written as
\begin{eqnarray}
\label{etf11}
\rho=\frac{k^2M_{\rm BH}}{4\pi r}\left\lbrack \cos(kr)
-\frac{\sin(kr)}{\tan(kR)}\right\rbrack.
\end{eqnarray}
It is plotted in Fig. \ref{rhoTF}. The total mass is given by
\begin{eqnarray}
\label{etf12}
\frac{M}{M_{\rm BH}}=\int_0^{kR} \left\lbrack \cos(x)
-\frac{\sin(x)}{\tan(kR)}\right\rbrack x\, dx.
\end{eqnarray}
Using the identities
\begin{eqnarray}
\label{etf13}
\int_0^{kR}\sin (x)x\, dx=\sin(kR)-kR \cos(kR), 
\end{eqnarray}
and
\begin{eqnarray}
\label{etf14}
\int_0^{kR}\cos (x)x\,
dx=\cos(kR)+kR \sin(kR) -1,
\end{eqnarray}
we obtain the exact mass-radius relation
\begin{eqnarray}
\label{etf15}
\frac{M}{M_{\rm BH}}=\frac{kR}{\sin(kR)}-1.
\end{eqnarray}
It can be compared to the mass-radius relation (\ref{tf1}) obtained from the
Gaussian
ansatz (see Fig. \ref{mrTFexact}). We note that $\rho$ is very different from a
Gaussian in the present
case.

\begin{figure}[h]
\scalebox{0.33}{\includegraphics{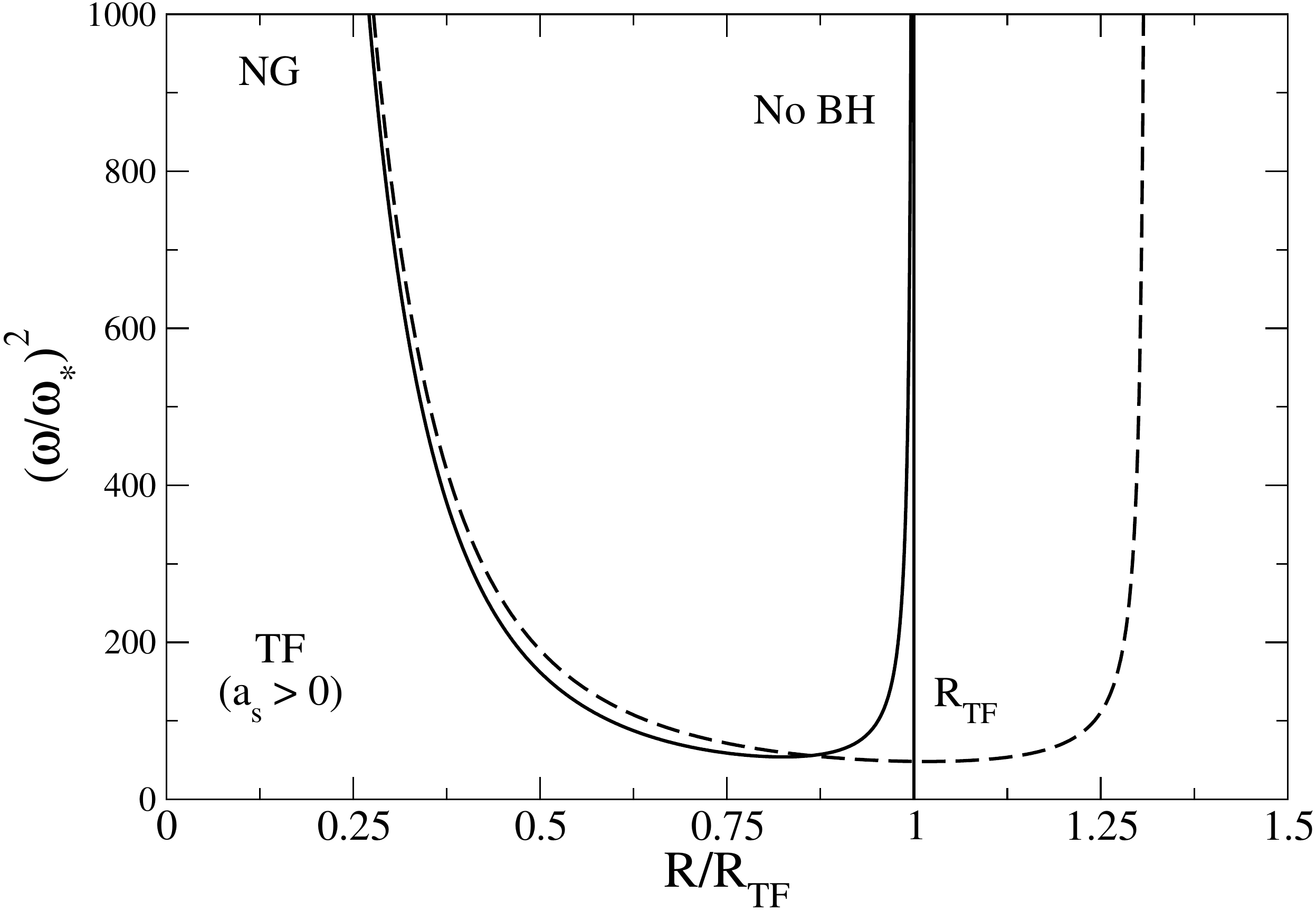}} 
\caption{Square pulsation of self-gravitating BECs with a repulsive
self-interaction ($a_s>0$) in the TF
limit ($\hbar=0$) in the presence of a central black hole as a function of the
radius. We have
normalized the radius by $R_{\rm TF}$  given by Eq. (\ref{intro2}) and
the pulsation by $\omega_*=(GM_{\rm BH}/R_{\rm TF}^3)^{1/2}$.
The solid line is
the exact relation from
Eq. (\ref{al8}) and the dashed line is the approximate relation
(\ref{tf1}) and (\ref{tf2}) obtained from the Gaussian ansatz (in that case,
$R$ on the figure represents the radius $R_{99}$ containing $99\%$ of the mass).
There is a minimum pulsation $\omega_{\rm min}=7.35\, \omega_*$
(the Gaussian ansatz gives $\omega_{\rm min}=6.95\, \omega_*$).}
\label{omega2TF}
\end{figure}

In the TF limit, the exact
equilibrium relations from Sec. \ref{sec_eer} reduce to
\begin{eqnarray}
\label{al1}
E_{\rm tot}=W+W_{\rm BH}+U,\qquad NE=2W+2U+W_{\rm BH},\qquad 3U+W+W_{\rm BH}=0.
\end{eqnarray}
We can determines the eigenenergy $E$ by applying the
relation [see Eq. (\ref{tigp4bb}) with
$Q=0$]
\begin{equation}
\label{al2}
\Phi+\frac{4\pi a_s\hbar^2}{m^3}\rho-\frac{GM_{\rm BH}}{r}=\frac{E}{m}
\end{equation}
at $r=R$, giving
\begin{equation}
\label{al3}
NE=-\frac{GM(M_{\rm BH}+M)}{R}.
\end{equation}
On the other hand, a direct calculation using Eq. (\ref{etf11}) gives
\begin{equation}
\label{al4}
W_{\rm BH}=-GM_{\rm BH}^2k\frac{1-\cos(kR)}{\sin(kR)},\quad
U=-\frac{1}{4}GkM_{\rm BH}^2\left\lbrack
\frac{1}{\tan(kR)}-\frac{kR}{\sin^2(kR)}\right\rbrack,\quad I=\frac{M_{\rm
BH}}{k^2}\left\lbrack
6+\frac{kR(k^2R^2-6)}{\sin(kR)}\right\rbrack.
\end{equation}
Finally, according to the results of Appendix \ref{sec_theorem}, we have
\begin{equation}
\label{al5}
W=-\frac{1}{4}\left (\frac{3GM^2}{R}+\frac{3GM_{\rm BH}M}{R}+W_{\rm BH}\right ).
\end{equation}
We can check that the relations of Eq. (\ref{al1}) are satisfied. On the other
hand, Eq. (\ref{a30}) reduces to 
\begin{equation}
\label{al6}
\omega^2=\frac{12U+2W+2W_{\rm BH}}{I}.
\end{equation}
Using the virial theorem from Eq. (\ref{al1}), we obtain
\begin{equation}
\label{al7}
\omega^2=\frac{6U}{I}=-\frac{2(W+W_{\rm BH})}{I}.
\end{equation}
Finally, using Eq. (\ref{al4}), we get
\begin{equation}
\label{al8}
\omega^2=-\frac{3}{2}GM_{\rm BH}k^3\frac{\sin(kR)\cos(kR)-kR}{\sin
(kR)\left\lbrack 6\sin(kR)+kR(k^2R^2-6)\right\rbrack}.
\end{equation}
The pulsation is plotted in Fig. \ref{omega2TF} as a function of the BEC radius.
It presents a minimum value $\omega_{\rm min}$.

For $M\rightarrow 0$ and $R\rightarrow 0$, we recover the nongravitational
limit of Appendix \ref{sec_ex}.
For $M\rightarrow +\infty$ and $R\rightarrow R_{\rm TF}$  we recover the no BH
limit of Appendix
\ref{sec_tfnobh} with the additional
relations
\begin{eqnarray}
\frac{M}{M_{\rm BH}}\sim \frac{1}{1-R/R_{\rm TF}},\qquad \left
(\frac{\omega}{\omega_*}\right )^2\sim
\frac{3\pi^2}{2(\pi^2-6)}\frac{1}{1-R/R_{\rm TF}}.
\end{eqnarray}
When $M_{BH}=0$, we recover the results of Appendix 
\ref{sec_tfnobh}.

\section{Gravitational TF model with a central BH}
\label{sec_gft}

In this Appendix, we generalize the results of Appendix \ref{sec_etf} and
consider a self-gravitating polytropic sphere of index
$\gamma>1$ surrounding a central BH. As in Appendix \ref{sec_etf}, we make the
TF approximation which amounts to neglecting the quantum potential.

\subsection{Exact results}
\label{sec_xr}

For a polytropic equation of state $P=K\rho^{\gamma}$ with $\gamma=1+1/n$, Eq.
(\ref{eq4}) takes the form
\begin{equation}
\label{xr1}
-K(n+1)\Delta\rho^{1/n}+\frac{\hbar^2}{2m^2}\Delta
\left (\frac{\Delta\sqrt{\rho}}{\sqrt{\rho}}\right )=4\pi G\rho+4\pi GM_{\rm
BH}\delta({\bf
r}).
\end{equation}
In the TF approximation  ($\hbar=0$), it reduces to
\begin{equation}
\label{xr2}
-K(n+1)\Delta\rho^{1/n}=4\pi G\rho+4\pi GM_{\rm
BH}\delta({\bf
r}).
\end{equation}
For $r\neq 0$, we get
\begin{equation}
\label{xr3}
-K(n+1)\Delta\rho^{1/n}=4\pi G\rho.
\end{equation}
As in the usual theory of self-gravitating polytropic spheres
\cite{chandrabook}, we introduce the variables ($\xi,\theta$) from the
relations
\begin{equation}
\label{xr4}
\rho=\rho_0\theta^n\qquad {\rm and}\qquad r=\left\lbrack
\frac{K(n+1)\rho_0^{1/n-1}}{4\pi
G}\right\rbrack^{1/2}\xi\equiv r_0\xi,
\end{equation}
but we stress that $\rho_0$ is {\it not} the central density (which is infinite
in the presence of a central point source). With these variables, Eq.
(\ref{xr3})
reduces to the Lane-Emden equation
\begin{equation}
\label{xr6}
\frac{1}{\xi^2}\frac{d}{d\xi}\left (\xi^2\frac{d\theta}{d\xi}\right )=-\theta^n.
\end{equation}
Integrating Eq. (\ref{xr2}) over a sphere of radius $r$, using
the Gauss-Ostrogradsky theorem to
convert a volume integral into a surface integral, and letting
$r\rightarrow 0$, we get 
\begin{eqnarray}
\label{xr7}
-K(n+1)\oint_{S_r}\nabla(\rho^{1/n})\cdot d{\bf S}=4\pi GM_{\rm
BH},
\end{eqnarray}
implying
\begin{eqnarray}
\label{xr8}
\frac{d\rho^{1/n}}{dr}\sim -\frac{GM_{\rm BH}}{K(n+1) r^2}.
\end{eqnarray}
Therefore, when $r\rightarrow 0$, the density behaves
as\footnote{We can also obtain this result from Eq. (\ref{tigp4bb}) which, in
the
TF
approximation,  reduces to
\begin{eqnarray}
\label{xr10}
(n+1)K\rho^{1/n}+\Phi-\frac{GM_{\rm BH}}{r}=\frac{E}{m}.
\end{eqnarray}
Taking the limit $r\rightarrow 0$, and using $\Phi\rightarrow 0$, we recover
Eq. (\ref{xr9}).} 
\begin{eqnarray}
\label{xr9}
\rho^{1/n}\sim \frac{GM_{\rm BH}}{K(n+1) r}.
\end{eqnarray}
It is convenient to
introduce the variable 
\begin{eqnarray}
\label{xr11}
u=\theta\xi.
\end{eqnarray}
In that case, the Lane-Emden equation (\ref{xr6}) is transformed into
\begin{equation}
\label{xr12}
\frac{d^2 u}{d\xi^2}=-\frac{u^n}{\xi^{n-1}}.
\end{equation}
On the other hand, the asymptotic behavior of the density close to the origin
[see Eq. (\ref{xr9})] leads to the boundary condition
\begin{equation}
\label{xr13}
u(0)=\frac{GM_{\rm BH}}{K(n+1)\rho_0^{1/n}r_0}=\frac{M_{\rm
BH}}{4\pi \rho_0 r_0^3},
\end{equation}
where we have introduced the radius $r_0$ defined by Eq. (\ref{xr4}).
We now choose the reference density  $\rho_0$ such that
\begin{equation}
\label{xr14}
u(0)=1. 
\end{equation}
This implies
\begin{equation}
\label{xr15}
\frac{M_{\rm
BH}}{4\pi \rho_0 r_0^3}=1,
\end{equation}
leading to
\begin{equation}
\label{xr16}
r_0=\left\lbrack \frac{K(n+1)}{4\pi G}\right\rbrack^{n/(3-n)}\left
(\frac{M_{\rm BH}}{4\pi}\right )^{(1-n)/(3-n)}
\end{equation}
and
\begin{equation}
\label{xr17}
\rho_0=\left\lbrack \frac{4\pi
G}{K(n+1)}\right\rbrack^{3n/(3-n)}\left
(\frac{M_{\rm BH}}{4\pi}\right )^{2n/(3-n)}.
\end{equation}
Using the Lane-Emden equation (\ref{xr6}) and the variable $u$ defined by Eq.
(\ref{xr11}), we find that the total mass of the
configuration
is given by\footnote{The integral converges for $r\rightarrow 0$
provided that $n<3$. We will make this assumption in the following.}
\begin{equation}
\label{xr18}
M=\int_0^R\rho 4\pi r^2\, dr=-4\pi\rho_0r_0^3\left\lbrack
\xi^2\frac{d\theta}{d\xi}\right\rbrack_0^{\xi_1}=-4\pi\rho_0r_0^3\left\lbrack
\xi u'-u\right\rbrack_0^{\xi_1}=-4\pi\rho_0r_0^3(\xi_1u'_1+1),
\end{equation}
where $\xi_1$ is the normalized distance at which the density vanishes
($\theta=u=0$) and $u'_1=u'(\xi_1)$.
Using (\ref{xr15}), we obtain
\begin{equation}
\label{xr20}
\frac{M}{M_{\rm BH}}=-\xi_1 u'_1-1.
\end{equation}
On the other hand, the (physical) radius of the configuration
is given by
\begin{equation}
\label{xr21}
\frac{R}{r_0}=\xi_1.
\end{equation}
The previous equations allow us to obtain the density profile and the
mass-radius relation for various polytropic index $n$.
To that purpose, one
has to solve the differential equation (\ref{xr12}) with the boundary condition
$u(0)=1$ and $u'(0)=a$ (for a given value of $a$) up to the normalized distance
$\xi_1$ at which $u$
vanishes. The density profile is then determined by Eqs.
(\ref{xr4}) and (\ref{xr11}) while the mass and the radius of the configuation
are given by Eqs. (\ref{xr20}) and (\ref{xr21}). By varying $a$ we can obtain
the complete mass-radius relation. In general, the
differential equation (\ref{xr12}) must be solved numerically except for the
particular index
$n=1$ (see Appendix 
\ref{sec_etf}). Application
of these results for different values of $n$ will be given in a forthcoming
paper \cite{forthcoming}. In the following section, we present approximate
analytical results obtained from the Gaussian ansatz.

{\it Remark:} In the electrostatic case, the previous equations with
$n=3/2$ correspond to
the TF theory of atoms in which a central charge $+Q$ is surrounded by a cloud
of opposite charges $-Ne$ in Coulombian interaction. In this analogy, the
central charge is the equivalent of the BH and the charged cloud is the
equivalent of the gravitational halo. The crucial difference\footnote{Another
important difference is
that, in the case of atoms, the number of charges $N$ is small making the mean
field approximation in general inaccurate. By contrast, for astrophysical
systems where
$N\gg 1$, the mean field approximation is excellent.}  is that the
charges $-e$ are mutually repulsive (the atom being stabilized by the attraction
of
the central charge $+Q$) while the gravitational particles are 
mutually attractive (the BH having the tendency to reinforce their attraction
and
possibly destabilize the system). This analogy
will be further developed in  a
forthcoming paper \cite{forthcoming}.

\subsection{Gaussian ansatz}
\label{sec_gatf}

Using a Gaussian ansatz, the mass-radius relation corresponding to Eq.
(\ref{xr1}) is [see Eq. (\ref{a13b})]:
\begin{eqnarray}
\label{gatf1}
-2\sigma\frac{\hbar^2M}{m^2R^3}+\nu\frac{GM^2}{R^2}-3\zeta\frac{
KM^\gamma}{R^{3\gamma-2}}+\lambda\frac{GM_{\rm BH}M}{R^2}=0.
\end{eqnarray}
In the TF approximation, it reduces to
\begin{eqnarray}
\label{gatf2}
R^{3\gamma-4}=\frac{3\zeta KM^{\gamma-1}}{\nu GM+\lambda GM_{\rm BH}}.
\end{eqnarray}
As an illustration, let us apply these results to  a system of nonrelativistic
self-gravitating fermions at $T=0$ surrounding a central object (black
hole). This could represent a model of fermionic dark matter halos.

\begin{figure}[h]
\scalebox{0.33}{\includegraphics{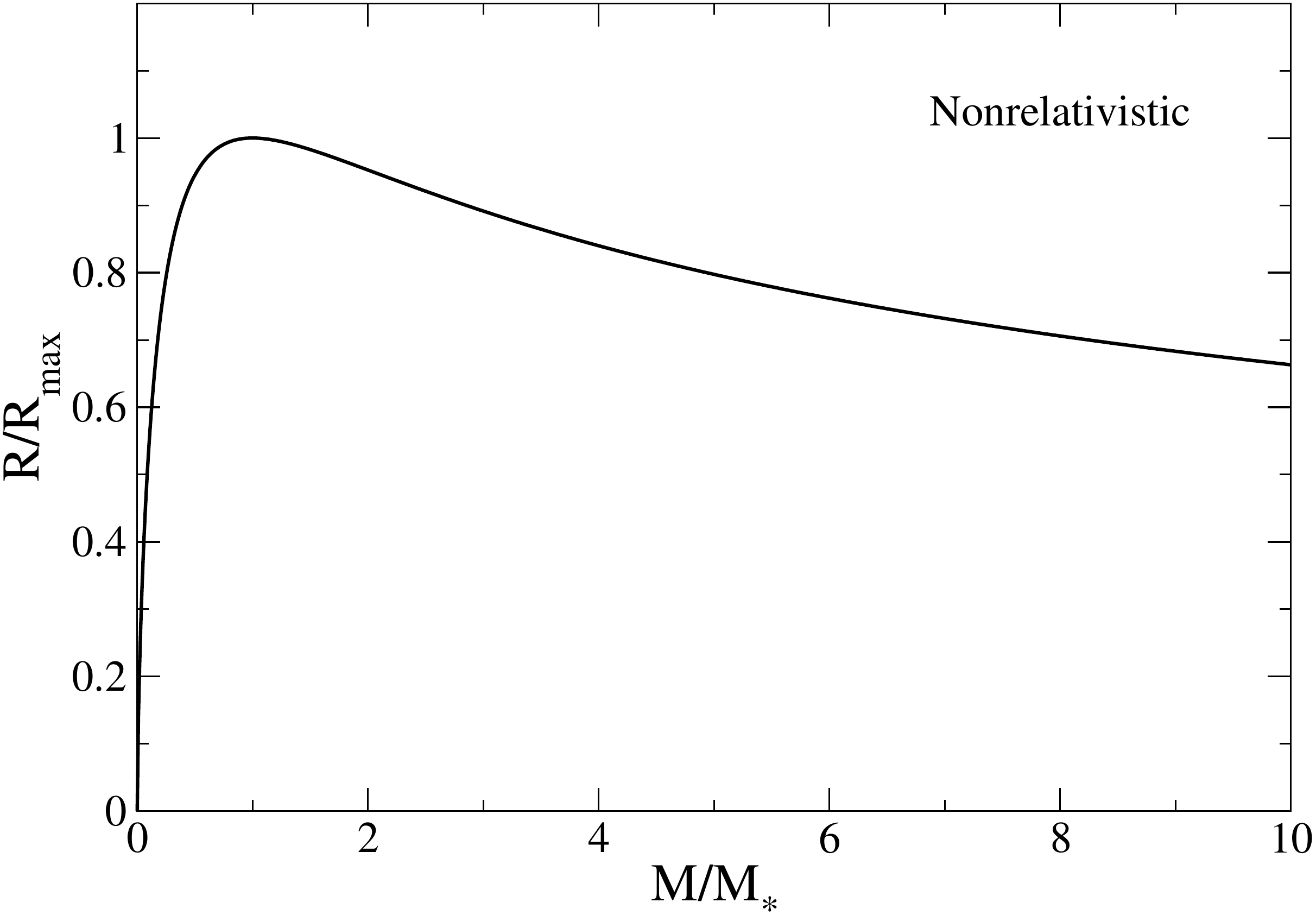}} 
\caption{Mass-radius relation (within the Gaussian ansatz) of  nonrelativistic
self-gravitating fermions at $T=0$ surrounding a central black hole.}
\label{nonrelat}
\end{figure}

We first consider nonrelativistic fermions at $T=0$ that are
described by an equation of state of the form \cite{chandrabook}:
\begin{eqnarray}
\label{gatf3}
P=\frac{1}{20}\left (\frac{3}{\pi}\right )^{2/3}\frac{h^2}{m^{8/3}}\rho^{5/3}.
\end{eqnarray}
This corresponds to a polytrope of index $\gamma=5/3$ (i.e. $n=3/2$).
In that case, the  mass-radius relation (\ref{gatf2}) becomes
\begin{eqnarray}
\label{gtf4}
R=\frac{3\zeta}{20}\left (\frac{3}{\pi}\right
)^{2/3}\frac{h^2}{Gm^{8/3}}\frac{M^{2/3}}{\nu M+\lambda M_{\rm BH}}.
\end{eqnarray}
When $M_{\rm BH}=0$, we recover the
standard mass-radius relation of nonrelativistic fermion stars (within the
Gaussian ansatz approximation):
\begin{eqnarray}
R=\frac{3\zeta}{20\nu}\left (\frac{3}{\pi}\right
)^{2/3}\frac{h^2}{Gm^{8/3}M^{1/3}}.
\end{eqnarray}
The prefactor is $0.0539$ (the exact prefactor is $0.114$ \cite{chandrabook} and
we recall that $R_{99}=2.38167 R$ for the Gaussian profile). This relation is
monotonic, the radius increasing as the mass decreases. When $M_{\rm BH}\neq 0$,
we find the existence
of a maximum
radius 
\begin{eqnarray}
\label{gatf5}
R_{\rm max}=\frac{\zeta}{20\lambda^{1/3}}\left (\frac{6}{\nu\pi}\right
)^{2/3}\frac{h^2}{Gm^{8/3}M_{\rm BH}^{1/3}}
\end{eqnarray}
corresponding to a mass
\begin{eqnarray}
\label{gatf6}
M_*=\frac{2\lambda}{\nu}M_{\rm BH}.
\end{eqnarray}
The prefactors are $0.0202$ and $5.66$ respectively. The mass-radius
relation (\ref{gtf4}) can be rewritten as
\begin{eqnarray}
\label{gatf7}
\frac{R}{R_{\rm max}}=\frac{3\left (\frac{M}{M_*}\right )^{2/3}}{2\left
(\frac{M}{M_*}\right )+1}.
\end{eqnarray}
It is plotted in Fig. \ref{nonrelat}.

\begin{figure}[h]
\scalebox{0.33}{\includegraphics{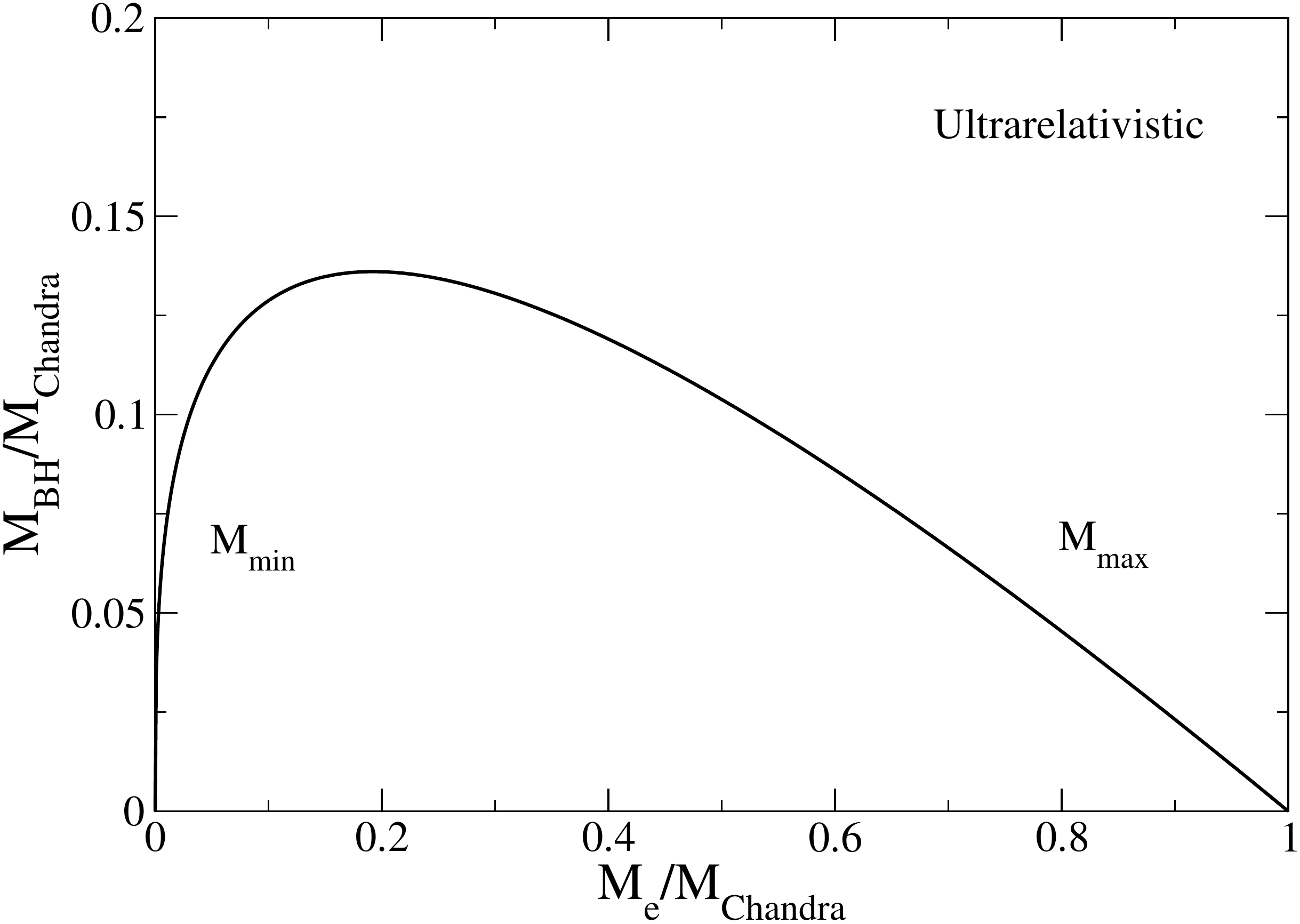}} 
\caption{Maximum and minimum masses (obtained in the ultrarelativistic
limit within the Gaussian ansatz) of self-gravitating fermions at $T=0$
surrounding a central black hole.}
\label{ultrarelat}
\end{figure}

Let us now consider ultra-relativistic fermions at $T=0$ that are
described by an equation of state of the form \cite{chandrabook}:
\begin{eqnarray}
\label{gatf8}
P=\frac{1}{8}\left (\frac{3}{\pi}\right )^{1/3}\frac{hc}{m^{4/3}}\rho^{4/3}.
\end{eqnarray}
This corresponds to a polytrope of index $\gamma=4/3$ (i.e. $n=3$).
In that case, the  mass-radius relation (\ref{gatf2}) becomes
\begin{eqnarray}
\label{gatf9}
\nu GM_e+\lambda GM_{\rm BH}= \frac{3\zeta}{8}\left (\frac{3}{\pi}\right
)^{1/3}\frac{hc}{m^{4/3}}M_e^{1/3}
\end{eqnarray}
The mass is independent of the radius. This equation actually determines a
maximum mass and a minimum mass as a function of the black hole mass. When
$M_{\rm BH}=0$, we recover the Chandrasekhar maximum mass (within the
Gaussian ansatz approximation):
\begin{eqnarray}
\label{gatf10}
M_{\rm Chandra}=\left (\frac{3\zeta}{8\nu}\right )^{3/2}\left
(\frac{3}{\pi}\right )^{1/2}\left
(\frac{hc}{G}\right )^{3/2}\frac{1}{m^2}.
\end{eqnarray}
The prefactor is $0.197$ which coincides (to the order of accuracy that we
consider) with the exact prefactor \cite{chandrabook}. We can then rewrite Eq.
(\ref{gatf9}) as
\begin{eqnarray}
\label{gatf11}
\frac{M_{\rm BH}}{M_{\rm Chandra}}=\frac{\nu}{\lambda}\left
\lbrack \left (\frac{M_e}{M_{\rm Chandra}}\right )^{1/3}-\frac{M_e}{M_{\rm
Chandra}}\right\rbrack.
\end{eqnarray}
The relation between the extremal masses $M_e$ and the BH mass $M_{\rm BH}$ is
plotted in Fig. \ref{ultrarelat}. The maximum mass and the minimum mass become
equal when
\begin{eqnarray}
\label{gatf13}
\frac{M_{\rm BH}}{M_{\rm
Chandra}}=\frac{\nu}{\lambda}\,\frac{2}{3\sqrt{3}}=0.136.
\end{eqnarray}
It that case, they have the value
\begin{eqnarray}
\label{gatf12}
\frac{M_{e}^*}{M_{\rm Chandra}}=\frac{1}{3\sqrt{3}}=0.192.
\end{eqnarray}
There is no possible equilibrium, for any mass $M$, when $M_{\rm
BH}>(2\nu/3\sqrt{3}\lambda)M_{\rm Chandra}=0.136\, M_{\rm Chandra}$.
A more complete
discussion of these results will be given elsewhere \cite{forthcoming}.


\begin{thebibliography}{99}



\bibitem{baldeschi}{\small M.R. Baldeschi, G.B. Gelmini, R. Ruffini, Phys. Lett.
B {\bf  122}, 221 (1983)}
\bibitem{khlopov}{\small M.Yu. Khlopov, B.A. Malomed, Ya.B. Zeldovich, Mon. Not.
R. astr. Soc. {\bf  215}, 575 (1985) }
\bibitem{membrado}{\small M. Membrado, A.F. Pacheco, J. Sanudo, Phys. Rev. A
{\bf  39}, 4207 (1989)}
\bibitem{sin}{\small S.J. Sin, Phys. Rev. D {\bf  50}, 3650 (1994)}
\bibitem{jisin}{\small S.U. Ji, S.J. Sin, Phys. Rev. D {\bf  50}, 3655 (1994)}
\bibitem{leekoh}{\small J.W. Lee, I. Koh, Phys. Rev. D {\bf  53}, 2236 (1996)}
\bibitem{schunckpreprint}{\small F.E. Schunck, [astro-ph/9802258]}
\bibitem{matosguzman}{\small T. Matos,
F.S. Guzm\'an, F. Astron. Nachr. {\bf 320}, 97 (1999)}
\bibitem{sahni}{\small V. Sahni, L. Wang
Phys. Rev. D {\bf 62}, 103517 (2000)}
\bibitem{guzmanmatos}{\small F.S. Guzm\'an,
T. Matos, Class. Quantum Grav.  {\bf 17}, L9 (2000)}
\bibitem{hu}{\small W. Hu, R. Barkana, A. Gruzinov, Phys. Rev. Lett. {\bf  85},
1158 (2000)}
\bibitem{peebles}{\small P.J.E. Peebles, Astrophys. J. {\bf 534}, L127 (2000)}
\bibitem{goodman}{\small J. Goodman, New Astronomy {\bf 5}, 103 (2000)}
\bibitem{mu}{\small T. Matos, L.A. Ure\~na-L\'opez,
Phys. Rev. D {\bf 63}, 063506 (2001)}
\bibitem{arbey1}{\small A. Arbey, J. Lesgourgues, P. Salati, Phys. Rev. D {\bf
64}, 123528 (2001)}
\bibitem{silverman1}{\small M.P. Silverman, R.L. Mallett, Class. Quantum Grav.
{\bf  18}, L103 (2001)}
\bibitem{matosall}{\small M. Alcubierre, F.S. Guzm\'an, T. Matos, D. N\'u\~nez,
L.A. Ure\~na-L\'opez, P.  Wiederhold, Class. Quantum. Grav. {\bf 19}, 5017
(2002)}
\bibitem{silverman}{\small M.P. Silverman, R.L. Mallett, Gen. Rel. Grav. {\bf
34}, 633 (2002)}
\bibitem{lesgourgues}{\small J. Lesgourgues,  A. Arbey, P. Salati, New Astron.
Rev. {\bf 46}, 791 (2002)}
\bibitem{arbey}{\small A. Arbey, J. Lesgourgues, P. Salati, Phys. Rev. D {\bf
68}, 023511 (2003)}
\bibitem{fm1}{\small T. Fukuyama, M. Morikawa, Prog. Theor. Phys.
{\bf 115}, 1047 (2006)}
\bibitem{bohmer}{\small C.G. B\"ohmer, T. Harko, J. Cosmol. Astropart. Phys.
{\bf 06}, 025 (2007)}
\bibitem{fm2}{\small T. Fukuyama, M. Morikawa, T. Tatekawa, J. Cosmol.
Astropart. Phys. {\bf 06}, 033 (2008)}
\bibitem{bmn}{\small A. Bernal, T. Matos, D. N\'u\~nez, Rev. Mex. Astron.
Astrofis.  {\bf 44}, 149 (2008)}
\bibitem{fm3}{\small T. Fukuyama, M. Morikawa, Phys. Rev. D {\bf 80}, 063520
(2009)}
\bibitem{sikivie}{\small P. Sikivie, Q. Yang, Phys. Rev. Lett. {\bf  103},
111301 (2009)}
\bibitem{mvm}{\small T. Matos, A. V\'azquez-Gonz\'alez,
J. Maga\~na, Mon. Not. R. Astron. Soc. {\bf 393}, 1359 (2009)}
\bibitem{lee09}{\small J.W. Lee, Phys. Lett. B {\bf 681}, 118 (2009)}
\bibitem{ch1}{\small T.P. Woo, T. Chiueh, Astrophys. J. {\bf 697}, 850 (2009)}
\bibitem{lee}{\small J.W. Lee, S. Lim, J. Cosmol. Astropart. Phys.  {\bf 01},
007 (2010)}
\bibitem{prd1}{\small P.H. Chavanis, Phys. Rev. D {\bf 84}, 043531 (2011)}
\bibitem{prd2}{\small P.H. Chavanis, L. Delfini, Phys. Rev. D {\bf 84}, 043532
(2011)}
\bibitem{prd3}{\small P.H. Chavanis, Phys. Rev. D {\bf 84}, 063518 (2011)}
\bibitem{briscese}{\small F. Briscese, Phys. Lett. B
{\bf 696}, 315 (2011)}
\bibitem{harkocosmo}{\small T. Harko, Mon. Not. R. Astron. Soc. {\bf 413}, 3095
(2011)}
\bibitem{harko}{\small T. Harko, J. Cosmol. Astropart. Phys. {\bf 05}, 022
(2011)}
\bibitem{abrilMNRAS}{\small A. Su\'arez, T. Matos, Mon. Not. R. Astron. Soc.
{\bf 416}, 87 (2011)}
\bibitem{aacosmo}{\small P.H. Chavanis, Astron. Astrophys. {\bf 537}, A127
(2012)}
\bibitem{velten}{\small H. Velten, E. Wamba, Phys. Lett. B {\bf 709}, 1
(2012)}
\bibitem{pires}{\small M.O.C. Pires, J.C.C. de Souza, J. Cosmol. Astropart.
Phys. {\bf 11} (2012) 024}
\bibitem{park}{\small C.-G. Park, J.-C. Hwang, H. Noh, Phys. Rev. D {\bf 86},
083535 (2012)}
\bibitem{rmbec}{\small V.H. Robles, T. Matos, Monthly Not. Roy. Astron. Soc.
{\bf 422},
282 (2012)}
\bibitem{rindler}{\small T. Rindler-Daller, P. R. Shapiro, Monthly Not. Roy.
Astron.
Soc. {\bf 422}, 135 (2012)}
\bibitem{chavharko}{\small P.H. Chavanis, T. Harko, Phys. Rev. D {\bf 86},
064011 (2012)}
\bibitem{lora}{\small V. Lora, J. Maga\~na, A. Bernal, F.J. S\'anchez-Salcedo,
E.K.
Grebel, J. Cosmol. Astropart. Phys.  {\bf  02}, 011 (2012)}
\bibitem{abrilJCAP}{\small J. Maga\~na, T. Matos, A. Su\'arez,
F. J. S\'anchez-Salcedo, JCAP {\bf 10}, 003 (2012)}
\bibitem{mhh}{\small G. Manfredi, P.A. Hervieux, F. Haas, Class. Quantum Grav.
{\bf 30}, 075006 (2013)}
\bibitem{lensing}{\small A.X. Gonz\'alez-Morales, A. Diez-Tejedor, L.A.
Ure\~na-L\'opez, O. Valenzuela, Phys. Rev. D {\bf 87}, 021301(R) (2013)}
\bibitem{glgr1}{\small F.S. Guzm\'an, F.D. Lora-Clavijo, J.J.
Gonz\'alez-Avil\'es, F.J. Rivera-Paleo, J. Cosmol. Astropart. Phys. {\bf 09}
(2013) 034}
\bibitem{ch2}{\small H.Y. Schive, T. Chiueh, T. Broadhurst, Nature Physics {\bf
10}, 496 (2014)}
\bibitem{ch3}{\small H.Y. Schive {\it et al.}, Phys. Rev. Lett. {\bf 113},
261302 (2014)}
\bibitem{shapiro}{\small B. Li, T. Rindler-Daller, P.R. Shapiro, Phys. Rev. D
{\bf 89}, 083536 (2014)}
\bibitem{bettoni}{\small D. Bettoni, M. Colombo, S. Liberati, JCAP
{\bf 02}, 004 (2014)}
\bibitem{lora2}{\small V. Lora, J. Maga\~na, JCAP
{\bf 09}, 011 (2014)}
\bibitem{mlbec}{\small P.H. Chavanis,  Eur. Phys. J. Plus {\bf 130}, 181 (2015)}
\bibitem{madarassy}{\small E.J.M. Madarassy, V.T. Toth,  Phys. Rev. D {\bf 91},
044041 (2015)}
\bibitem{abrilph}{\small A. Su\'arez, P.H. Chavanis,  Phys. Rev. D {\bf 92},
023510 (2015)}
\bibitem{playa}{\small A. Su\'arez, P.H. Chavanis, J. Phys.: Conf. Series {\bf
654}, 012088 (2015)}
\bibitem{stiff}{\small P.H. Chavanis,  Phys. Rev. D {\bf 92},
103004 (2015)}
\bibitem{guth}{\small A.H. Guth, M.P. Hertzberg, C. Prescod-Weinstein,  Phys.
Rev. D {\bf 92},
103513 (2015)}
\bibitem{souza}{\small J.C.C. de Souza, M. Ujevic, Gen. Relat. Grav. {\bf 47},
100 (2015)}
\bibitem{freitas}{\small R.C. de Freitas, H. Velten, Eur. Phys. J. C {\bf 75},
597 (2015)}
\bibitem{alexandre}{\small J. Alexandre,  Phys. Rev. D {\bf 92},
123524 (2015)}
\bibitem{schroven}{\small K. Schroven, M. List, C. L\"ammerzahl,  Phys. Rev. D
{\bf 92}, 124008 (2015)}
\bibitem{pop}{\small D. Marsh, A.R. Pop, Monthly Not. Roy. Astron. {\bf 451},
2479 (2015)}
\bibitem{eby}{\small J. Eby, C. Kouvaris, N.G. Nielsen, L.C.R. Wijewardhana, 
JHEP {\bf 02}, 028 (2016)}
\bibitem{cembranos}{\small J.A.R. Cembranos, A.L. Maroto, S.J. N\'u\~nez
Jare\~no, JHEP {\bf 03}, 013 (2016)}
\bibitem{braaten}{\small E. Braaten, A. Mohapatra, H. Zhang, Phy. Rev. Lett.
{\bf 117}, 121801 (2016)}
\bibitem{davidson}{\small S. Davidson, T. Schwetz, Phys. Rev. D {\bf 93},
123509 (2016)}
\bibitem{schwabe}{\small B. Schwabe, J. Niemeyer, J. Engels, Phys. Rev. D {\bf
94}, 043513 (2016)}
\bibitem{fan}{\small J. Fan, Phys. Dark Univ. {\bf 14}, 84 (2016)}
\bibitem{calabrese}{\small E. Calabrese, D.N. Spergel, Monthly Not. Roy. Astron.
Soc. {\bf 460}, 4397 (2016)}
\bibitem{marsh}{\small D. Marsh, Phys. Rep. {\bf 643}, 1 (2016)}
\bibitem{bectcoll}{\small P.H. Chavanis,  Phys. Rev. D {\bf 94},
083007 (2016)}
\bibitem{cotner}{\small E. Cotner, Phys. Rev. D {\bf 94}, 063503 (2016)}
\bibitem{chavmatos}{\small P.H. Chavanis, T. Matos, Eur. Phys. J. Plus {\bf
132}, 30 (2017)}
\bibitem{helfer}{\small T. Helfer {\it et al.}, JCAP {\bf 03}, 055 (2017)}
\bibitem{hui}{\small L. Hui, J. Ostriker, S. Tremaine, E. Witten, Phys. Rev. D
{\bf 95}, 043541 (2017)}
\bibitem{tkachevprl}{\small D.G. Levkov, A.G.  Panin, I.I.
Tkachev, Phys. Rev. Lett. {\bf 118}, 011301 (2017)}
\bibitem{abrilphas}{\small A. Su\'arez, P.H. Chavanis,  Phys. Rev. D {\bf 95},
063515 (2017)}
\bibitem{shapironew}{\small B. Li, T. Rindler-Daller, P.R. Shapiro, Phys. Rev.
D {\bf 96}, 063505 (2017)}
\bibitem{ggp}{\small P.H. Chavanis, Eur. Phys. J. Plus {\bf 132}, 248
(2017)}
\bibitem{moczetal}{\small P. Mocz {\it et al.}, Mon. Not.
R. astr. Soc. {\bf  471}, 4559 (2017)}
\bibitem{eby2}{\small J. Eby, M. Ma, P. Suranyi, L.C.R. Wijewardhana, 
JHEP {\bf 01}, 066 (2018)}
\bibitem{phi6}{\small P.H. Chavanis, Phys. Rev. D {\bf 98}, 023009 (2018)}
\bibitem{abriljeans}{\small A. Su\'arez, P.H. Chavanis, arXiv:1710.10486}
\bibitem{zhang}{\small J. Zhang, Y.L. Sming Tsai, J.L. Kuo, K. Cheung, M.C.
Chu, Astrophys. J. {\bf 853}, 51 (2018)}
\bibitem{moczchavanis}{\small P. Mocz, L. Lancaster, A. Fialkov, F. Becerra,
P.H. Chavanis, Phys. Rev. D {\bf 97}, 083519 (2018)}
\bibitem{desjacques}  {\small V. Desjacques, A. Kehagias, A. Riotto,
Phys. Rev. D {\bf 97}, 023529 (2018)}
\bibitem{pdu}{\small P.H. Chavanis, Phys. Dark Univ.  {\bf 22}, 80 (2018)}
\bibitem{predictive}{\small P.H. Chavanis, arXiv:1810.08948}
\bibitem{cusp}{\small B. Moore, T. Quinn, F. Governato, J. Stadel, G. Lake,
MNRAS {\bf 310}, 1147 (1999)}
\bibitem{satellites1}{\small G. Kauffmann, S.D.M. White, B.
Guiderdoni, Mon. Not. R. astr. Soc. {\bf 264}, 201 (1993)}
\bibitem{satellites2}{\small A. Klypin, A.V.
Kravtsov, O. Valenzuela, Astrophys. J. {\bf 522}, 82 (1999)}
\bibitem{satellites3}{\small B. Moore, S.
Ghigna, F. Governato, G. Lake, T. Quinn,
J. Stadel, P. Tozzi, Astrophys. J. Letter
{\bf 524}, L19 (1999)}
\bibitem{satellites4}{\small  M. Kamionkowski,
A.R. Liddle, Phys. Rev. Lett. {\bf 84}, 4525 (2000)}
\bibitem{tbtf}{\small M. Boylan-Kolchin, J. S. Bullock, M. Kaplinghat,
MNRAS {\bf 415}, L40 (2011)}
\bibitem{axiverse}{\small A. Arvanitaki, S. Dimopoulos, S. Dubovsky, N. Kaloper,
J.  March-Russell,  Phys. Rev. D {\bf 81}, 123530 (2010)}
\bibitem{revuebec}{\small F. Dalfovo, S. Giorgini, L.P. Pitaevskii, S.
Stringari, Rev. Mod. Phys. {\bf 71}, 463 (1999)}
\bibitem{madelung}{\small E. Madelung, Z. Phys. {\bf 40}, 322 (1927)}
\bibitem{poincare}  {\small H. Poincar\'e, Acta Math. {\bf 7}, 259 (1885)}
\bibitem{chandrabook}{\small S. Chandrasekhar, An Introduction to the Study of
Stellar Structure (Dover, 1958)}
\bibitem{sulem}{\small C. Sulem, P.L. Sulem,  The Nonlinear Schr\"odinger
Equation (Springer, 1999)}
\bibitem{ledoux}{\small P. Ledoux, C.L. Pekeris, Astrophys. J. {\bf 94}, 124
(1941)}
\bibitem{csledoux}{\small P.H. Chavanis, C. Sire, Phys. Rev. E {\bf 73}, 066103
(2006)}
\bibitem{prd}{\small P.H. Chavanis, Phys. Rev. D {\bf 76},
023004 (2007)}
\bibitem{lang}{\small P.H. Chavanis, C. Sire, Phys. Rev. E {\bf 69}, 016116
(2004)}
\bibitem{forthcoming}{\small P.H. Chavanis, in preparation}






\end{thebibliography}
\end{document}